\documentclass[preprint,aps,shownopacs,preprintnumbers,amsmath,amssymb,nofootinbib,epsf,epsfig]
{revtex4}

\usepackage{amstext,amssymb}
\usepackage{amsmath}
\usepackage{graphicx}
\usepackage[hyperfootnotes=true]{hyperref}
\usepackage{color}
\usepackage{array,multirow}
\usepackage{slashed} 
\usepackage{multirow}
\newcommand{\be}{\begin{equation}}
\newcommand{\ee}{\end{equation}}
\newcommand{\bea}{\begin{eqnarray}}
\newcommand{\eea}{\end{eqnarray}}
\newcommand{\nn}{\nonumber}

\def\s1{\hat s}
\def\para{\parallel}

\def\U1mt{U(1)_{L_\mu-L_\tau}}

\def\ol{\overline}
\def\nl{\nonumber\\}

\usepackage{slashed}
\global\long\def\d{\partial}

\begin{document}
\title{\bf Exploring dark matter, neutrino mass and $R_{K^{(*)},\phi}$ anomalies in  $L_{\mu}-L_{\tau}$ model }
\author{Shivaramakrishna Singirala$^a$}
\email{krishnas542@gmail.com}
\author{Suchismita Sahoo$^b$}
\email{suchismita8792@gmail.com}

\author{Rukmani Mohanta$^a$}
\email{rmsp@uohyd.ernet.in}
\affiliation{$^a$School of Physics,  University of Hyderabad, Hyderabad-500046,  India\\
$^b$Theoretical Physics Division, Physical Research Laboratory, Ahmedabad-380009, India}


\begin{abstract}
We investigate Majorana dark matter in a new variant of $U(1)_{L_{\mu}-L_{\tau}}$ gauge extension of Standard Model, where the scalar sector is enriched with an inert doublet and a $(\bar{3},1,1/3)$ scalar leptoquark. We compute the WIMP-nucleon cross section in leptoquark portal and the relic density mediated by inert doublet components, leptoquark and the new $Z^{\prime}$ boson. We constrain the parameter space consistent with Planck limit on relic density, PICO-60 and LUX bounds on spin-dependent direct detection cross section.  Furthermore, we constrain the new couplings from the present experimental data on ${\rm Br}(\tau \to \mu \nu_\tau \bar \nu_\mu)$, ${\rm Br}( B \to X_s \gamma)$,  ${\rm Br}(\bar B^0 \to \bar K^0 \mu^+ \mu^-)$, ${\rm Br}(B^+ \to K^+ \tau^+ \tau^-)$ and $B_s-\bar{B_s}$ mixing, which occur at one-loop level in the presence of $Z^\prime$  and leptoquark. Using the allowed parameter space, we estimate the form factor independent $P_{4,5}^\prime$ observables and the lepton non-universality parameters $R_{K}$, $R_{K^*}$ and $R_\phi$.  We also briefly discuss about the neutrino mass generation at one-loop level and the viable parameter region to explain current neutrino oscillation data.
\end{abstract}

\maketitle
\flushbottom
\section{Introduction} 

Though the experimental measured values of various physical observables are in excellent agreement with the  Standard Model (SM) predictions,  there are many open unsolved problems like the matter-antimatter asymmetry,  hierarchy problem and the dark matter (DM) content of the universe  etc., which make ourselves believe that there is something beyond the SM. In this regard, the study of rare semileptonic $B$ decay processes provide an ideal testing ground to critically test the SM and to look for possible extension of it.  Although, so far we have not observed any clear indication of new physics (NP) in the $B$ sector, there are several physical observables associated with flavor changing neutral current (FCNC) $b \to s l^+ l^-$ processes which have $(2-4)\sigma$ \cite{Aaij:2017vbb, Aaij:2014ora, Aaij:2013aln, Aaij:2013qta, Aaij:2014pli, Langenbruch:2015dqz} discrepancies. Especially, the observation of $3\sigma$ anomaly in the $P_5^\prime$ angular observables  \cite{Aaij:2013qta} and the decay rate \cite{Aaij:2014pli} of  $ B^0 \to  K^{*0} \mu^+ \mu^-$ processes have attracted a lot of attention in recent times. 
The decay rate of $B_s \to \phi \mu^+ \mu^-$ has also $3\sigma$ deviation compared to its SM prediction \cite{Aaij:2013aln}. 
Furthermore, the LHCb Collaboration has observed the violation of lepton  universality  in $B^+ \to K^+ l^+ l^-$ process  in the low $q^2 \in [1,6]~{\rm GeV}^2$ region \cite{Aaij:2014ora}
\bea
 R_K^{\rm Expt} = \frac{{\rm Br}(B^+ \to K^+ \mu^+ \mu^-)}{{\rm Br}(B^+ \to K^+ e^+ e^-)}=0.745^{+0.090}_{-0.074} \pm 0.036, 
\eea
 which has a $2.6\sigma$ deviation from the corresponding SM result \cite{Bobeth:2007dw}
 \bea
 R_K^{\rm SM} = 1.0003 \pm 0.0001.
 \eea  
  In addition, an analogous lepton non-universality (LNU)  parameter $(R_{K^*})$ has also been observed in  $ B^0 \to   K^{*0} l^+ l^-$ processes \cite{Aaij:2017vbb}
\bea
R_{K^*}^{\rm Expt} =\frac{{\rm Br}( B^0 \to   K^{*0} \mu^+ \mu^-)}{{\rm Br}( B^0 \to  K^{*0} e^+ e^-)}&=& 0.66^{+0.11}_{-0.07} \pm 0.03, ~~~~q^2 \in [0.045, 1.1]~{\rm GeV}^2, \nn \\
&=& 0.69^{+0.11}_{-0.07} \pm 0.05, ~~~~q^2 \in [1.1, 6]~{\rm GeV}^2,
\eea
which correspond to the deviation of $2.2 \sigma$ and $2.4 \sigma$ from their SM predictions \cite{Capdevila:2017bsm}
\bea 
 R_{K^*}^{\rm SM}\big |_{q^2 \in [0.045,1.1]~ {\rm GeV^2}}=0.92\pm 0.02, ~~~~R_{K^*}^{\rm SM}\big |_{q^2 \in [1.1, 6]~ {\rm GeV^2}}=1.00\pm 0.01. 
 \eea
 
To resolve the above $b \to s l l$ anomalies, we extend the SM gauge group $SU(3)_C \times SU(2)_L \times U(1)_Y$ with a local $U(1)_{L_\mu-L_{\tau}}$ symmetry.  The anomaly free $L_{\mu}-L_{\tau}$ gauge extensions \cite{He:1990pn,He:1991qd} are captivating with minimal new particles and parameters, rich in phenomenological perspective. The model is quite simple in structure, suitable to study the phenomenology of DM, neutrino and also the flavor anomalies \cite{Crivellin:2015mga, Patra:2016shz,Biswas:2016yan,Kamada:2018zxi,Bauer:2018egk}.  It is well explored in dark matter context in literature \cite{Patra:2016shz,Biswas:2016yan,Kamada:2018zxi,Bauer:2018egk}, in the gauge and scalar portals.  In the literature \cite{Crivellin:2015lwa, Das:2017ski, Das:2017deo, Das:2017flq}, the DM, neutrino and flavor phenomenology are also investigated in several  $U(1)$ extended models. The approach of adding color triplet particles to shed light on the flavor sector thereby connecting with dark sector is interesting. Leptoquarks (LQ) are not only advantageous in addressing the flavor anomalies, but also act as a mediator between the visible and dark sector. Few works were already done with this motivation \cite{Baek:2017sew,Mandal:2018czf,Arcadi:2017kky,Allahverdi:2017edd}.

Leptoquarks are hypothetical color triplet gauge particles, with either spin-0 (scalar) or spin-1 (vector), which connect the quark and lepton sectors and thus, carry both  baryon and lepton numbers simultaneously.  They can arise from various extended standard model scenarios \cite{Georgi:1974sy, Georgi:1974my, Langacker:1980js, Fritzsch:1974nn, Pati:1974yy, Pati:1973uk, Pati:1973rp,   Shanker:1981mj, Shanker:1982nd, Schrempp:1984nj, Gripaios:2009dq, Kaplan:1991dc}, which treat quarks and leptons on equal footing, such as the grand unified theories (GUTs) \cite{Georgi:1974sy, Fritzsch:1974nn, Langacker:1980js, Georgi:1974my}, color $SU(4)$ Pati-Salam model \cite{Pati:1974yy, Pati:1973uk, Pati:1973rp, Shanker:1981mj, Shanker:1982nd}, extended technicolor model \cite{Schrempp:1984nj, Gripaios:2009dq} and the composite models of  quark and lepton \cite{Kaplan:1991dc}.  In this article, we study a new version of $U(1)_{L_\mu-L_\tau}$ gauge extension of SM with a $(\bar{3},1,1/3)$ scalar LQ (SLQ) and an inert doublet, to study the phenomenology of dark matter, neutrino mass generation and compute the flavor observables on a single platform. The SLQ mediates the annihilation channels contributing to relic density and also plays a crucial role in direct searches as well, providing a spin-dependent WIMP-nucleon cross section which is quite sensitive to the recent and ongoing direct detection experiments such as PICO-60 and LUX. The $Z^\prime$ gauge boson of extended $U(1)$ symmetry and the SLQ  also play an important role in settling the known issues of flavor sector. In this regard, we would  like to investigate whether the observed anomalies in the rare leptonic/semileptonic decay processes mediated by $b \to s l^+ l^-$  transitions, can be explained in the present framework. We  analyze the implications of the model on both the DM and flavor sectors, in particular on   $ B \to  V l^+ l^-~(V=K^*, \phi)$ decay modes. In literature \cite{Alok:2017sui, Becirevic:2017jtw, Hiller:2017bzc, DAmico:2017mtc, Becirevic:2016yqi, Bauer:2015knc, Li:2016vvp, Calibbi:2015kma, Freytsis:2015qca, Dumont:2016xpj, Dorsner:2016wpm, Varzielas:2015iva, Dorsner:2011ai, Davidson:1993qk, Saha:2010vw,  Mohanta:2013lsa, Sahoo:2015fla, Sahoo:2015pzk, Sahoo:2015qha, Sahoo:2015wya, Kosnik:2012dj, Chauhan:2017ndd, Becirevic:2018afm, Angelescu:2018tyl},  there were many attempts being made to explain the observed anomalies of rare $B$ decays in the scalar leptoquark model. 

The paper is structured as follows. We describe the particle content, relevant Lagrangian and interaction terms, pattern of symmetry breaking in section-II. We derive the mass eigenstates of the new fermions and the scalar spectrum in section-III. We then provide a detailed study of DM phenomenology in prospects of relic density and direct detection observables in section-IV. The mechanism of generating light neutrino mass at one-loop level, consistent with the current oscillation data is illustrated in section V.  Section-VI contains the additional constraint on the new parameters obtained from  the existing anomalies of the flavor sector, like   ${\rm Br}(\tau \to \mu \nu_\tau \bar \nu_\mu)$, ${\rm Br}( B \to X_s \gamma)$,  ${\rm Br}(\bar B^0 \to \bar K^0 \mu^+ \mu^-)$, ${\rm Br}( B^+ \to K^+ \tau \tau)$ and $B_s-\bar{B_s}$ mixing. We then investigate the impact of additional $U(1)_{L_\mu-L_\tau}$ gauge symmetry on the $R_{K^{(*)}}$, $R_\phi$ LNU parameters and optimized $P_{4,5}^\prime$ observables in section-VII.  We summarize our  findings in Section-VIII.
%
\section{New $L_{\mu}-L_{\tau}$ model with a scalar leptoquark}
We study the well known anomaly free $U(1)_{L_{\mu}-L_{\tau}}$ extension of SM containing three additional neutral fermions $N_{e}, N_{\mu}, N_{\tau}$, with $L_{\mu}-L_{\tau}$ charges $0,1$ and $-1$ respectively. A scalar singlet $\phi_2$, charged $+2$ under the new $U(1)$ is added to spontaneously break the local $U(1)_{L_{\mu}-L_{\tau}}$ gauge symmetry. We also introduce an inert doublet $\eta$ and a scalar leptoquark  $S_{1}(\bar{3},1,1/3)$ with $L_{\mu}-L_{\tau}$ charges $0$ and $-1$ to the scalar content of the model. We impose an additional $Z_2$ symmetry under which all the new fermions, $\eta$ and the leptoquark are odd and rest are even. The particle content and their corresponding charges are displayed in Table. \ref{mutau_model}\,.
\begin{table}[htb]
\begin{center}
\begin{tabular}{|c|c|c|c|c|}
	\hline
			& Field	& $ SU(3)_C \times SU(2)_L\times U(1)_Y$	& $U(1)_{L_{\mu}-L_{\tau}}$	& $Z_2$\\
	\hline
	\hline
	Fermions	& $Q_L \equiv(u, d)^T_L$			& $(\textbf{3},\textbf{2},~ 1/6)$	& $0$	& $+$\\
			& $u_R$							& $(\textbf{3},\textbf{1},~ 2/3)$	& $0$ & $+$	\\
			& $d_R$							& $(\textbf{3},\textbf{1},~-1/3)$	& $0$	& $+$\\
			& $e_L \equiv(\nu_e,~e)^T_L$	& $(\textbf{1},\textbf{2},~  -1/2)$	&  $0$	& $+$\\
			& $e_R$							& $(\textbf{1},\textbf{1},~  -1)$	&  $0$	& $+$\\
						& $\mu_L \equiv(\nu_\mu,~\mu)^T_L$	& $(\textbf{1},\textbf{2},~  -1/2)$	&  $1$	& $+$\\
			& $\mu_R$							& $(\textbf{1},\textbf{1},~  -1)$	&  $1$	& $+$\\
						& $\tau_L \equiv(\nu_\tau,~\tau)^T_L$	& $(\textbf{1},\textbf{2},~  -1/2)$	&  $-1$	& $+$\\
			& $\tau_R$							& $(\textbf{1},\textbf{1},~  -1)$	&  $-1$	& $+$\\
			& $N_{e}$						& $(\textbf{1},\textbf{1},~   0)$	&  $0$	& $-$\\
			& $N_{\mu}$						& $(\textbf{1},\textbf{1},~   0)$	&  $1$	& $-$\\
			& $N_{\tau}$						& $(\textbf{1},\textbf{1},~   0)$	&  $-1$	& $-$\\
	\hline
	Scalars	& $H$							& $(\textbf{1},\textbf{2},~ 1/2)$	&   $0$	& $+$\\
		& $\eta$							& $(\textbf{1},\textbf{2},~ 1/2)$	&   $0$	& $-$\\
			& $\phi_2$						& $(\textbf{1},\textbf{1},~   0)$	&  $2$	& $+$\\  
			& $S_1$						& $(\bar{\textbf{3}},\textbf{1},~   1/3)$	&  $-1$	& $-$\\    
	\hline
	\hline
\end{tabular}
\caption{Fields and their charges of the proposed $U(1)_{L_{\mu}-L_{\tau}}$ model.}
\label{mutau_model}
\end{center}
\end{table}

The Lagrangian of the present model can be written as
\begin{align}
{\cal L} &={\cal L}_{\rm SM}  -{1 \over 4} Z'_{\mu\nu} Z^{'\mu\nu} -g_{\mu\tau}\overline{\mu}_L \gamma^\mu \mu_L Z_\mu^\prime 
            - g_{\mu\tau} \overline{\mu}_R \, \gamma^\mu \mu_R Z_\mu^\prime  +g_{\mu\tau}\overline{\tau}_L  \gamma^\mu \tau_L Z_\mu^\prime 
           + g_{\mu\tau} \overline{\tau}_R \, \gamma^\mu \tau_R  Z_\mu^\prime \nl
            &+\overline{N}_{e} i \slashed{\d}\,N_{e}+ \overline{N}_{\mu} \left(i \slashed{\d} - g_{\mu\tau} \,Z_\mu^\prime \gamma^\mu \right)N_{\mu}	 +  \overline{N}_{\tau} \left(i \slashed{\d} + g_{\mu\tau} \,Z_\mu^\prime \gamma^\mu \right)N_{\tau}  - \frac{f_\mu}{2}\left({\ol{N_{\mu}^c}} N_{\mu}\phi_2^{\dagger}+{\rm h.c.}\right) \nl &- \frac{f_\tau}{2}\left({\ol{N_{\tau}^c}} N_{\tau}\phi_2 + {\rm h.c.} \right) -\frac{1}{2}M_{ee}\ol{N_{e}^c} N_{e} -\frac{1}{2}M_{\mu\tau}(\ol{N_{\mu}^c} N_{\tau} + \ol{N_{\tau}^c} N_{\mu}) - \sum_{q=d,s,b} (y_{q R}\; \ol{d_{qR}^c} S_1 N_{\mu} + {\rm{h.c.}})  \nl
&- \sum_{l=e,\mu,\tau} (Y_{\beta l} (\ol{\ell_L})_\beta \tilde \eta N_{lR} + {\rm{h.c}})+ \left|\left(i \d_\mu - \frac{g}{2} \boldsymbol{\tau}^a\cdot\bold{W}_\mu^a  -\frac{g^{\prime}}{2}B_\mu\right) \eta \right|^2 
+ \left| \left(i \d_\mu -\frac{g^{\prime}}{3}B_\mu + g_{\mu\tau} \,Z_\mu^\prime  \right) S_1\right|^2
\nl &
+ \left| \left(i \d_\mu -2 g_{\mu\tau} \,Z_\mu^\prime  \right) \phi_2\right|^2 
- V(H,\eta,\phi_2,S_1),
\label{eq:Lag}
\end{align}
where the scalar potential $V$ is
\begin{align}
V(H,\eta,\phi_2,S_1) &=  \mu^2_H  H^\dagger H + \lambda_H (H^\dagger H)^2  + \mu_{\eta}(\eta^{\dagger}\eta)+ \lambda_{H\eta}(H^{\dagger}H)(\eta^{\dagger}\eta) + \lambda_{\eta}(\eta^{\dagger}\eta)^2 + \lambda'_{H\eta}(H^{\dagger}\eta)(\eta^{\dagger}H) \nn\\
      +& \frac{\lambda''_{H\eta}}{2}\left[(H^{\dagger}\eta)^2 + {\rm h.c.}\right] 
 + \mu^2_{2} (\phi^\dagger_2 \phi_2) + \lambda_{2} (\phi^\dagger_2 \phi_2)^2 
      +\mu^2_{S} ({S_1}^\dagger {S_1})  +\lambda_{S} ({S_1}^\dagger {S_1})^2 \nonumber \\
      +&  \left[\lambda_{H2} (\phi^\dagger_2 \phi_2) + \lambda_{HS} (S^\dagger_1 S_1)\right](H^\dagger H)+  \lambda_{S2}(\phi^\dagger_2 \phi_2) ({S_1}^\dagger {S_1}) +\lambda_{\eta2}(\phi^\dagger_2 \phi_2) (\eta^\dagger \eta) \nn\\
      +& \lambda_{S\eta}({S_1}^\dagger S_1) (\eta^\dagger \eta).
\label{eq:potential}
\end{align}
The gauge symmetry $SU(2)_L \times U(1)_Y\times U(1)_{L_\mu-L_\tau}$ is spontaneously broken to $SU(2)_L \times U(1)_Y$ by assigning a VEV $v_2$ to the complex singlet $\phi_2$. Then the SM Higgs doublet breaks the SM gauge group to low energy theory by obtaining a VEV $v$. The new neutral gauge boson $Z^{\prime}$ associated with the $U(1)$ extension absorbs the massless pseudoscalar in $\phi_2$ to become massive. The neutral components of the fields $H$ and $\phi_2$ can be written in terms of real scalars and pseudoscalars as
\begin{align}
&H^0 =\frac{1}{\sqrt{2} }(v+h)+  \frac{i}{\sqrt{2} } A^0\,, \nonumber \\
& \phi_2 = \frac{1}{\sqrt{2} }(v_2+h_2)+  \frac{i}{\sqrt{2} } A_2\,.\end{align}
The inert doublet is denoted by $\eta = \begin{pmatrix}
		 \eta^+		\\
		 \eta^0	\\
	\end{pmatrix}$, with $\eta^0 = \displaystyle{\frac{\eta_e+i\eta_o}{\sqrt{2}}}$. The masses of its charged and neural components are given by
\begin{eqnarray}
M_{\eta^+}^2 &=& \mu_{\eta}^2  + \frac{ \lambda_{ H\eta}}{2} v^2 +  \frac{ \lambda_{ \eta2}}{2} v^2_2\;, \nn\\
M_{\eta_e}^2 &=& \mu_{\eta}^2  +  \frac{ \lambda_{\rm \eta2}}{2} v^2_2 +  \left(\lambda_{H\eta} + \lambda'_{H\eta} + \lambda''_{H\eta}\right)\frac{v^2}{2}\;, \nn\\
M_{\eta_o}^2 &=& \mu_{\eta}^2  +  \frac{ \lambda_{\rm \eta2}}{2} v^2_2 + \left(\lambda_{H\eta} + \lambda'_{H\eta} - \lambda''_{H\eta}\right)\frac{v^2}{2}.
\end{eqnarray}
The masses obtained by the colored scalar and the gauge boson $Z^{\prime}$ are
\begin{eqnarray}
M_{S_1}^2 &=& 2\mu_S^2 + \lambda_{HS}v^2 + \lambda_{S2}v_2^2\;,\nn\\
M_{Z^\prime} &=& 2v_2 g_{\mu\tau}.
\end{eqnarray}
In the whole discussion of the results, we consider  the benchmark values for the masses of the scalar spectrum as $(M_{S_1},M_{\eta^+}, M_{\eta_{e,o}}) = (1.2,2,1.5)$ TeV.
\section{Mixing in the fermion and scalar sector}
The fermion and scalar mass matrices take the form
\begin{align}
	M_N
	=
	\begin{pmatrix}
		 \frac{1}{\sqrt{2}}f_{\mu}v_2	& M_{\mu\tau}	\\
		 M_{\mu\tau}	& \frac{1}{\sqrt{2}}f_{\tau}v_2				\\
	\end{pmatrix} \;,
    \quad M_S
	=
	\begin{pmatrix}
		 2 \lambda_H v^2   & {\lambda}_{H2} {v}v_2  \\
 {\lambda}_{H2} {v}v_2  & 2 \lambda_{2} v^2_2				\\
	\end{pmatrix} \;.
\end{align}
One can diagonalize the above mass matrices by $U_{\alpha(\zeta)}^T M_{N(S)} U_{\alpha(\zeta)} = {\rm{diag}}~[M_{N_{-}(H_{1})},M_{N_{+}(H_{2})}]$, where \begin{align}
U_\theta
	=
	\begin{pmatrix}
		 \cos{\theta}	& \sin{\theta}	\\
		 -\sin{\theta}	& \cos{\theta}	\\
	\end{pmatrix},
	\label{massMatrix1}
\end{align}
with $\zeta = \frac{1}{2}\tan^{-1}\left(\displaystyle{\frac{\lambda_{H2} v v_2}{\lambda_2 v^2_2 - \lambda_H v^2}}\right)$ and $\alpha = \frac{1}{2}\tan^{-1}\left(\displaystyle{\frac{2M_{\mu\tau}}{(f_\tau - f_{\mu})(v_2/\sqrt{2})}}\right)$.\\\\
We denote the scalar mass eigenstates as $H_1$ and $H_2$, with $H_1$ is assumed to be observed Higgs at LHC with $M_{H_1} = 125.09$ GeV and $v= 246$ GeV. The mixing parameter $\zeta$ is taken minimal to stay with LHC limits on Higgs decay width.
We indicate $N_-$ and $N_+$ to be the fermion mass eigenstates, with the lightest one ($N_-$) as the probable dark matter candidate in the present work. 
\section{Dark matter phenomenology}
\subsection{Relic abundance}
\begin{figure}[thb]
\begin{center}
\includegraphics[width=0.3\linewidth]{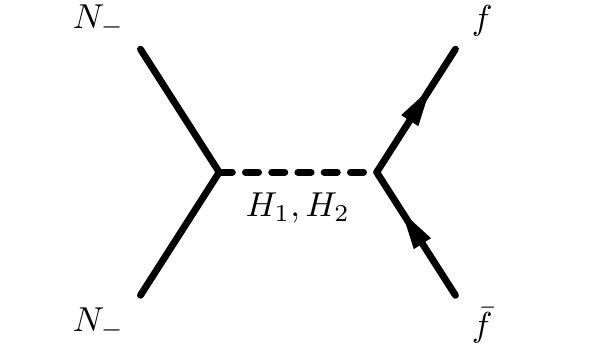}
\vspace{0.01 cm}
\includegraphics[width=0.3\linewidth]{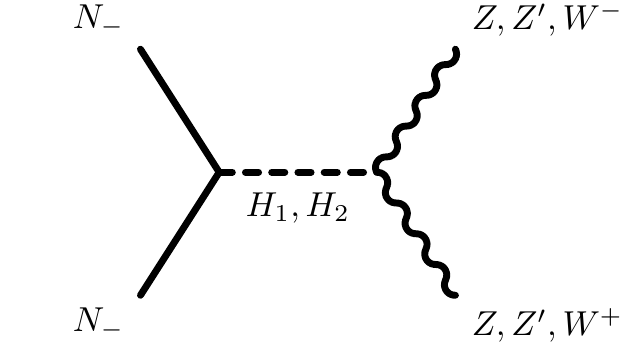}
\vspace{0.01 cm}
\includegraphics[width=0.3\linewidth]{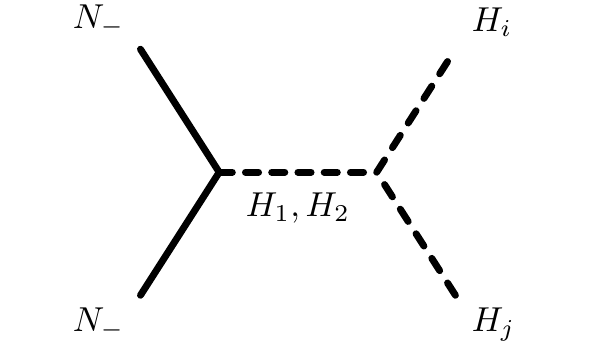}
\vspace{0.01 cm}
\includegraphics[width=0.3\linewidth]{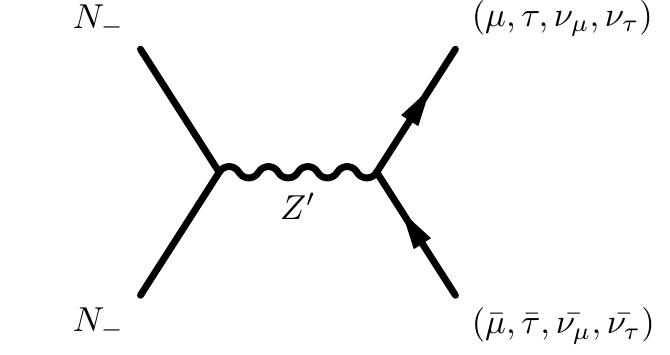}
\vspace{0.01 cm}
\includegraphics[width=0.3\linewidth]{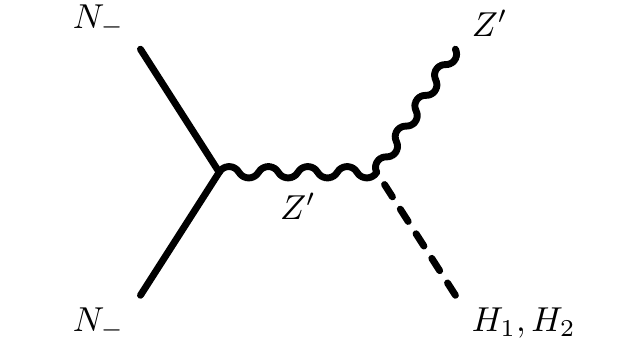}\\
\includegraphics[width=0.3\linewidth]{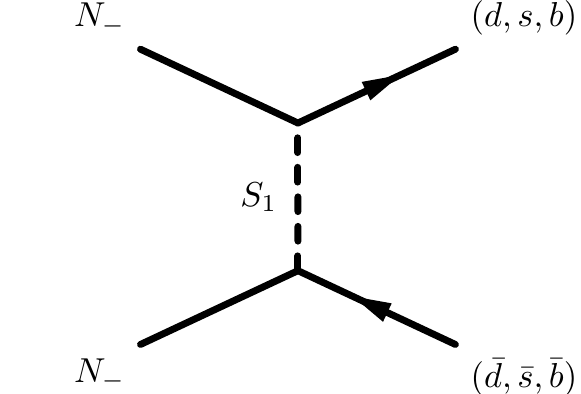}
\vspace{0.01 cm}
\includegraphics[width=0.3\linewidth]{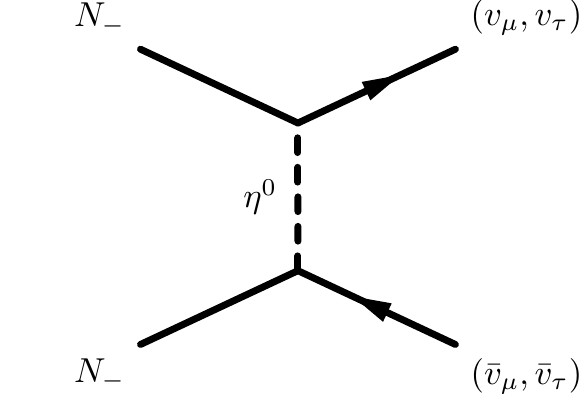}
\vspace{0.01 cm}
\includegraphics[width=0.3\linewidth]{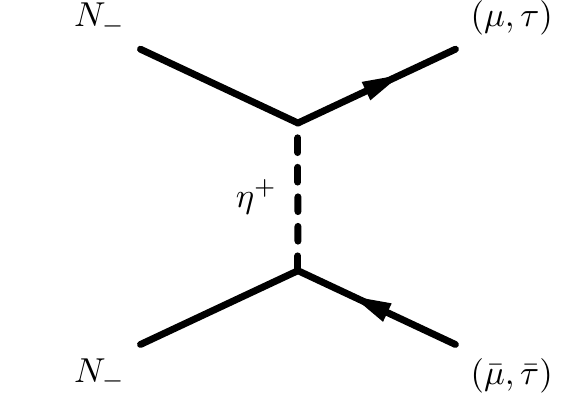}
\caption{Feynman diagrams contributing to relic density.}
\label{relicfeyn}
\end{center}
\end{figure}

The model allows the dark matter ($N_-$) to have gauge and scalar mediated annihilation channels. The possible contributing diagrams are provided in Fig. \ref{relicfeyn} which are mediated by $(H_{1},H_{2},\eta^+,\eta^0,S_1,Z^{\prime})$. Majorana DM in $H_{1,2}$ portal (upper row in Fig. \ref{relicfeyn}\;) has already been well explored in literature \cite{Singirala:2017cch,Nanda:2017bmi}. Here, we focus on $(Z^{\prime},S_1,\eta)$-mediated channels (middle and bottom rows in Fig. \ref{relicfeyn}\;) contributing to DM observables, which we later make connection with radiative neutrino mass as well as flavor observables. 
\begin{figure}[thb]
\begin{center}
\includegraphics[width=0.48\linewidth]{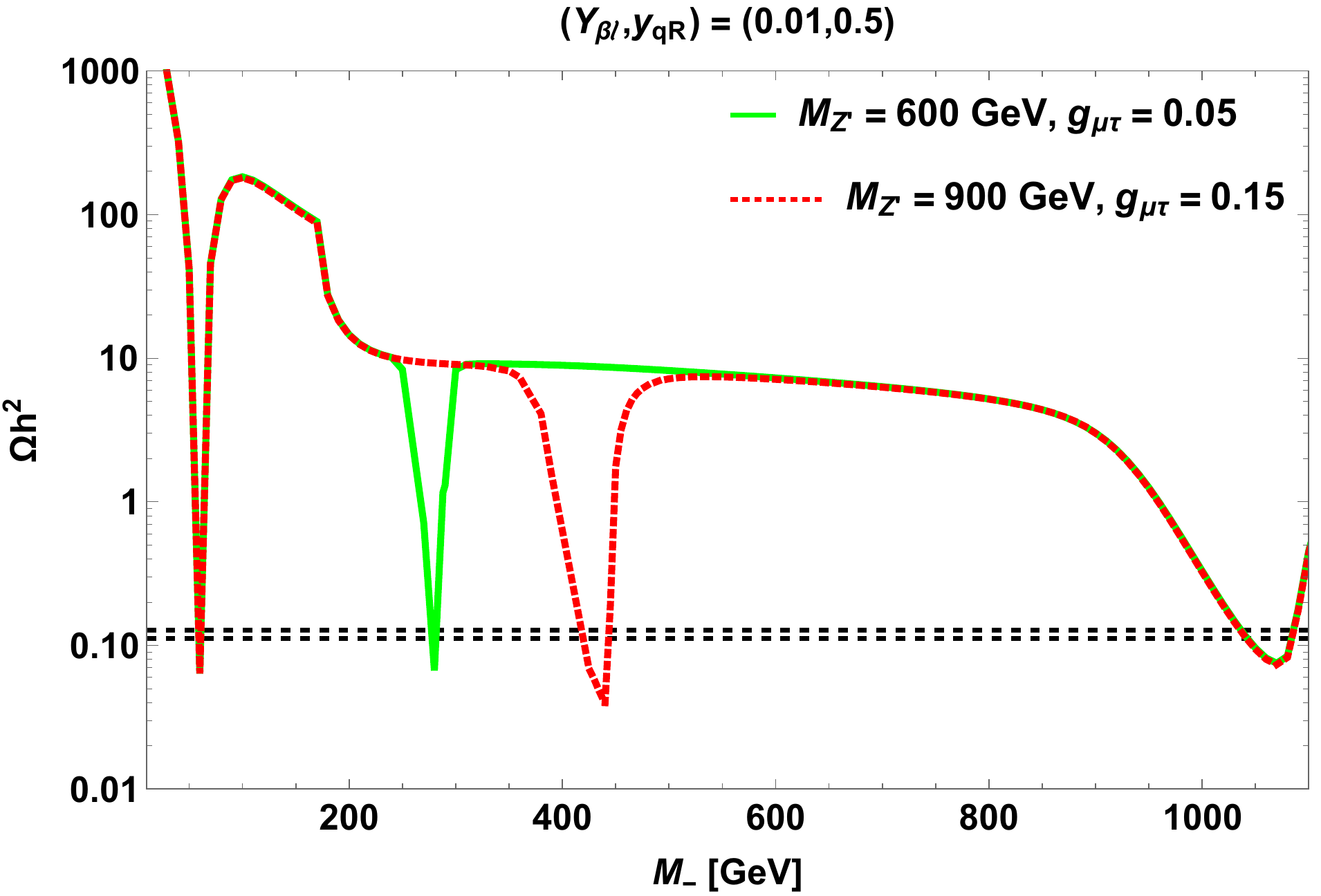}
\vspace{0.1 cm}
\includegraphics[width=0.48\linewidth]{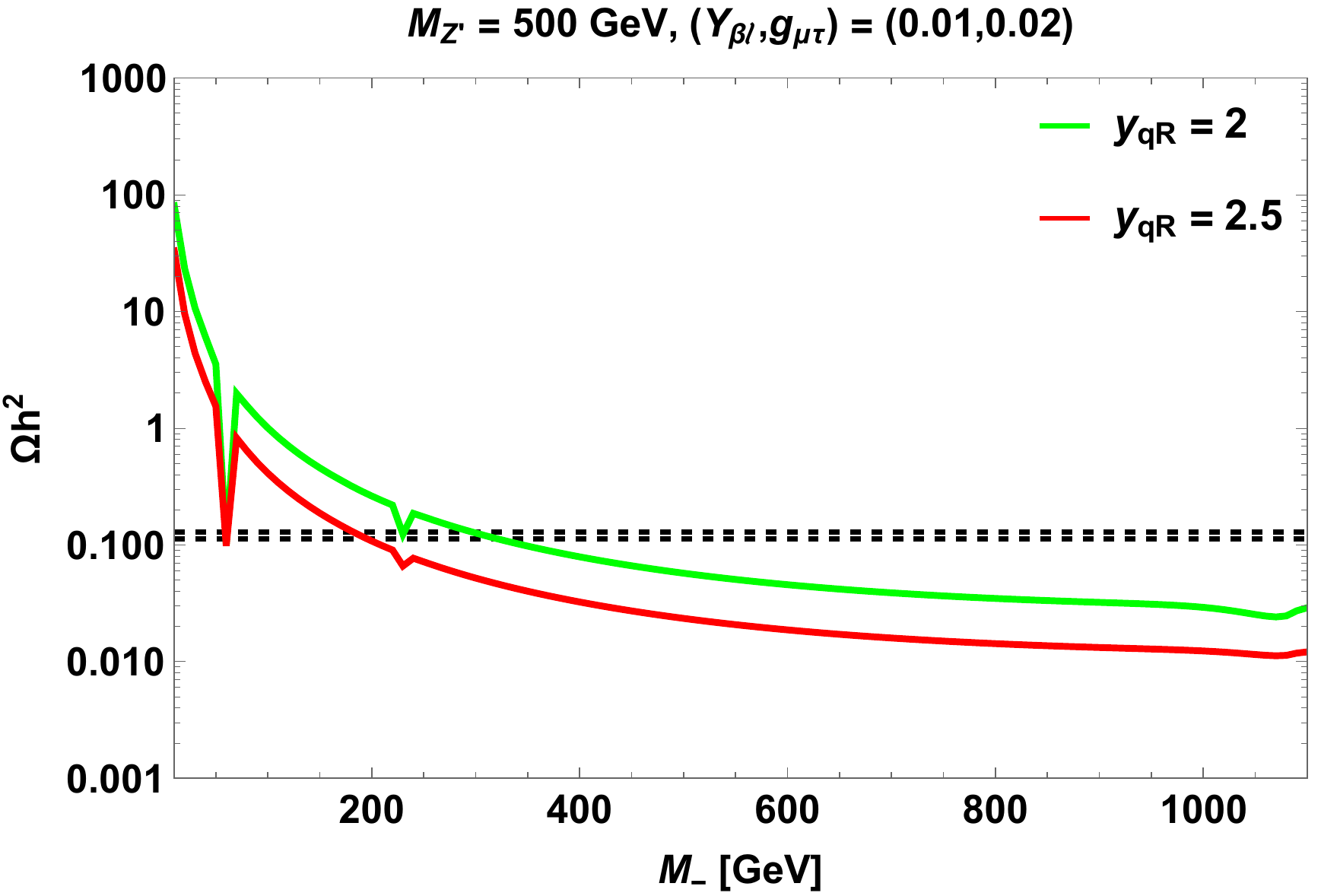}
\caption{Behavior of relic density plotted against DM mass with $M_{H_2} = 2.2$ TeV, shown with varying $M_{Z^\prime}$ and  $g_{\mu\tau}$ (left panel) and $y_{qR}$ (right panel). Black horizontal dotted lines denote the $3\sigma$ range of Planck limit \citep{Aghanim:2018eyx}.}
\label{relic_curve}
\end{center}
\end{figure}
The relic abundance of dark matter is computed by
\begin{equation}
\label{eq:relicdensity}
\Omega h^2 = \frac{1.07 \times 10^{9} ~{\rm{GeV}}^{-1}}{  M_{\rm{Pl}}\; {g_\ast}^{1/2}}\frac{1}{J(x_f)}\;.
\end{equation}
Here the Planck mass, $M_{\rm{Pl}}=1.22 \times 10^{19} ~\rm{GeV}$ and $g_\ast = 106.75$ denotes the total number of effective relativistic degrees of freedom. The function $J(x_f)$ reads as
\begin{equation}
J(x_f)=\int_{x_f}^{\infty} \frac{ \langle \sigma v \rangle (x)}{x^2} dx.
\end{equation}
The thermally averaged annihilation cross section $\langle \sigma v \rangle$ is given by the expression
\begin{equation}
 \langle\sigma v\rangle (x) = \frac{x}{8 M_{-}^5 K_2^2(x)} \int_{4 M_{-}^2}^\infty \hat{\sigma} \times ( s - 4 M_{-}^2) \ \sqrt{s} \ K_1 \left(\frac{x \sqrt{s}}{M_{-}}\right) ds,
\end{equation}
where $K_1$, $K_2$ denote the modified Bessel functions, $x = M_{-}/T$, with $T$ is the temperature and $\hat  \sigma$ is the dark matter annihilation cross section. The analytical expression for the freeze out parameter $x_f$ is 
\begin{equation}
x_f= \ln \left( \frac{0.038 \ g \ M_\text{Pl} \ M_{-} \ \langle\sigma v\rangle (x_f) }{({g_\ast x_f})^{1/2}} \right).
\end{equation}
Here $g$ represents the number of degrees of freedom of the dark matter particle $N_{-}$.

As seen from the left panel of Fig. \ref{relic_curve}, the relic density with $s$-channel contribution is featured to meet the Planck limit \citep{Aghanim:2018eyx} near the resonance in propagator ($H_1,H_2,Z^\prime$), i.e., near $M_- = \frac{M_{\rm prop}}{2}$.  We restrict our discussion to the mass region (in GeV), $100 \leq M_{Z^\prime}\leq 1000$, $80 \leq M_-\leq 1000$ and also $H_2$ is considered to be sufficiently large such that its resonance doesn't meet the Planck limit below $1$ TeV region of DM mass. Now, in this mass range of DM, the channels mediated by $(Z^\prime,\eta,S_1)$ drive the relic density observable, where the gauge coupling $g_{\mu\tau}$ controls the $s$-channel contribution, while $Y_{\beta l},~ y_{qR}$ are relevant in $t$-channel contributions. 
Hence, the relevant parameters in our investigation are $(M_-,g_{\mu\tau},M_{Z^\prime},Y_{\beta l},y_{qR})$. The effect of these parameters on the relic abundance is made transparent in Fig. \ref{relic_curve}\,, where we have considered $Y_{\beta l} \sim 10^{-2}$, in order to explain neutrino mass at one loop level. Left panel shows the variation of relic density with varying gauge parameters $g_{\mu \tau}$ and $M_{Z^\prime}$, right panel depicts the behavior with varying $y_{qR}$ parameter. No significant constraint on $M_{Z^\prime}$, $g_{\mu\tau}$ parameters is observed, however relic density has an appreciable footprint on $M_--(y_{qR})^2$ parameter space, which will be discussed in the next section.
\subsection{Direct searches}
\begin{figure}[thb]
\begin{center}
\includegraphics[width=0.35\linewidth]{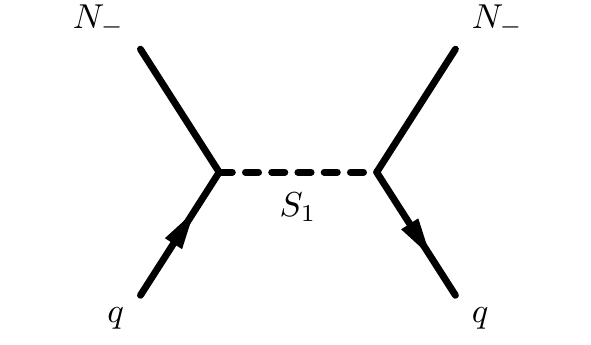}
\includegraphics[width=0.3\linewidth]{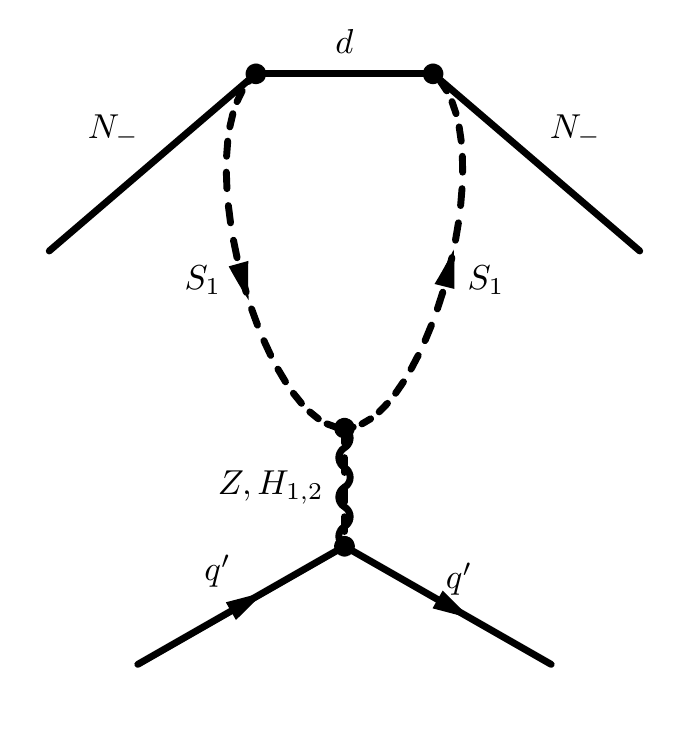}
\vspace{0.01 cm}
\includegraphics[width=0.3\linewidth]{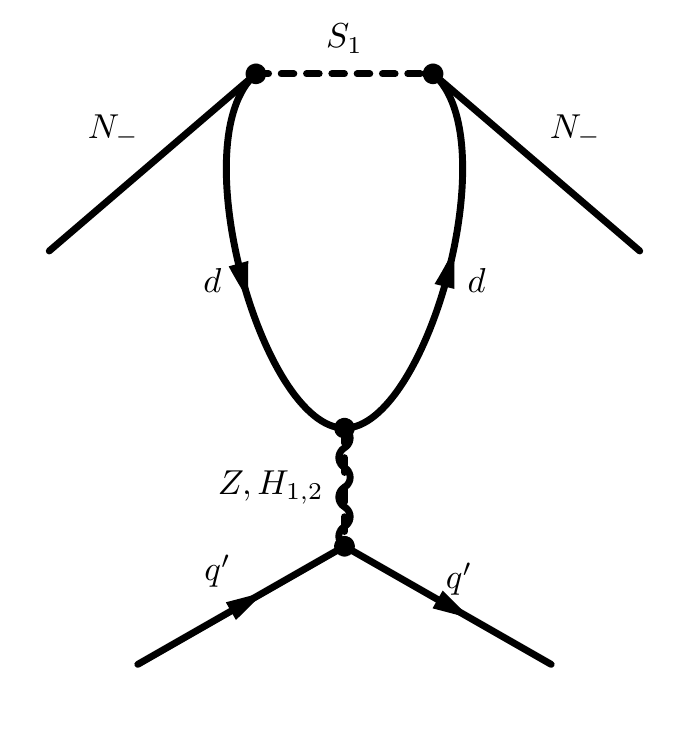}
\caption{Diagrams involving SLQ relevant in direct detection study.}
\label{DD_penguin}
\end{center}
\end{figure}

\begin{figure}[thb]
\begin{center}
\includegraphics[width=0.48\linewidth]{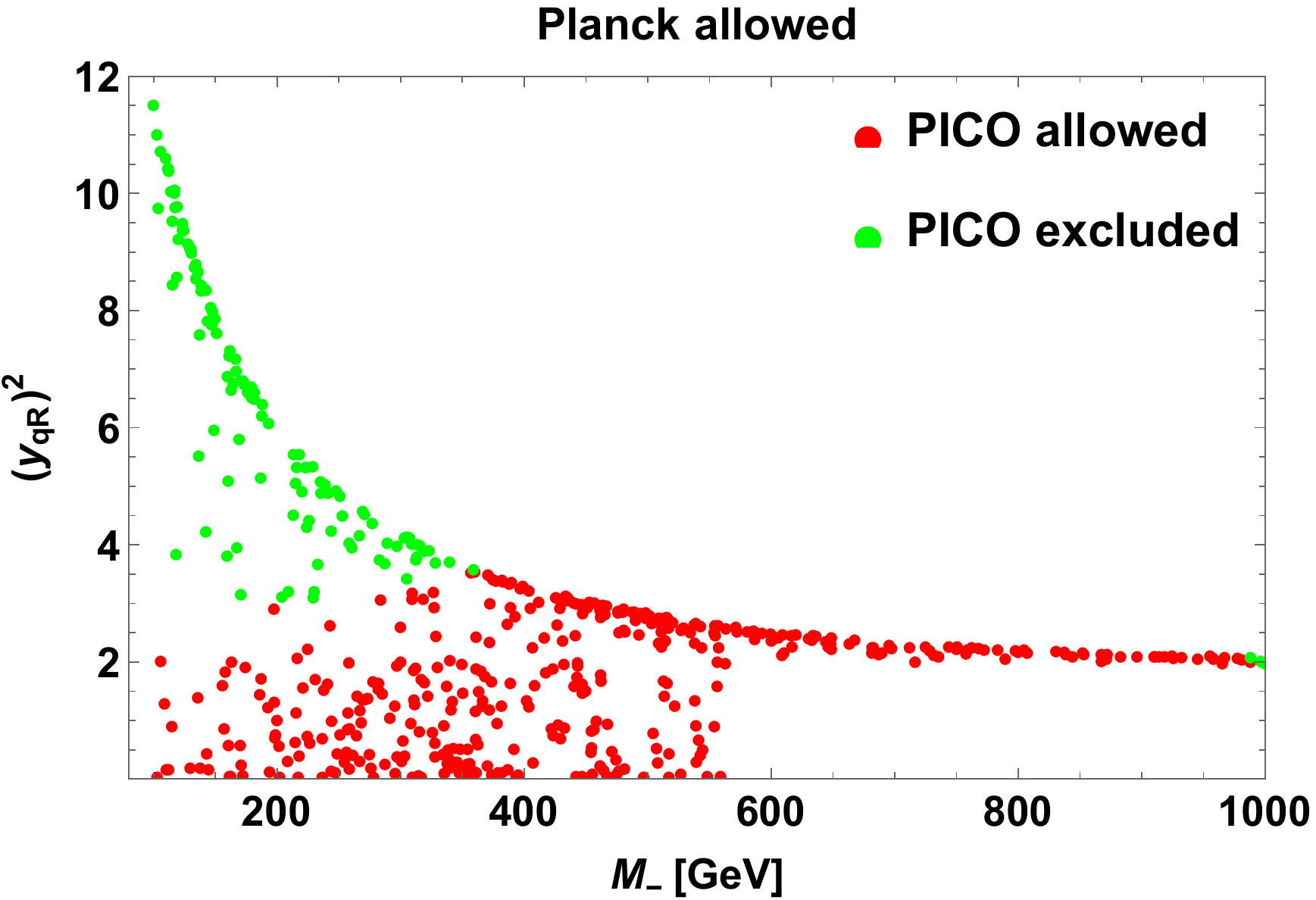}
\vspace{0.01 cm}
\includegraphics[width=0.48\linewidth]{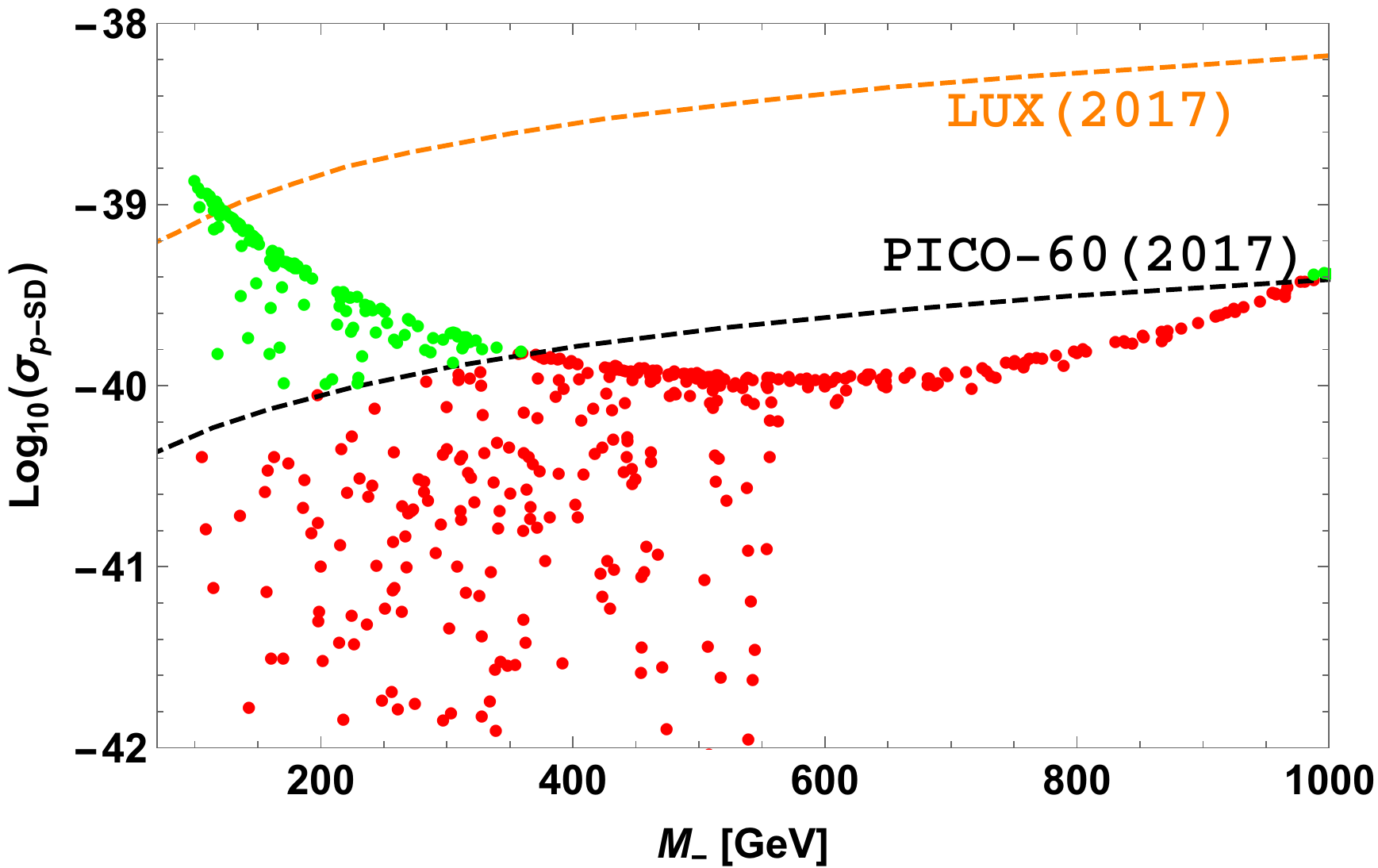}
\caption{Left panel depicts the $M_--(y_{qR})^2$ parameter space consistent upto $3\sigma$ level of Planck limit \citep{Aghanim:2018eyx} on relic density. Right panel gives the SD WIMP-proton cross section as a function of DM mass. Dashed lines represent the recent  bounds obtained from PICO-60 \cite{Amole:2017dex} and LUX \cite{Akerib:2017kat}. Green (red) data points in both the panels represent Planck allowed and PICO excluded (Planck and PICO allowed).}
\label{DDscatter}
\end{center}
\end{figure}

\begin{figure}[thb]
\begin{center}
\includegraphics[width=0.48\linewidth]{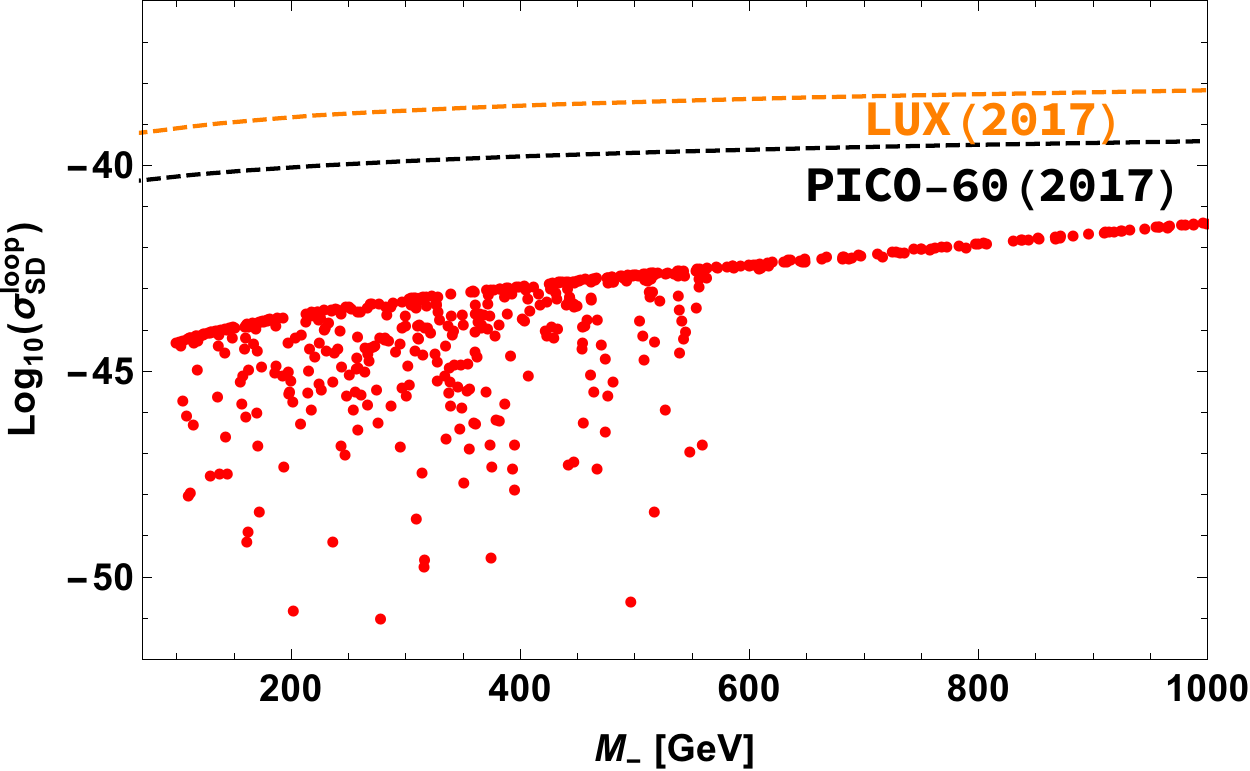}
\includegraphics[width=0.48\linewidth]{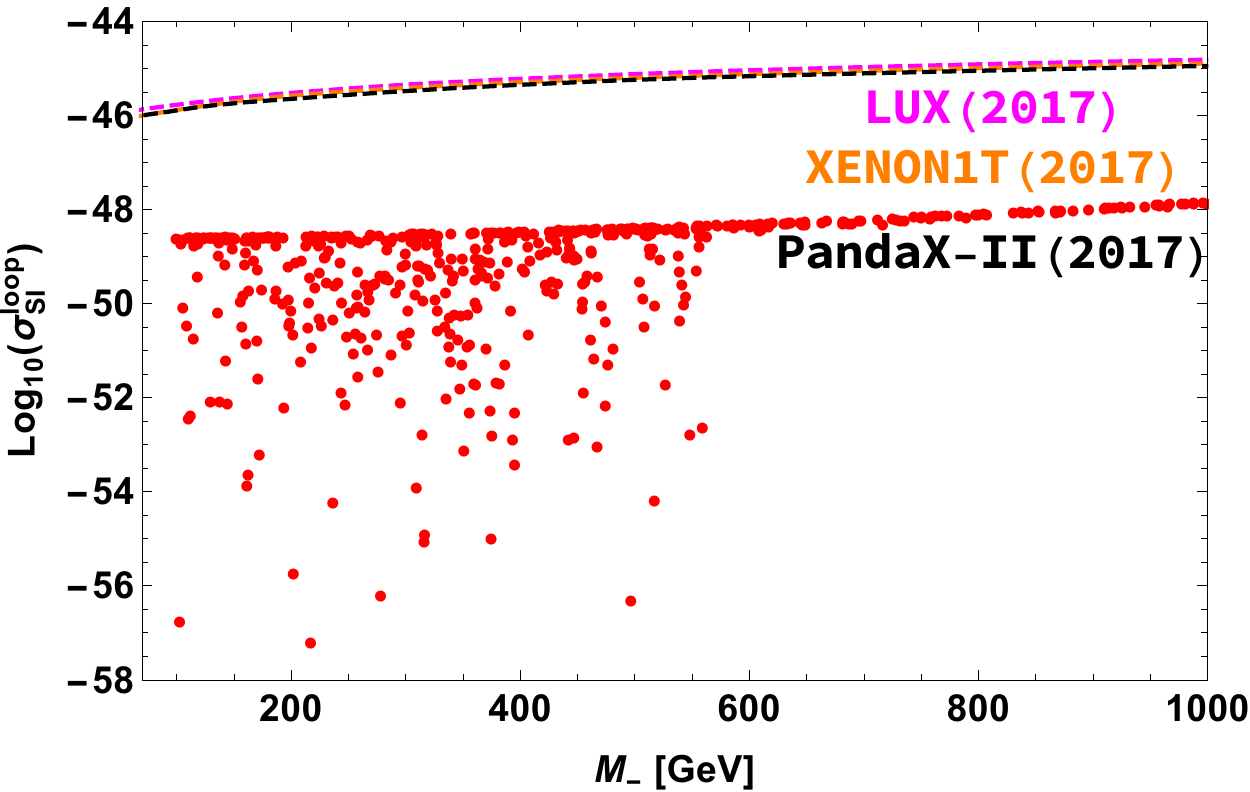}
\caption{One-loop SD and SI contributions projected in the left and right panels respectively. Stringent upper limits of PICO-60 \cite{Amole:2017dex} and LUX \cite{Akerib:2017kat} are given in the left panel, PandaX-II \cite{Cui:2017nnn}, XENON1T \cite{Aprile:2017iyp} and LUX \cite{Akerib:2016vxi} bounds are provided in the right panel.}
\label{DD_loopscatter}
\end{center}
\end{figure}

Moving to direct searches, the $Z^\prime$ mediated WIMP-nucleon interaction is not possible at tree-level as the $Z'$ boson does not couple to quarks. 
The $t$-channel scalar ($H_1,H_2$) exchange can give spin-independent (SI) contribution, but it doesn't help our purpose of study. In the scalar portal, one can obtain contribution from spin-dependent (SD) interaction mediated by SLQ, of the form
\begin{equation}
\mathcal{L^{\rm SD}_{\rm eff}} \simeq \frac{y_{qR}^2\cos^2\alpha}{4(M_{S_1}^2-M_-^2)} \overline{N_-}\gamma^\mu\gamma^5 N_- \overline{q}\gamma_\mu\gamma^5 q\,.
\end{equation}
The s-channel process is depicted in the left most panel of Fig. \ref{DD_penguin} and the corresponding cross section is given by \cite{Agrawal:2010fh}
\begin{equation}
\sigma_{\rm SD} = \frac{ \mu_r^2}{\pi} \frac{\cos^4\alpha}{(M_{S_1}^2 - M_-^2)^2}\left[y_{dR}^2\Delta_d + y_{sR}^2\Delta_s\right]^2 J_n(J_n+1),
\end{equation}
where the angular momentum $J_n = \frac{1}{2}$, reduced mass $\mu_r = \frac{M_- M_n}{M_-+M_n}$ with $M_n \simeq 1$ GeV for nucleon. The values of quark spin functions $\Delta_{q}$ are provided in \citep{Agrawal:2010fh}. Now, it is obvious that it can constrain the parameters $M_-$ and $(y_{qR})^2$. Fig. \ref{DDscatter}, left panel displays $M_--(y_{qR}^2)$ parameter space (green and red regions) remained after imposing Planck \cite{Aghanim:2018eyx} $3\sigma$ limit on current relic density. Here, the region shown in green turns out to be excluded by most stringent PICO-60 \cite{Amole:2017dex} limit on SD WIMP-proton cross section, as seen from the right panel.
Apart from tree-level, one-loop penguin diagrams (middle and right panels of Fig. \ref{DD_penguin}) involving SLQ,  mediated by SM neutral  gauge boson ($Z$) and the neutral scalars ($H_{1}, H_2$) can provide SD and SI contributions respectively. The effective interaction Lagrangian relevant for SD cross section is given by \cite{Ibarra:2016dlb,Herrero-Garcia:2018koq}
\begin{equation}
\mathcal{L^{\rm SD-loop}_{\rm eff}} \simeq \xi_{q^\prime}\overline{N_-}\gamma^\mu\gamma^5 N_- \overline{q^\prime}\gamma_\mu\gamma^5 q^\prime\,,
\end{equation}
where
\bea
\xi_{q^\prime} = \sum_{q=d,s,b}\frac{y_{qR}^2 \cos^2\alpha~ a_{q^\prime}}{32\pi^2 M_Z^2} \left[(v_q + a_q) G\left(\frac{M_-^2}{M_{S_1}^2}\right)\right]\,.
\eea
And the loop function takes the form
\bea
G(x)= -1+\frac{2\left( x+(1-x)
{\rm{ln}}(1-x) \right)}{x^2}\,.
\eea
The corresponding WIMP-nucleon cross section is given by 
\begin{equation}
\sigma^{\rm loop}_{\rm SD} = \frac{16 \mu_r^2}{\pi} J_n (J_n + 1)(\xi_u \Delta_u + \xi_d \Delta_d + \xi_s \Delta_s)^2\,.
\end{equation}
In the above expression, $v_q = \displaystyle{\frac{g}{2\cos \theta_w}} \left(-\frac{1}{2}+\frac{2}{3}\sin^2 \theta_w \right)$, $a_q = -\displaystyle{\frac{g}{4\cos \theta_w}}$ and $a_{q^\prime} = \displaystyle{\frac{g}{4\cos \theta_w}}\left(-\displaystyle{\frac{g}{4\cos \theta_w}} \right)$ for $q^\prime = u ~(d,s)$. For SI contribution, the effective interaction term is 
\begin{equation}
\mathcal{L^{\rm SI-loop}_{\rm eff}} \simeq \Lambda_{q^\prime}\overline{N_-}N_-\overline{q^\prime}q^\prime \,.
\end{equation}
and the corresponding cross-section are given by 
\begin{equation}
\sigma^{\rm loop}_{\rm SI} = \frac{4 \mu_r^2}{\pi} M_n^2 \left(\frac{\Lambda_{q^\prime}}{m_{q^\prime}}\right)^2 f_n^2\;,
\end{equation}
where 
\begin{eqnarray}
\Lambda_{q^\prime} = -\sum_{q=d,s,b}\frac{y_{qR}^2 \cos^2\alpha}{16\pi^2} \left[ \frac{\lambda_{HS}v\cos \zeta - \lambda_{S2}v_2\sin \zeta}{M_{H_1}^2} + \frac{\lambda_{HS}v\sin \zeta + \lambda_{S2}v_2\cos \zeta}{M_{H_2}^2}\right]G_1\left(\frac{M_-^2}{M_{S_1}^2}\right) \frac{m_{q^\prime}}{v M_{-}}\,.\nn
\end{eqnarray}
Here $\displaystyle{G_1(x) = \frac{x+(1-x)
{\rm{ln}}(1-x)}{x}}$. For proton, the typical value of the scalar form factor $f_n$ is $\sim 0.3$. Fig. \ref{DD_loopscatter}, left and right panels project the one-loop SD and SI contributions respectively for the parameter space consistent with Planck data. We see that these contributions are well below the current upper limits set by direct detection collaborations such as PICO-60 \cite{Amole:2017dex}, LUX \cite{Akerib:2017kat} for SD and PandaX-II \cite{Cui:2017nnn}, XENON1T \cite{Aprile:2017iyp}, LUX \cite{Akerib:2016vxi} for SI cases. Thus they do not have any impact on the range of model parameters.
%
\section{Radiative neutrino mass}
\begin{figure}[thb]
\begin{center}
\includegraphics[width=0.4\linewidth]{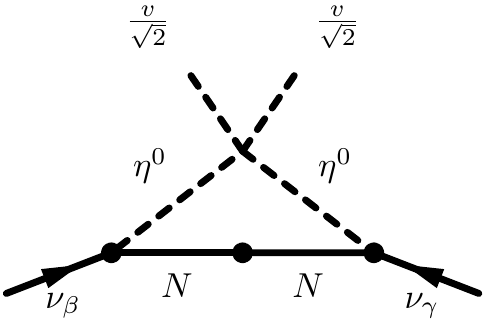}
\caption{Radiative generation of neutrino mass.}
\label{numass}
\end{center}
\end{figure}

\begin{figure}[thb]
\begin{center}
\includegraphics[width=0.48\linewidth]{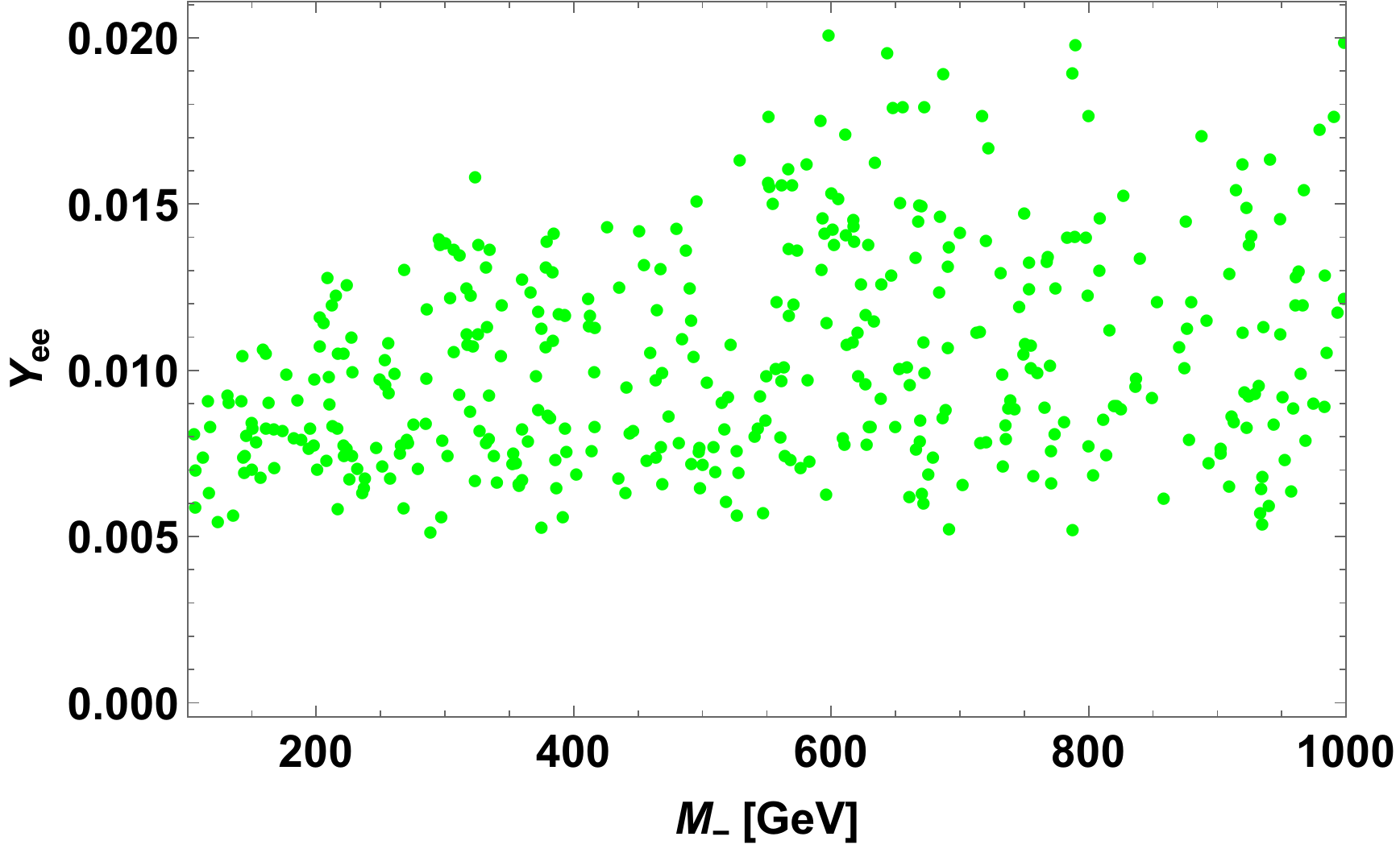}
\quad
\includegraphics[width=0.48\linewidth]{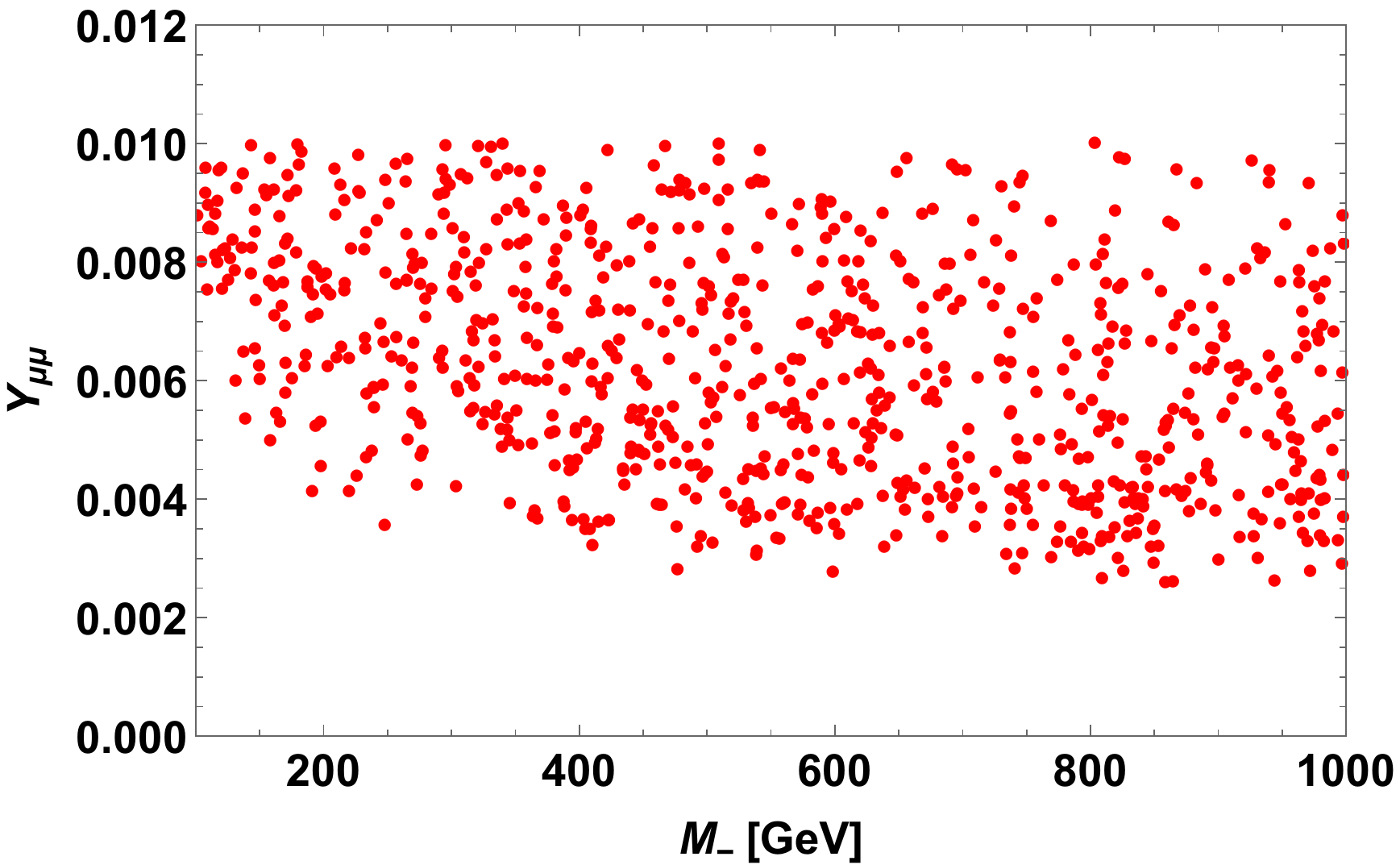}
\includegraphics[width=0.48\linewidth]{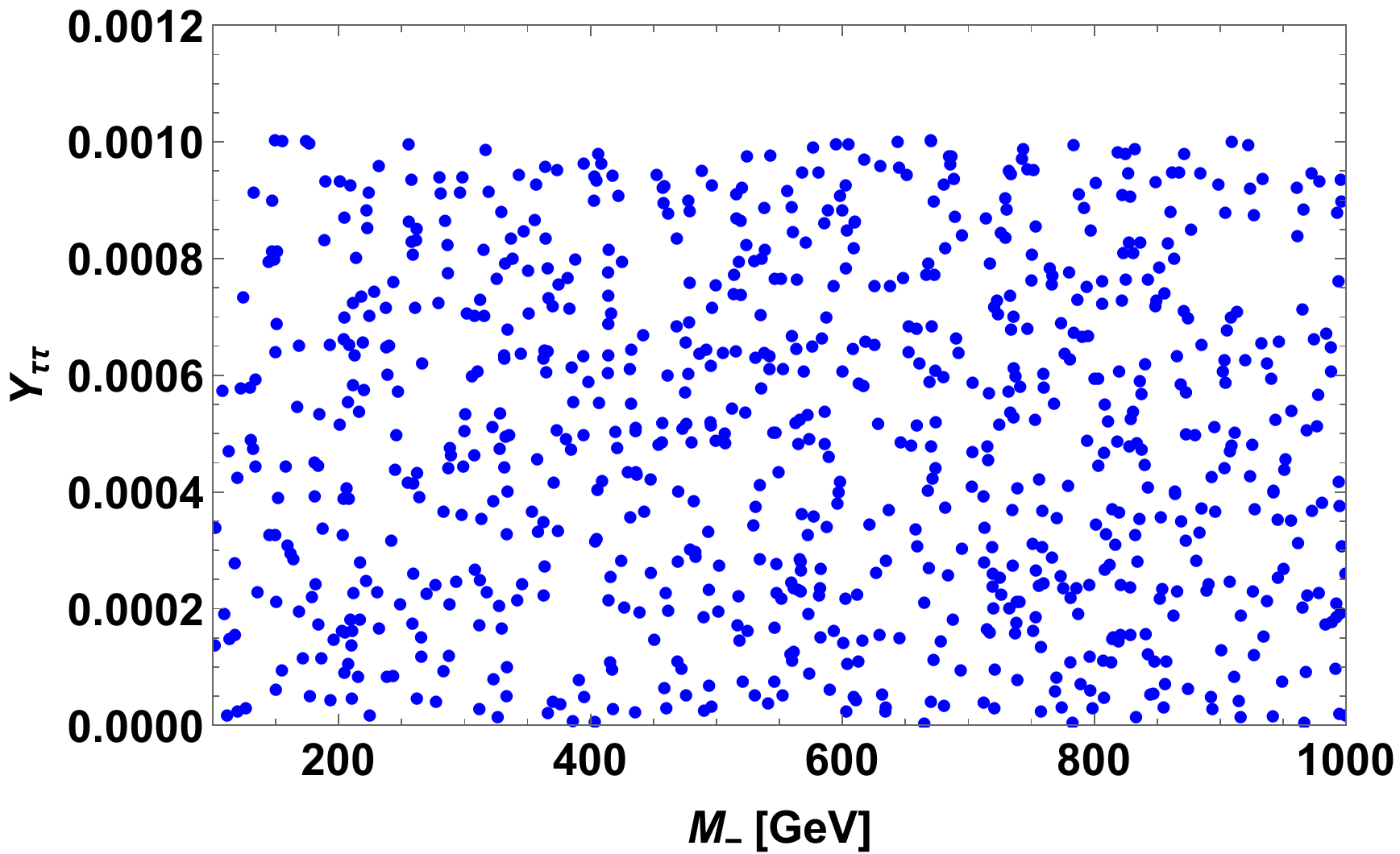}
\caption{Allowed parameter space consistent with the current data from neutrino sector in $Y_{ll}-M_-$ plane.}
\label{numass-1}
\end{center}
\end{figure}

The light neutrino mass can be generated at one-loop level  from the Yukawa interaction with inert doublet in Eqn. \ref{eq:Lag}.
and the corresponding diagram is shown in Fig. \ref{numass}\,. Assuming $m_0^2 = (M^2_{\eta_e} + M^2_{\eta_o})/{2}$ is much greater than $M^2_{\eta_e} -M^2_{\eta_o} = \lambda''_{H\eta}v^2,$ the expression for the radiatively generated neutrino mass \cite{Ma:2006km} is given by
\begin{equation}
({\cal M}_\nu)_{\beta\gamma} = {\lambda''_{H\eta} v^2  \over16 \pi^{2}} 
\sum_{l=e,\mu,\tau} {Y_{ \beta l} Y_{\gamma l} M_{Dl} \over m_0^{2} - M^{ 2}_{Dl}} \left[ 
1 +{M^{ 2}_{Dl} \over m_0^{2}-M^{ 2}_{Dl}} \ln { M^{ 2}_{Dl}\over m_0^{2}  }  \right].\label{nu-mass}
\end{equation}
Here $M_{Dl} = (U^T M_N U)_{l} = {\rm diag}(M_{ee},M_-,M_+)$ and the fermion mass eigenstates $N_{Dl} = U^{\dagger}_{lm} N_{m}$. The light neutrino mass matrix 
(\ref{nu-mass}) can be expressed as
\bea
({\cal M}_\nu)_{\beta\gamma} \equiv (Y^T \Lambda Y)_{ \beta \gamma}\;, \label{mass}
\eea
where the matrix $\Lambda $ is defined as $\Lambda = {\rm diag}(\Lambda_1,
\Lambda_2, \Lambda_3)$, with
\bea
\Lambda_l= {\lambda''_{H\eta} v^2  \over16 \pi^{2}}  {M_{Dl} \over m_0^{2} - M^{ 2}_{Dl}} \left[ 
1 +{M^{ 2}_{Dl} \over m_0^{2}-M^{ 2}_{Dl}} \ln { M^{ 2}_{Dl}\over m_0^{2}  }  \right].
\eea
Generation-wise, the $L_\mu-L_\tau$ charges of SM leptons match with the charges of new fermions $N_l$ ($0,+1,-1$ respectively). The Yukawa interaction term is written using an inert doublet $\eta$ with vanishing  $L_\mu-L_\tau$ charge, and thus the Yukawa matrix $Y_{\beta l}$ takes diagonal form. Hence, 
the neutrino mass matrix  (\ref{mass}) is diagonal, by model construction i.e., ${\cal M}_\nu = {\rm diag}(M_1,M_2,M_3)$.
%
%
%
In order to find the viable region for model parameters consistent with the current neutrino oscillation data, i.e., $6.93 \leq \Delta M_{\rm sol}^2~[ {10^{-5}}\;{\rm eV^2}] \leq 7.97 $,  $2.37 \leq \Delta M_{\rm atm}^2~[ {10^{-3}}\;{\rm eV^2}] \leq 2.63 $ \cite{Capozzi:2016rtj}, and cosmological  bound on the sum of  the light neutrino mass, $\sum_i M_i < 0.23$ eV \cite{Ade:2015xua}, we perform  a scan by varying the parameters  in the following range (where the masses are considered in GeV):
\bea
&&1000 \leq M_{ee} \leq  3000, ~~~100 \leq M_{-} \leq  1000, ~~~2000 \leq M_{+} \leq  5000, ~~~1000 \leq m_0 \leq  2000\;,\nn\\
&&0.0001 \leq Y_{ee}, Y_{\mu \mu}, Y_{\tau \tau} \leq 0.05\;.
\eea
The allowed parameter space in $Y_{ll}-M_-$ plane, satisfying the above constraints from the neutrino sector is shown in Figure \ref{numass-1} . Thus, the proposed model gives viable region of parameter space consistent with recent oscillation data in the context of radiative light neutrino mass matrix.

\section{Flavor Phenomenology}
The general effective Hamiltonian responsible for  the quark level transition $ b \to s l^+ l^-$ is given by \cite{Bobeth:1999mk, Bobeth:2001jm} 
\bea
{\cal H}_{\rm eff} &=& - \frac{ 4 G_F}{\sqrt 2} V_{tb} V_{ts}^* \Bigg[\sum_{i=1}^6 C_i(\mu) O_i +\sum_{i=7,9,10} \Big ( C_i(\mu) O_i
+ C_i^\prime(\mu) O_i^\prime \Big )
\Bigg]\;,\label{ham}
\eea
where $G_F$ is the Fermi constant and  $V_{qq^\prime}$ denote the Cabibbo-Kobayashi-Maskawa (CKM) matrix elements. The  $C_i$'s stand for  the Wilson coefficients evaluated at the renormalized scale $\mu = m_b$ \cite{Hou:2014dza}, where the sum over $i$ includes the current-current
operators $(i = 1, 2)$ and the QCD-penguin operators $(i = 3, 4, 5, 6)$. The quark level operators mediating leptonic/semileptonic processes are given as 
\bea
O_7^{(\prime)} &=&\frac{e}{16 \pi^2} \Big[\bar s \sigma_{\mu \nu}
\left (m_s P_{L(R)} + m_b P_{R(L)} \right ) b\Big] F^{\mu \nu}, \nn\\
O_9^{(\prime)}&=& \frac{\alpha_{\rm em}}{4 \pi} (\bar s \gamma^\mu P_{L(R)} b)(\bar l \gamma_\mu l)\;,~~~~~~~ O_{10}^{(\prime)}= \frac{\alpha_{\rm em}}{4 \pi} (\bar s \gamma^\mu 
P_{L(R)} b)(\bar l \gamma_\mu \gamma_5 l)\;, 
\eea
where  $\alpha_{\rm em}$ denotes  the fine-structure constant and $P_{L,R} = (1\mp \gamma_5)/2$ are the chiral operators.
The primed operators are absent in the SM, but can exist in the proposed $L_\mu-L_\tau$  model.

The previous section has discussed the available new parameter space consistent with the DM observables which are within their respective experimental limits. However, these parameters can be further constrained from the quark and lepton sectors, to be presented in the subsequent sections.

\subsection{$B_s-\bar B_s$ mixing}

In this subsection,  we discuss the constraint on the new parameters from the mass difference  between the $B_s$ meson mass eigenstates ($\Delta M_s$),
which characterizes the $B_s - \bar B_s$ mixing phenomena. In the SM, $B_s-\bar B_s$ mixing proceeds to an excellent approximation
through the box diagram with internal top quark and $W$ boson exchange. The effective Hamiltonian describing the $\Delta B=2$
transition  is given by \cite{Inami:1980fz} 
\bea  \label{ham2}
{\cal H}_{\rm eff}=\frac{G_F^2}{16 \pi^2}~ \lambda_t^2~ M_W^2 S_0(x_t)\eta_B
(\bar s b)_{V-A}(\bar s b)_{V-A}\;, 
\eea
 where  $\lambda_t=V_{tb} V_{ts}^*$, $\eta_B$ is the QCD correction factor and  the loop function $S_0(x_t)$ is given by  \cite{Inami:1980fz} 
 \be S_0(x_t)=\frac{4 x_t -11 x_t^2
+x_t^3}{4(1-x_t)^2} - \frac{3}{2} \frac{\log x_t x_t^3}{(1-x_t)^3}\;,
\ee 
 with $x_t=m_t^2/M_W^2$. 
Using Eqn. (\ref{ham2}),  the $B_s - \bar B_s$ mass difference in the SM is given as 
 \bea
\Delta M_s^{\rm SM} = 2 |M_{12}^{\rm SM}|  =\frac{\langle \bar B_s|{\cal H}_{eff}| B_s \rangle}{
M_{B_s}}  = \frac{G_F^2}{6
\pi^2} M_W^2~ \lambda_t^2~ \eta_B~ \hat B_s f_{B_s}^2 M_{B_s} S_0(x_t)\;.
\label{sm}
 \eea 
The SM predicted  value of $\Delta M_s$ by using the input parameters from \cite{Tanabashi:2018oca,Charles:2015gya} is 
\bea \label{SM-Bs}
\Delta M_s^{\rm SM} = (17.426\pm 1.057)~ {\rm ps^{-1}},
\eea
and  the corresponding experimental value is  \cite{Tanabashi:2018oca}
 \bea
\Delta M_s^{\rm Expt} = 17.761 \pm 0.022~ {\rm ps^{-1}}.\label{Exp-Bs}
\eea
Even though the theoretical prediction is in good agreement with the experimental $B_s - \bar B_s$  oscillation data, it does not completely rule out the possibility of new physics.
\begin{figure}[thb]
\begin{center}
\includegraphics[width=0.65\linewidth]{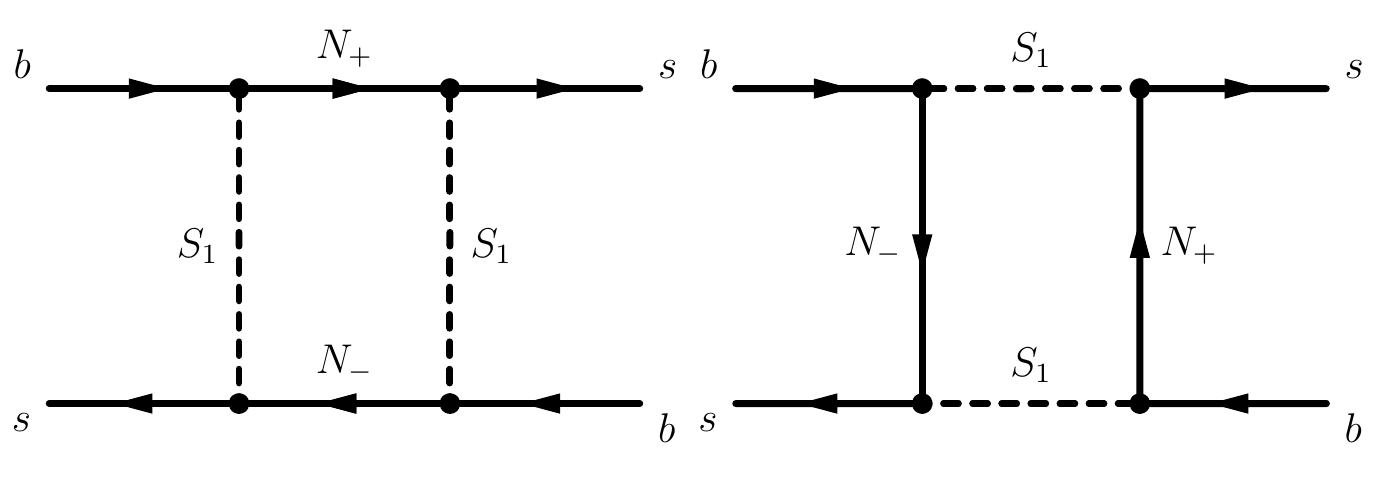}\\
\includegraphics[width=0.65\linewidth]{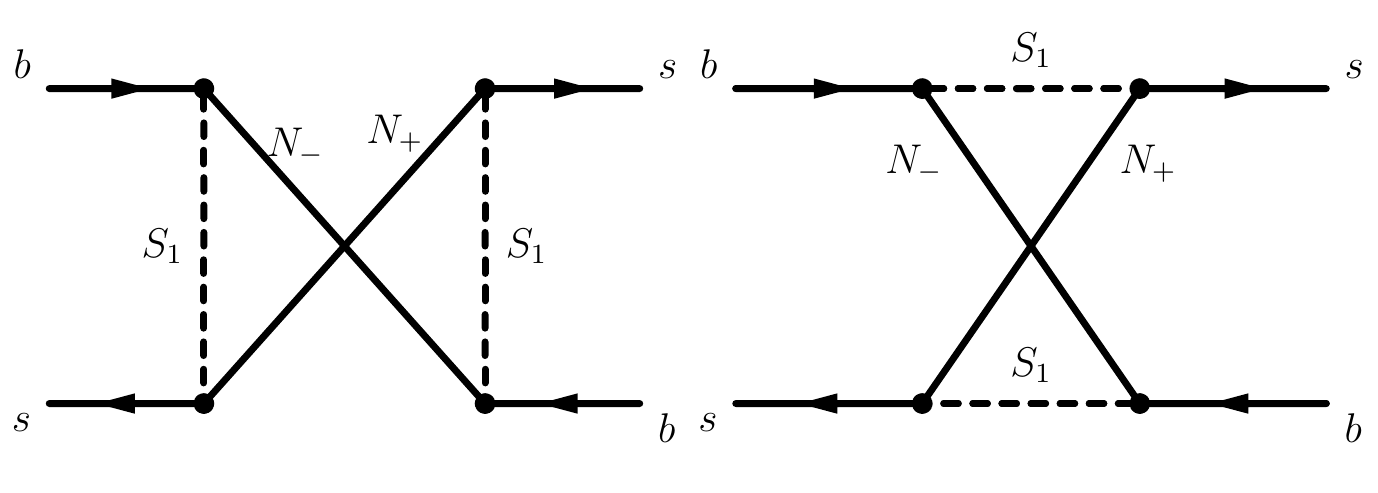}\\
\caption{Box diagrams of $B_s-\bar B_s$ mixing  with leptoquark  in the loop.}
\label{box-Bs}
\end{center}
\end{figure}

The box diagrams for $B_s-\bar B_s$ mixing in the presence of singlet SLQ and $N_\pm$ are shown in Fig. \ref{box-Bs}\,.   The effective Hamiltonian in the presence NP is given by 
\bea \label{ham3}
{\cal H}_{\rm eff}=\frac{(y_{s R} y_{b R})^2}{128\pi^2 M_{S_1}^2} \cos^2 \alpha \sin^2 \alpha  C_{B_s}^{\rm NP}\left(\bar s b\right)_{V+A}\left(\bar s b\right)_{V+A}\;,
\eea 
where
\bea
C_{B_s}^{\rm NP}&=&2k\left(\chi_-, \chi_-,1\right)+ 4k\left(\chi_-, \chi_+,1\right)+2k\left(\chi_+, \chi_+,1\right)+\chi_- j\left(\chi_-, \chi_-,1\right)\nn \\&+&2\sqrt{\chi_-\chi_+}j\left(\chi_-, \chi_+,1\right)+\chi_+ j\left(\chi_+, \chi_-,1\right),
\eea
with $\chi_\mp=M_\mp^2/M_{S_1}^2$ and the loop functions $k\left(\chi_\pm, \chi_\mp,1\right),~j\left(\chi_\pm, \chi_\mp,1\right)$ are given in Appendix A.  
 Using Eqn. (\ref{ham3}), the mass difference of $B_s-\bar B_s$ mixing due to the exchange of $S_1$ and $N_\pm$ is found to be
 \bea
\Delta M_s^{\rm NP}  = \frac{(y_{s R} y_{b R})^2}{48  \pi^2 M_{S_1}^2} \cos^2 \alpha \sin^2 \alpha  C_{B_s}^{\rm NP} \eta_B \hat B_{B_s} f_{B_s}^2 M_{B_s}\;.
\label{LQ}
 \eea 
Including the NP contribution arising due to the SLQ exchange,  the  total mass difference can be written as
 \bea \label{Bs-delM-f}
\Delta M_s = \Delta M_s^{\rm SM}  \left [1 + \frac{C_{B_s}^{\rm NP}\cos^2 \alpha \sin^2 \alpha }{8 G_F^2 V_{tb}^2 V_{ts}^{*2} M_W^2 S_0(x_t)}
 \left (\frac{(y_{s R} y_{b R})^2}{M_{S_1}^2}  \right )\right ]\;.
\eea 
Using Eqns. (\ref{SM-Bs}) and (\ref{Exp-Bs}) in (\ref{Bs-delM-f}), one can put bound on  $(y_{qR})^2$ and $M_-$ parameters. 
\subsection{$B \to K l^+ l^-$ process}
\begin{figure}[thb]
\begin{center}
\includegraphics[width=0.32\linewidth]{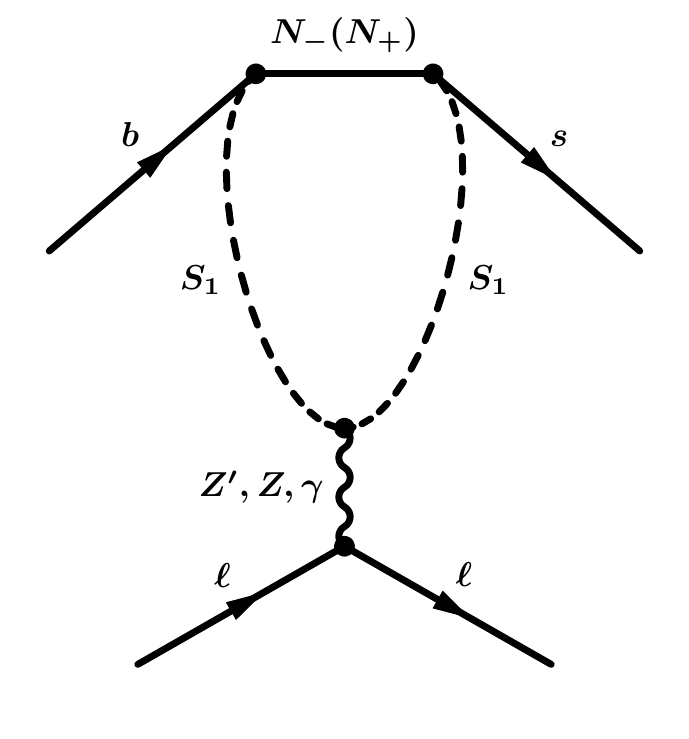}
\vspace{0.01 cm}
\includegraphics[width=0.32\linewidth]{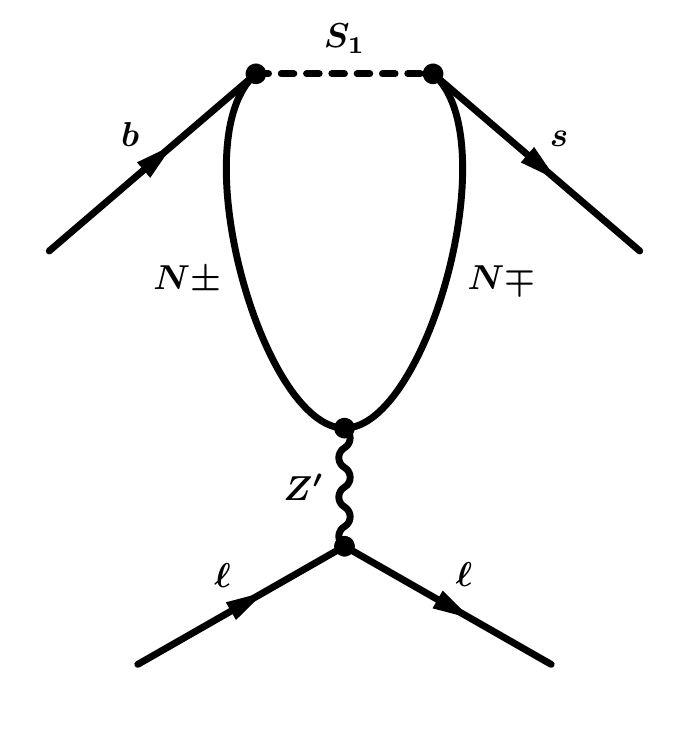}
\caption{Penguin diagram of $ b \to s ll$ processes, where $\l=\mu,\tau$ with leptoquark in the loop.}
\label{penguin}
\end{center}
\end{figure}

The rare semileptonic $B \to K l^+ l^-$ process is mediated via $b \to s l^+ l^-$ quark level transitions. In the current framework, the $b \to s l^+l^-$  transitions can occur via the $Z^\prime$ exchanging one-loop penguin diagrams shown in Fig. \ref{penguin}\,.
The penguin diagram in the left panel is suppressed by the factor, $m_b/M_{\pm}$, thus providing negligible contribution. Furthermore, due to the zero hypercharge of dark matter fermion, the diagrams with SM neutral bosons ($Z,\gamma$) replacing $Z^\prime$ in the right panel of Fig. \ref{penguin}\, are not possible in the present framework.
The matrix elements of the various hadronic currents between the initial $B$ meson and $K$ meson in the final state are related to the form factors  $f_{+, 0}$ as follows \cite{Bobeth:2007dw, Ball:2004ye}
\bea \label{ff}
 &&\langle{K}\left(p_K\right)|\bar{s}\gamma^\mu b|{B}\left(p_B\right)\rangle = f_+\left(q^2\right) \left(p_B+p_K\right)^\mu + \left[f_0\left(q^2\right) - f_+\left(q^2\right)\right]\frac{M^2_B - M^2_K}{q^2}q^\mu\,,
 \eea
where $p_B~(p_K)$ and $M_B~ (M_K)$ denote the 4-momenta and mass of the  $B~(K)$ meson and $q^2$ is the momentum transfer. By using Eqn. (\ref{ff}), the transition amplitude of $B \to K \mu^+ \mu^-$ process is given by 
\bea \label{amp}
\mathcal{M}=\frac{1}{2^{5}\pi^2}\frac{y_{b R} y_{s R} g_{\mu \tau}^2}{M_{Z^\prime}^2} \mathcal{V}_{sb}(\chi_-, \chi_+) [\bar{u}(p_B)\gamma^\mu (1+\gamma_5)u(p_K))][\bar{v}(p_2)\gamma_\mu u(p_1))],
\eea
where $p_1$ and $p_2$ are the four momenta of charged leptons and the loop function $\mathcal{V}_{sb}(\chi_-, \chi_+)$ is given in Appendix A  \cite{Baek:2017sew, Hisano:1995cp}.
Now comparing this amplitude (\ref{amp}) with the amplitude obtained from the effective Hamiltonian (\ref{ham}), we obtain a new Wilson coefficient  associated with the right-handed semileptonic electroweak penguin operator $\mathcal{O}_9^\prime$ as 
\bea \label{new-C9p}
C_9^{\prime \rm NP } =\frac{\sqrt{2} }{2^4\pi G_F \alpha_{\rm em} V_{tb}V_{ts}^*}\frac{y_{b R} y_{s R} g_{\mu \tau}^2}{{M_{Z^\prime}}^2}\mathcal{V}_{sb}\left(\chi_-, \chi_+\right)\,.
\eea
The  differential branching ratio of $B \to K ll$  process with respect to $q^2$ is given by 
 \bea
 \frac{d {\rm Br}}{d q^2 }= \tau_B \frac{G_F^2 \alpha_{\rm em}^2 |V_{tb}V_{ts}^*|^2}{2^8 \pi^5 M_B^3}\sqrt{\lambda(M_B^2, M_K^2, q^2)} \beta_l f_+^2 \Big ( a_l(q^2)+\frac{c_l(q^2)}{3} \Big )\;,
 \eea
 where
 \bea
 a_l(q^2)&=&  q^2|F_P|^2+ \frac{\lambda(M_B^2, M_K^2, q^2)}{4}
 (|F_A|^2+|F_V|^2) \nn \\ &&+ 2 m_l (M_B^2-M_K^2+q^2) {\rm Re}(F_P F_A^*) +4 m_l^2 M_B^2 |F_A|^2 \;,\nn\\
  c_l(q^2)&=& -   \frac{\lambda(M_B^2, M_K^2, q^2)}{4} \beta_l^2 
 \left (|F_A|^2+|F_V|^2\right ),
 \eea
 with
\bea
F_V & =&\frac{2 m_b}{M_B} C_7^{\rm eff}+ C_9^{\rm eff} +  C_9^{\prime \rm NP}, ~~~~~F_A = C_{10},\nn\\
F_P&=&  m_l C_{10} \Big[ \frac{M_B^2 -M_K^2}{q^2}\Big(\frac{f_0(q^2)}{f_+(q^2)}-1\Big) -1 \Big]\;,
\eea 
and  
\bea
\lambda(a,b,c) =a^2+b^2+c^2-2(ab+bc+ca),~~~~~\beta_l = \sqrt{1-4 m_l^2/q^2}\;.
\eea
For numerical estimation, we have used the  lifetime  and  masses of particles from \cite{Tanabashi:2018oca} and the form factors are taken from \cite{Colangelo:1996ay}. 
The  experimental limit on the  branching ratios of $\bar B^0 \to \bar K^0 \mu^+ \mu^-$ and  $B^+ \to K^+ \tau^+ \tau^-$  processes are \cite{Tanabashi:2018oca}
\bea
&&{\rm Br}(\bar B^0 \to \bar K^0 \mu^+ \mu^-)\big |^{\rm Expt} = (3.39\pm 0.34)\times 10^{-7}\,,\\
&&{\rm Br}(B^+ \to K^+ \tau^+ \tau^-)\big |^{\rm Expt} < 2.5\times 10^{-3},
\eea
while their  predicted values in the  SM   are
\bea
&&{\rm Br}(\bar B^0 \to \bar K^0 \mu^+ \mu^-)\big |^{\rm SM} =(1.82\pm 0.15)\times 10^{-7}\,,\\
&&{\rm Br}(B^+ \to K^+ \tau^+ \tau^-)\big |^{\rm SM} =(1.56\pm 0.125)\times 10^{-7}.
\eea 
Since $Z^\prime$ doesn't couple to electron, the branching ratio of $B \to K e^+ e^-$ process is considered to be SM like.  The   decay modes $\bar B^0 \to \bar K^0 \mu^+ \mu^- $ and $B^+ \to K^+ \tau^+ \tau^-$ can put constraint on all the four parameters, i.e., $(y_{qR})^2, ~g_{\mu \tau}, ~M_{Z^\prime}$ and $M_-$. 

\subsection{$B \to X_s \gamma$ process}
\begin{figure}[thb]
\begin{center}
\includegraphics[width=0.4\linewidth]{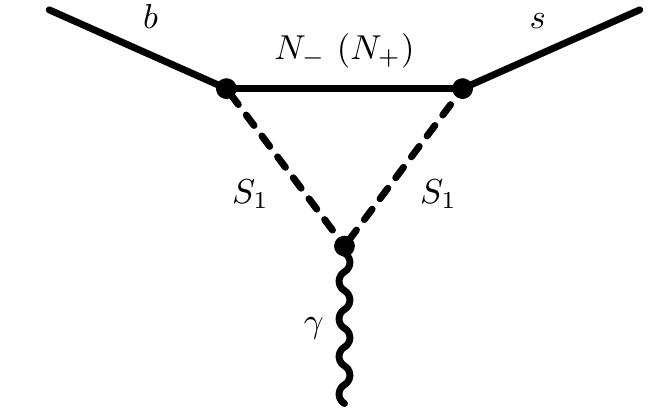}
\caption{ Feynman  diagram of $ b \to s \gamma$  processes in the presence of scalar leptoquark.}
\label{bsgam}
\end{center}
\end{figure}

The $ B \to X_s \gamma$ process involves $b \to s \gamma$ quark level transition, the experimental limit on the corresponding branching ratio is given by  \cite{Amhis:2016xyh}
\bea \label{Exp-bsgamma}
{\rm Br}( B \to X_s \gamma)\big |^{\rm Expt}_{ E_\gamma > 1.6~{\rm GeV}}=(3.32\pm 0.16)\times 10^{-4}.
\eea
 Fig. \ref{bsgam} represents the one loop penguin diagram of $b \to s \gamma$ process mediated by SLQ and $N_\pm$. 

Including the NP contribution, the total branching ratio  of $ B \to X_s \gamma$ is given by 
\bea \label{tot-bsgamma}
{\rm Br}( B \to X_s \gamma) ={\rm Br}( B \to X_s \gamma)\big |^{\rm SM} \Bigg ( 1+\frac{C_7^{\gamma \prime \rm NP}}{C_7^{\gamma \rm SM}} \Bigg )^2,
\eea
where the predicted SM branching ratio  is \cite{Misiak:2015xwa}
\bea \label{SM-bsgamma}
{\rm Br}( B \to X_s \gamma)\big |^{\rm SM}_{E_\gamma > 1.6~{\rm GeV}}=(3.36\pm 0.23)\times 10^{-4}.
\eea
The new $C_7^{\gamma\prime \rm NP}$ Wilson coefficient obtained from Fig. \ref{bsgam} is given by
\bea \label{c7new}
C_7^{\gamma \prime \rm NP}= -\frac{\sqrt{2}/3}{8 G_F V_{tb} V_{ts}^*}\frac{y_{bR}y_{sR}}{M_{S_1}^2} \Big( J_1(\chi_-)\cos^2 \alpha +J_1 (\chi_+)\sin^2 \alpha \Big),
\eea
where the loop functions $J_1(\chi_{\pm})$ are given as  \cite{Baek:2017sew}
\bea
J_1(\chi_\pm)=\frac{1-6\chi_\pm+3\chi_\pm^2+2\chi_\pm^3-6\chi_\pm^2 \log \chi_\pm}{12 (1-\chi_\pm)^4}\,.
\eea  
 Using Eqns. (\ref{Exp-bsgamma}\,, \ref{SM-bsgamma}\,, \ref{c7new}) in (\ref{tot-bsgamma}), the parameters $(y_{qR})^2$ and $M_-$ can be constrained.
\subsection{$\tau \to \mu \nu_\tau \bar \nu_\mu$ process}
\begin{figure}[thb]
\begin{center}
\includegraphics[width=0.45\linewidth]{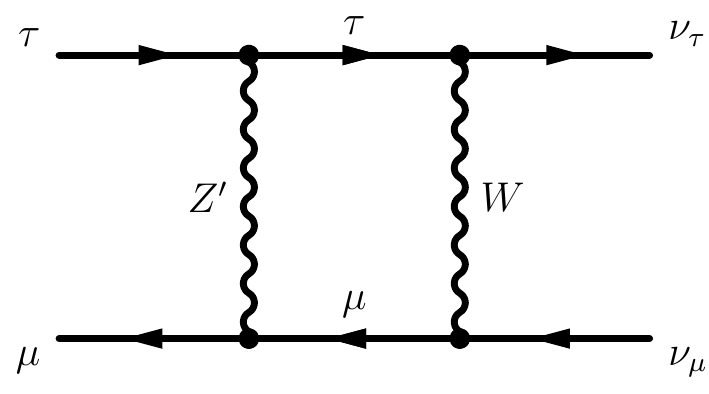}
\caption{One loop box  diagram of  $\tau \to \mu \nu_\tau \bar \nu_\mu$  processes.}
\label{box-tau}
\end{center}
\end{figure}

In the presence of $Z^\prime$ boson, the $\tau \to \mu \nu_\tau \bar \nu_\mu$ process can occur via  box diagram as shown in  Fig. \ref{box-tau}\,. There are four possible one-loop box diagrams with the $Z^\prime$ connected to the lepton legs.
The total branching ratio of this process is given by \cite{Altmannshofer:2014cfa}
\bea
{\rm Br}(\tau \to  \mu \nu_\tau \bar \nu_\mu) ={\rm Br}(\tau \to  \mu \nu_\tau \bar \nu_\mu)\big |^{\rm SM} \Bigg ( 1+ \frac{3g_{\mu \tau}^2}{4\pi^2} \frac{\log (M_W^2/M_{Z^\prime}^2)}{1-M_{Z^\prime}^2/M_W^2} \Bigg )^2, 
\eea
where 
the branching ratio in the SM is given by \cite{Altmannshofer:2014cfa}
\bea
{\rm Br}(\tau \to \mu \nu_\tau \bar \nu_\mu)\big |^{\rm SM}=(17.29\pm 0.032)\%.
\eea
Now comparing the theoretical result with  the experimental measured value  \cite{Tanabashi:2018oca}
\bea
{\rm Br}(\tau \to \mu \nu_\tau \bar \nu_\mu)\big |^{\rm Expt}=(17.39\pm 0.04)\%,
\eea
one can put bounds on $M_{Z^\prime}-g_{\mu \tau}$ parameter space.

\begin{figure}[thb]
\begin{center}
\includegraphics[width=0.48\linewidth]{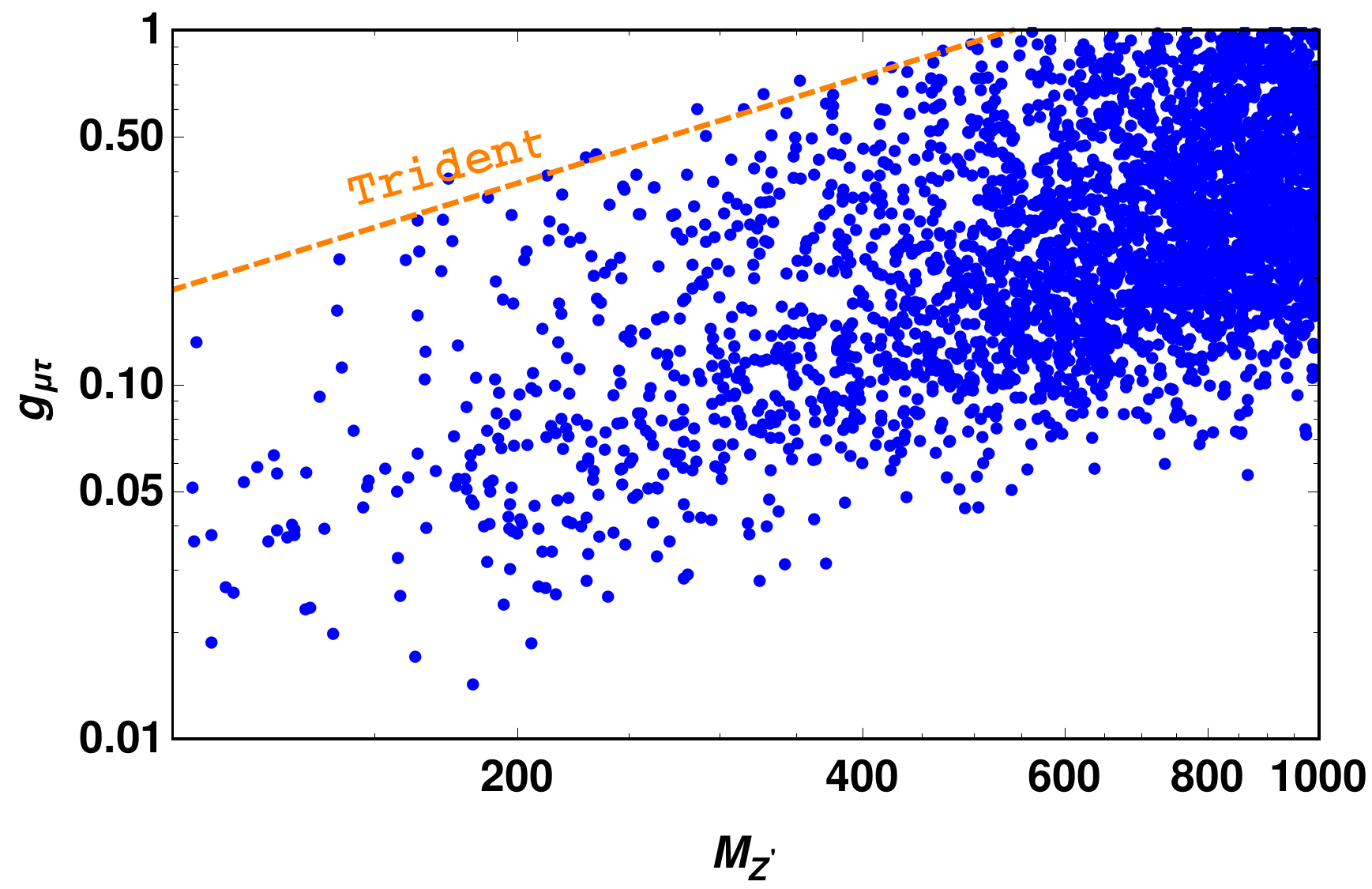}
\quad
\includegraphics[width=0.48\linewidth]{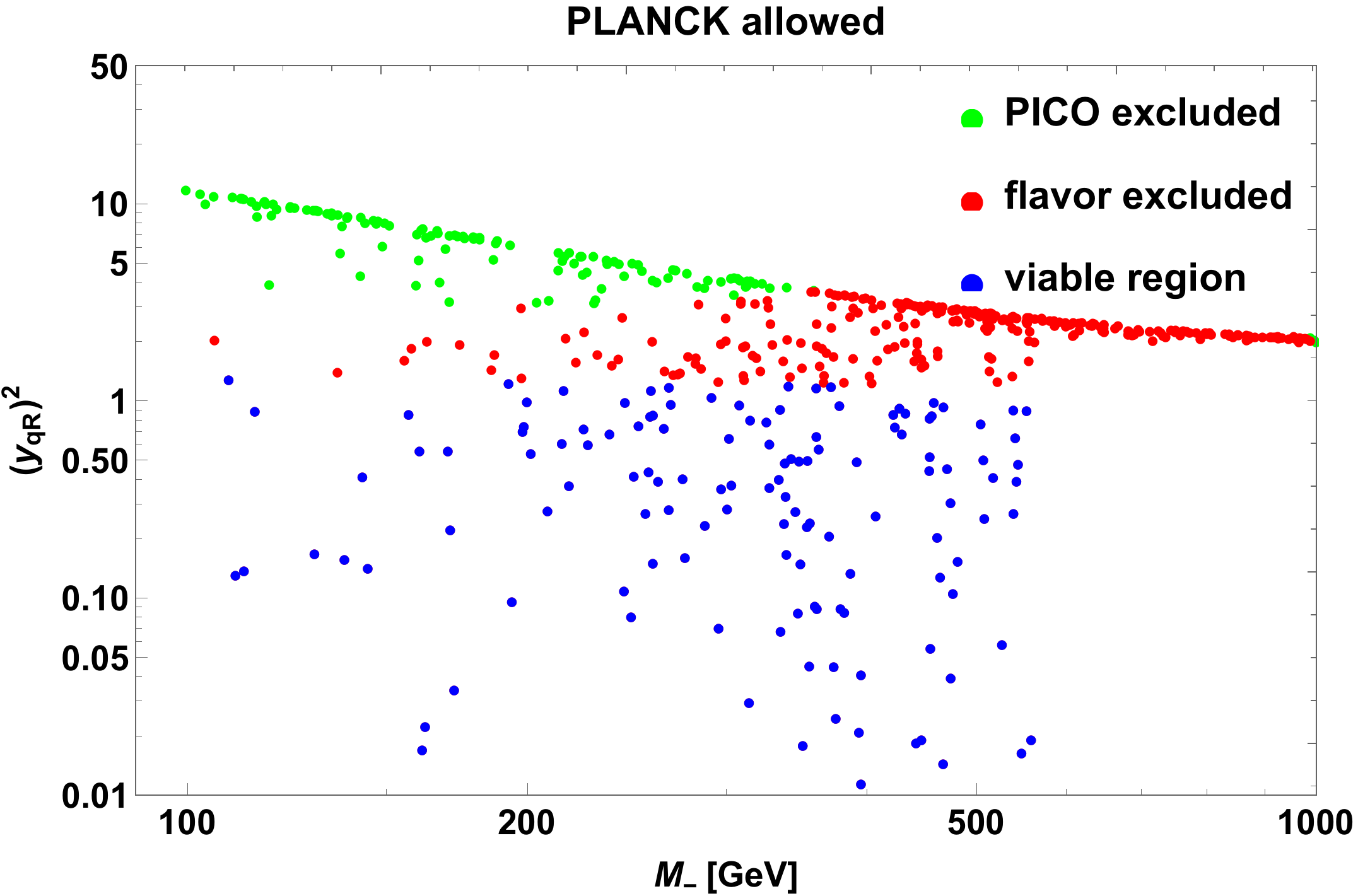}
\caption{Left panel projects the constraint on $g_{\mu\tau}$ and $M_{Z^\prime}$ obtained from  ${\rm Br}(B^+ \to K^+ \tau^+ \tau^-)$, ${\rm Br}(\bar B^0 \to \bar K^0 \mu^+ \mu^-)$ and  ${\rm Br}(\tau \to \mu \nu_\tau \bar \nu_\mu)$ observables. Dashed orange line denotes the neutrino trident bound \cite{Mishra:1991bv,Altmannshofer:2014pba}. In the right panel, blue data points denote the allowed parameter space obtained from ${\rm Br}(\bar B^0 \to \bar K^0 \mu^+ \mu^-)$, $B_s -\bar{B_s}$ mixing, ${\rm Br}(B^+ \to K^+ \tau^+ \tau^-)$,  Br($B\to X_s\gamma)$ experimental data, which are also consistent with Planck \citep{Aghanim:2018eyx} and PICO-60 limit \cite{Amole:2017dex}. Here, green (red) data points denote PICO-60 and flavor excluded (PICO-60 allowed and flavor excluded) region.}
\label{pspace}
\end{center}
\end{figure}
\begin{table}[htb]
\centering
\begin{tabular}{|c|c|c|c|c|}
\hline
Parameters ~&~DM-I ~&~DM-II~&~ DM+Flavor\\
\hline
\hline
$M_-$ [GeV] ~&~$103-560$~&~$561-988$~ &~$103-560$\\
\hline
$(y_{qR})^2$~&~$0-3.51$~&~$1.94-2.56$&$0-1.26$\\
\hline
\end{tabular}
\caption{Predicted allowed range of parameters $M_-$ and $(y_{qR})^2$. Here DM-I and DM-II represent two regions in Fig. \ref{DDscatter} consistent with only DM observables, DM+Flavor denotes the region favored by both the dark matter and flavor studies.}\label{allowed par}
\end{table}

Now correlating the theoretical predictions of ${\rm Br}(\bar B^0 \to \bar K^0 \mu^+\mu^-)$, ${\rm Br}(B^+ \to K^+ \tau^+\tau^-)$ and ${\rm Br}(\tau \to \mu \nu_\tau \bar \nu_\mu)$ with the  corresponding  $3\sigma$ experimental data,  we compute  the $M_{Z^\prime}-g_{\mu\tau}$ allowed parameter space.  Since $Z^\prime$ does not couple to quarks, these gauge parameters couldn't be constrained from $b \to s \gamma$ decay modes and $B_s-\bar B_s$ oscillation data.  The constraint on $M_- - (y_{qR})^2$ parameter space is obtained from ${\rm Br}(\bar B^0 \to \bar K^0 \mu^+\mu^-)$,  ${\rm Br}(B^+ \to K^+ \tau^+\tau^-)$, ${\rm Br}( B \to X_s \gamma)$ and $B_s-\bar B_s$ mixing results.  In addition, the branching ratio of  rare semileptonic $ B \to   K \nu_l \bar \nu_l$  process can   play a vital role in restricting these parameters. Though the proposed model can allow $b \to s \nu_l \bar \nu_l$ decay modes, but the contributions of $\mu$ and $\tau$ leptons cancel with each other in the leading order of NP  due to  their equal and opposite  $L_\mu-L_\tau$ charges. Since there is no  $Z^\prime\mu\tau$ coupling, the neutral and charged lepton flavor violating decay processes like $B \to K^{(*)} \mu^\mp \tau^\pm$, $\tau^- \to \mu^- \gamma$, $\tau \to\mu \mu \mu$ do not play any role.  It should be noted that, the model considered here provides only additional contribution to $b \to s ll$ transitions through $C_9^{\prime {\rm NP}}$ coefficient, hence the $B_s \to l^+l^-$ processes  could not provide any strict bound on the new parameters.  In this analysis, we consider that  the $y_{qR}$ coupling is perturbative, i.e., $|y_{qR}| \lesssim \sqrt{4\pi}$. Left (right) panel in Fig. \ref{pspace} denotes the parameter space in the plane of $M_{Z^\prime}-g_{\mu\tau}$ ($M_--(y_{qR})^2$) consistent with DM and flavor studies. From left panel, one can see that the obtained parameter space survives the lower limit imposed on the ratio $M_{Z^\prime}/g_{\mu\tau}$ by neutrino trident production \cite{Mishra:1991bv,Altmannshofer:2014pba}, i.e., $540$ GeV. It is also noted that the allowed region favored by the $(g-2)_\mu$ anomaly is completely excluded by the constraint from the neutrino trident production \cite{Altmannshofer:2014cfa}.  In the right panel of Fig. \ref{pspace}\,, we redisplay $M_--(y_{qR})^2$ parameter space of Fig. \ref{DDscatter} after a combined analysis made by imposing the DM and flavor experimental limits, with the surviving region shown in blue color.  In Table \ref{allowed par}, we report the allowed region of the parameters $M_-$ and $(y_{qR})^2$ which are consistent with only DM studies (DM-I,II), both DM and flavor sectors (DM+Flavor). 

\begin{figure}[thb]
\begin{center}
\includegraphics[width=0.48\linewidth]{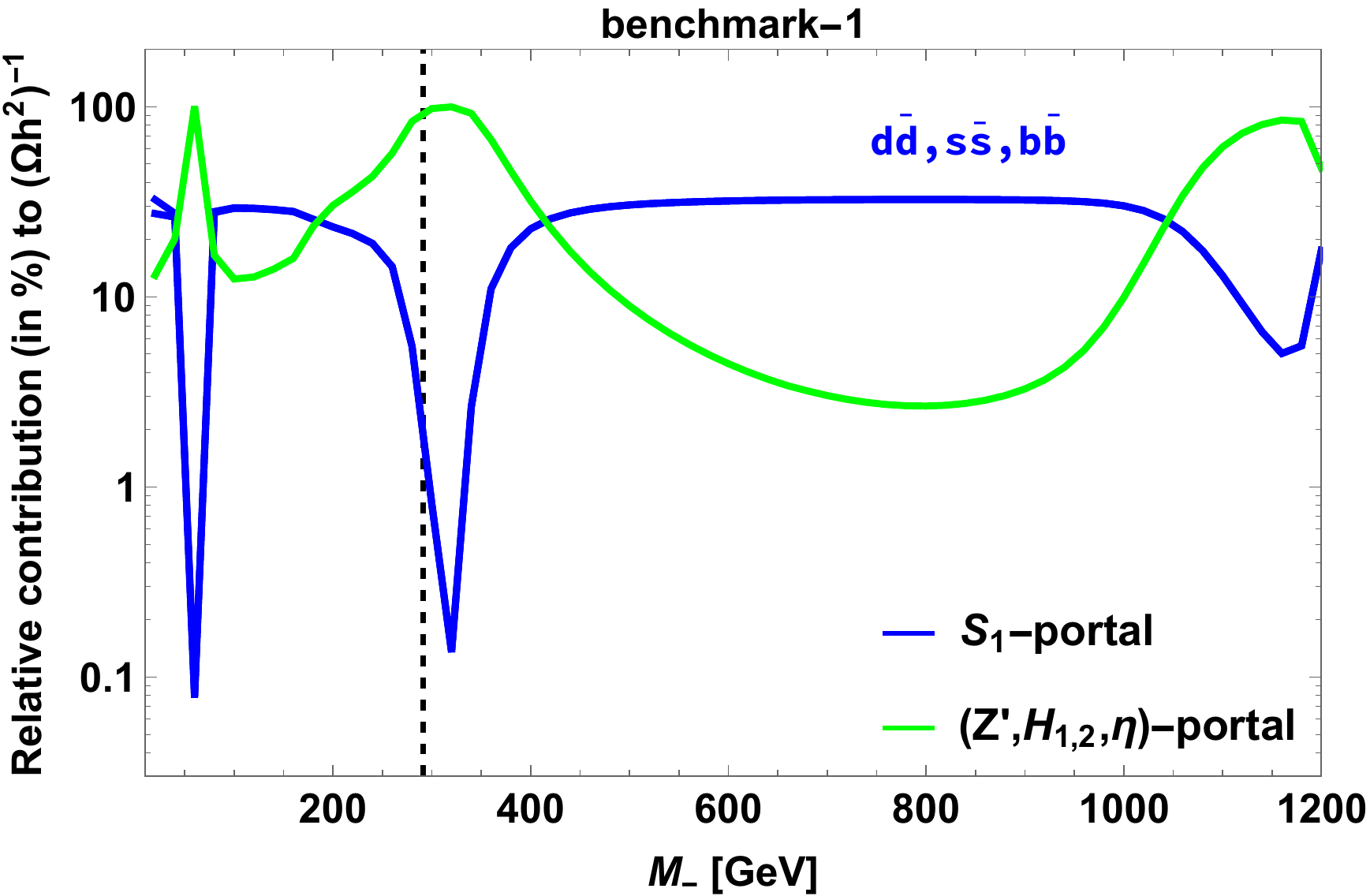}
\quad
\includegraphics[width=0.48\linewidth]{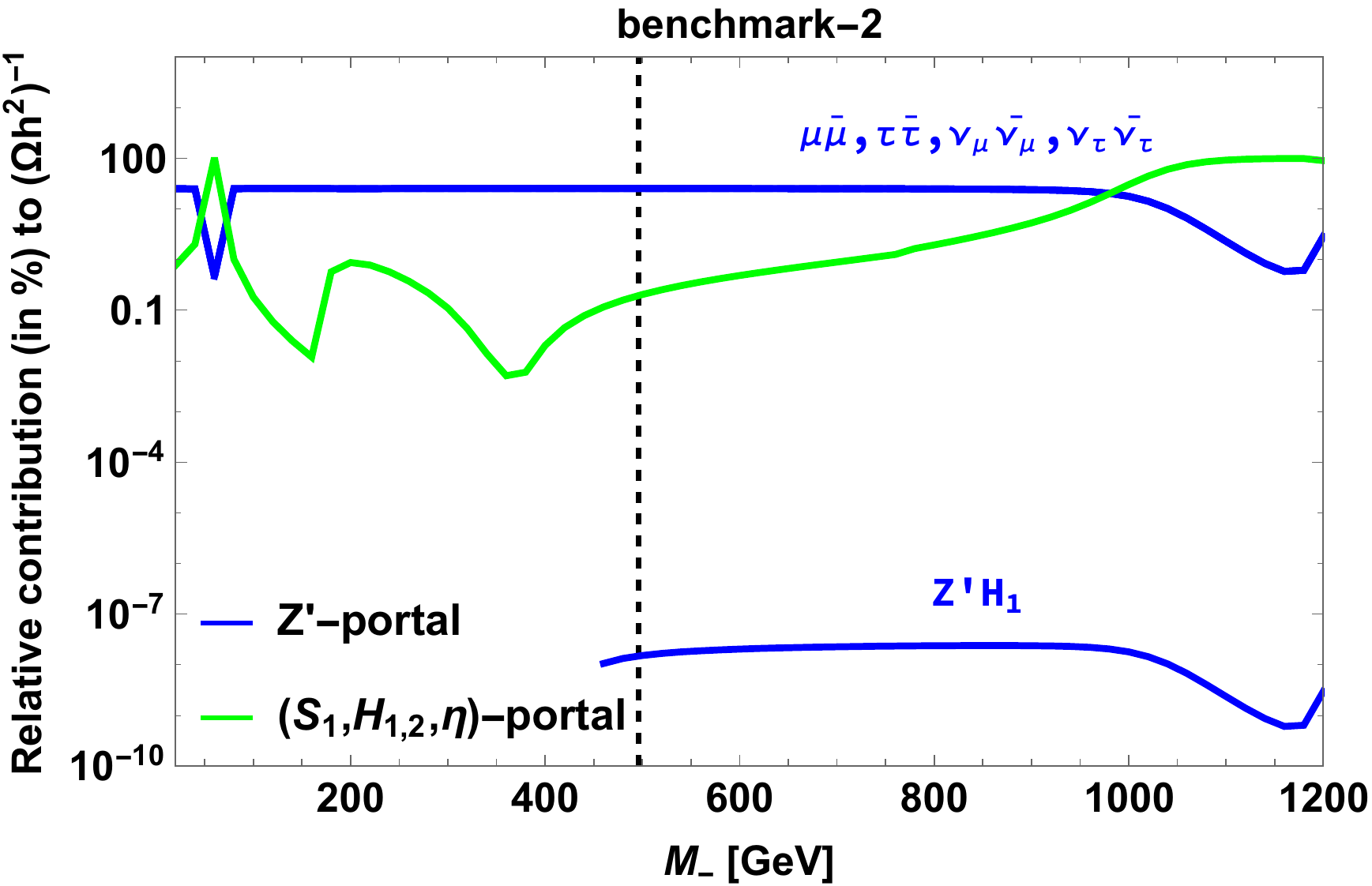}
\caption{Left and right panels depict the relative contribution of annihilation channels for the benchmark values of Table. \ref{bench_contri}. Vertical dashed line represents value of DM mass corresponding to the benchmark.}
\label{relative_con}
\end{center}
\end{figure}

To illustrate the relative contribution of annihilation channels to relic density for the parameter space depicted in Fig. \ref{pspace}, we choose two benchmark values (Table. \ref{bench_contri}), particularly the composition of maximum $(y_{qR})^2$ and minimum $g_{\mu\tau}$ (benchmark-1) and vice-versa (benchmark-2). For these values, we show the relative contribution of each $S_1$-portal ($Z^\prime$-portal) channel in the left (right) panel of Fig. \ref{relative_con}. For benchmark-1 i.e., maximal $(y_{qR})^2$, the contribution (blue curve) from SLQ-portal channels - $d\bar{d},s\bar{s},b\bar{b}$ ($\sim 32\%$ each) dominate over $(Z^\prime,H_{1,2},\eta)$-portal contribution (green curve) for almost whole DM mass region except near the resonances in propagators  $Z^\prime,H_{1}$ and $H_2$. Similarly, for benchmark-2 i.e., maximal $g_{\mu\tau}$, the $Z^\prime$-portal channels - $\mu\bar{\mu}, \tau\bar{\tau},\nu_{\mu}\bar{\nu}_{\mu},\nu_{\tau}\bar{\nu}_{\tau}$ ($\sim 24\%$ each) provide dominant contribution (blue curve) over the rest (green curve) with exception to the region near resonances of the propagators $H_1$ and $H_2$. The contribution of the channel with $Z^{\prime} H_1$ in the final state is negligible due to $\zeta^2$ (Higgs mixing) factor, the process with $Z^{\prime} H_2$ as final state particles is not kinematically allowed ($M_{H_2} = 2.4$ TeV) for the displayed DM mass range. In the rest of the parameter region of Fig. \ref{pspace}, the dominant contribution however, depends on all the four parameters listed in Table. \ref{bench_contri}.
\begin{table}[htb]
\centering
\begin{tabular}{|c|c|c|c|c|}
\hline
S.No ~&~$(y_{qR})^2$ ~&~$g_{\mu\tau}$~&~ $M_{Z^\prime}$ [GeV] ~&~ $M_-$ [GeV]\\
\hline
\hline
1.~&~$1.016$~&~$0.336$~ &~$669.84$~&~$291.22$ \\
\hline
2.~&~$0.0003$~&~$0.94$&$763.29$~&~$496.7$\\
\hline
\end{tabular}
\caption{Sample benchmark values chosen from the allowed paramter space of Fig. \ref{pspace}.}\label{bench_contri}
\end{table}

%

\section{Implication on  $R_K$ and  $B_{(s)} \to K^*(\phi) \mu^+ \mu^-$ Processes} 
The constrained parameter space discussed in the previous  section can have an impact on $R_K$ and  the observables of $B \to V l^+ l^-$ process, where $V=K^*, \phi$ are the vector mesons, which subsequently decay into $K^* \to K \pi$ and $\phi \to K^+ K^-$ states. The $B \to V$ hadronic matrix elements of the local quark bilinear operators can be parametrized as \cite{Ali:1999mm, Wirbel:1985ji}
\bea
 &&  \langle V \left(k \right)| \bar{s}\gamma _\mu (1-\gamma_5) b | B\left(p\right)\rangle =    \epsilon_{\mu \nu \alpha \beta} \epsilon^{*\nu } p^\alpha 
q^\beta \frac{2V(q^2)}{M_B + M_{V}} - i \epsilon^*_\mu (M_B + M_{V}) A_1(q^2)\nn \\ 
&& \hspace*{2 cm} + i (\epsilon^* \cdotp q)(2p-q)_\mu \frac{A_2(q^2)}{M_B + M_{V}} + 
i \frac{2M_{V}}{q^2} (\epsilon^* \cdotp q) \left[ A_3(q^2) - A_0(q^2)\right] q_\mu\;,\label{kstar}
  \eea
where 
\begin{equation}
 A_3(s) = \frac{(M_B + M_{V})}{2M_{V}} A_1(s) - \frac{(M_B - M_{V})}{2M_{V}} A_2(s), 
\end{equation} 
$q^2$ is the momentum transfer between the  $B$ and $V$ mesons, i.e., $q_\mu = p_\mu - k_\mu$ and $\epsilon_\mu$ is the polarization vector of the 
$V$ meson. The full angular differential decay distribution for the processes $B^0 \to (K^{*0} \to K^- \pi^+)l^+ l^-$  and $B_s \to (\phi \to K^+ K^-) l^+ l^-$ in terms of $q^2$, $\theta_l$, $\theta_{V}$ and $\phi$ variables is given as  \cite{Bobeth:2008ij, Egede:2010zc, Egede:2008uy}
 \begin{equation}
 \frac{d^4\Gamma}{dq^2~ d\cos\theta_l ~d\cos\theta_{V}~ d\phi} = \frac{9}{32\pi} J\left(q^2, \theta_l, \theta_{V}, \phi\right)\;,
 \end{equation}
  where 
\bea
 J\left(q^2, \theta_l, \theta_{V}, \phi\right) &= & J^s_1 \sin^2\theta_{V} + J^c_1 \cos^2\theta_{V} + \left(J^s_2 \sin^2\theta_{V} 
+ J^c_2 \cos^2\theta_{V}\right) \cos2\theta_l \nn\\
& +& J_3 \sin^2\theta_{V} \sin^2\theta_l \cos2\phi + J_4 \sin2\theta_{V} \sin2\theta_l \cos\phi 
  +  J_5 \sin2\theta_{V} \sin\theta_l \cos\phi\nn\\
& +&(J_6^s \sin^2\theta_{V} +J_6^c \cos^2\theta_{V})\cos\theta_l
 + J_7 \sin2\theta_{V} \sin\theta_l \sin\phi 
\nn\\ 
& +&  J_8 \sin2\theta_{V} \sin2\theta_l \sin\phi + J_9 \sin^2\theta_{V} \sin^2\theta_l \sin2\phi\;,
\eea
 $\theta_l$ is  the angle between $l^-$ and 
$B_{(s)}$ in the dilepton frame, $\theta_{V}$ is defined as the angle between $K^-$ and $B_{(s)}$ in the $K^-\pi^+~(K^-K^+)$ frame, the angle between the normal of the $K^-\pi^+~(K^-K^+)$ and the dilepton  planes is given by $\phi$.  The complete expressions for $J_i^{s(c)},~i=1,2,...,9$ as a function of transversity amplitudes are given in the  Appendix B \cite{Altmannshofer:2008dz}. The   transversity  amplitudes written in terms of the form factors  and Wilson coefficients are as follows \cite{Altmannshofer:2008dz}
\bea
A_{\perp L,R}&=& N\sqrt{2  \lambda}\Big [ \left ( (C_9^{\rm eff}+C_9^{\prime \rm NP}) \mp C_{10} \right )
\frac{V(q^2)}{M_B+M_{V}} + \frac{2 m_b}{q^2} C_7 T_1(q^2) \Big]\;,\nn\\
A_{\para L,R}&=& -N \sqrt{2} (M_B^2 -M_{V}^2)\Big[ \left ((C_9^{\rm eff}-C_9^{\prime \rm NP}) \mp C_{10} \right ) \frac{A_1(q^2)}{M_B - M_{V}} + \frac{2 m_b}{q^2}C_7 T_2(q^2) \Big]\;,\nn\\
A_{0 L,R}& =& - \frac{N}{2 M_{V} \sqrt{s}} \Big[ \left (C_9^{\rm eff}-C_9^{\prime \rm NP}) \mp C_{10} \right ) \nn\\
&& \times \left ( (M_B^2 -M_{V}^2 -q^2)(M_B + M_{V}) A_1(q^2)-\lambda \frac{A_2(q^2)}{M_B + M_{V}} \right )\nn\\
&&+ 2 m_B C_7 \left ( (M_B^2 +3 M_{V}^2 -q^2) T_2(q^2) - \frac{\lambda}{M_B^2 - M_{V}^2} \right ) \Big]\;,\nn\\
A_t&=& 2N \sqrt{\frac{\lambda}{q^2}}   C_{10}  A_0(q^2),
\eea
where
\bea
N= V_{tb}V_{ts}^* \left [ \frac{G_F^2 \alpha_{\rm em}^2}{3 \cdot 2^{10} \pi^5 M_B^3} q^2 \beta_l \sqrt{\lambda} \right ]^{1/2}\;,~~~~~~\lambda = \lambda(M_{V}^2, M_B^2, q^2). 
\eea

The dilepton invariant mass spectrum for ${B} \rightarrow  V l^+ l^-$ decay after integration over all angles \cite{Bobeth:2008ij} is given by
 \bea
 \frac{d\Gamma}{dq^2} = \frac{3}{4} \left(J_1 - \frac{J_2}{3}\right), 
 \eea
where $J_i = 2J_i^s + J_i^c$.  The most interesting observables in these decay modes are the lepton non-universality parameter defined as 
\bea
R_{V} = \frac{{\rm Br}( B \to V \mu^+ \mu^-)}{{\rm Br}( B \to V e^+ e^-)},
\eea
and the form factor independent (FFI) observables \cite{DescotesGenon:2012zf}
\bea
P_4^\prime = \frac{J_4}{\sqrt{-J_2^s J_2^c}}, ~~~~~
P_5^\prime = \frac{J_5}{2\sqrt{-J_2^s J_2^c}}.
\eea
\begin{figure}[htb]
\centering
\includegraphics[scale=0.43]{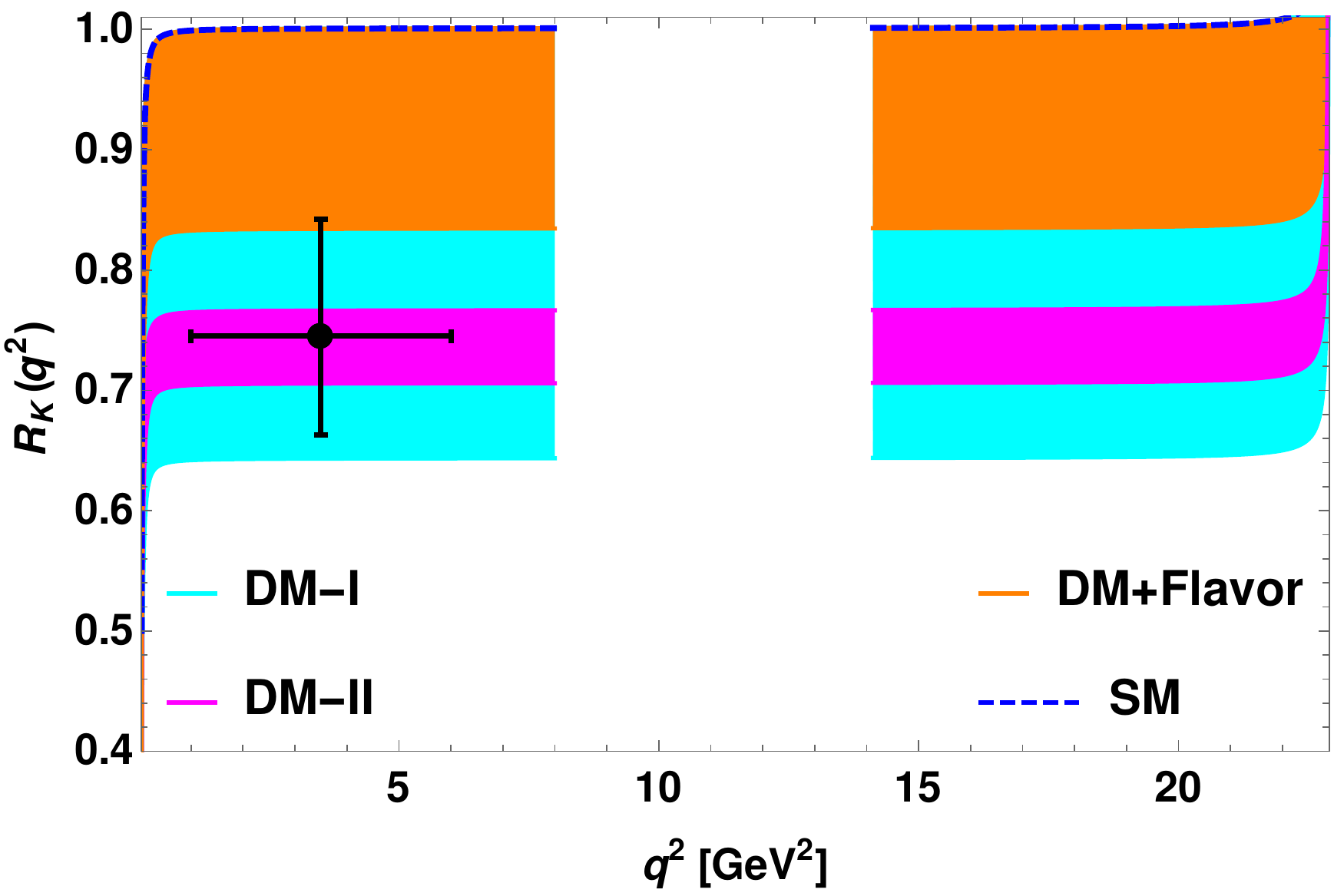}
\quad
\includegraphics[scale=0.6]{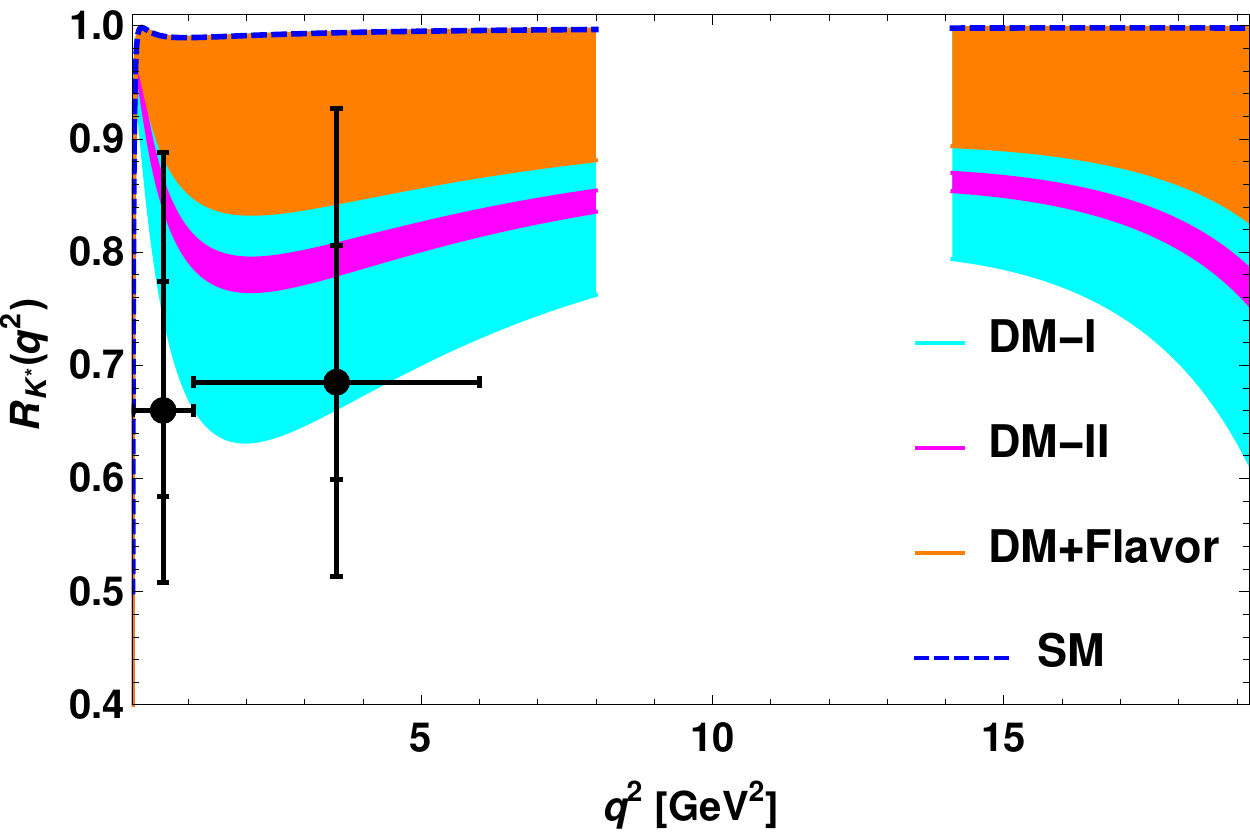}
\quad
\includegraphics[scale=0.6]{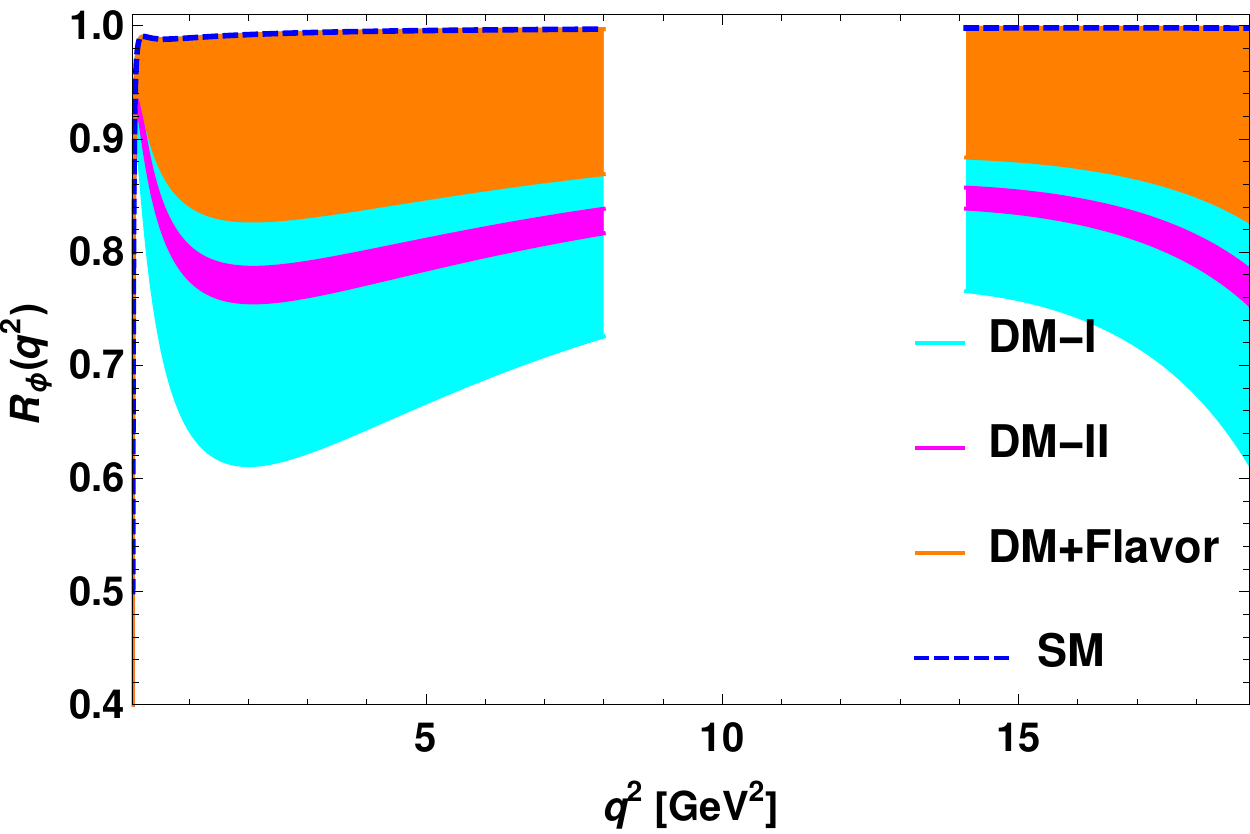}
\caption{The $q^2$ variation of  $R_{K}$ (top-left panel),  $R_{K^*}$ (top-right panel) and $R_\phi$ (bottom panel) LNU parameters in the $L_\mu-L_\tau$ model. Here the blue dashed lines represent the SM prediction, the cyan (magenta) bands stand for the NP contribution from the dark matter studies i.e., DM-I (DM-II). Orange bands are due to the contribution from both the flavor and DM sectors (DM+Flavor). The experimental data points (with 2$\sigma$ error bars) \cite{Aaij:2017vbb} are shown in black lines .}
\label{RKstar}
\end{figure}

\begin{figure}[htb]
\centering
\includegraphics[scale=0.6]{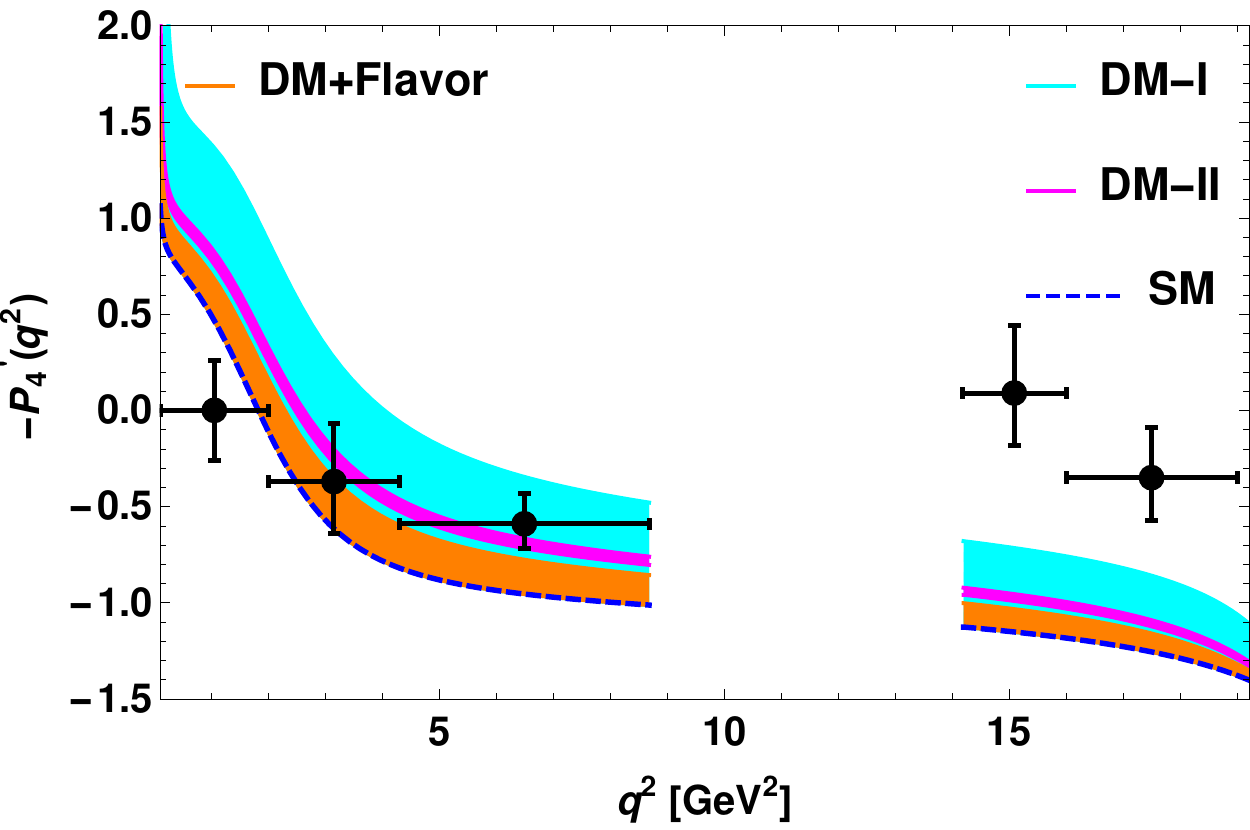}
\quad
\includegraphics[scale=0.6]{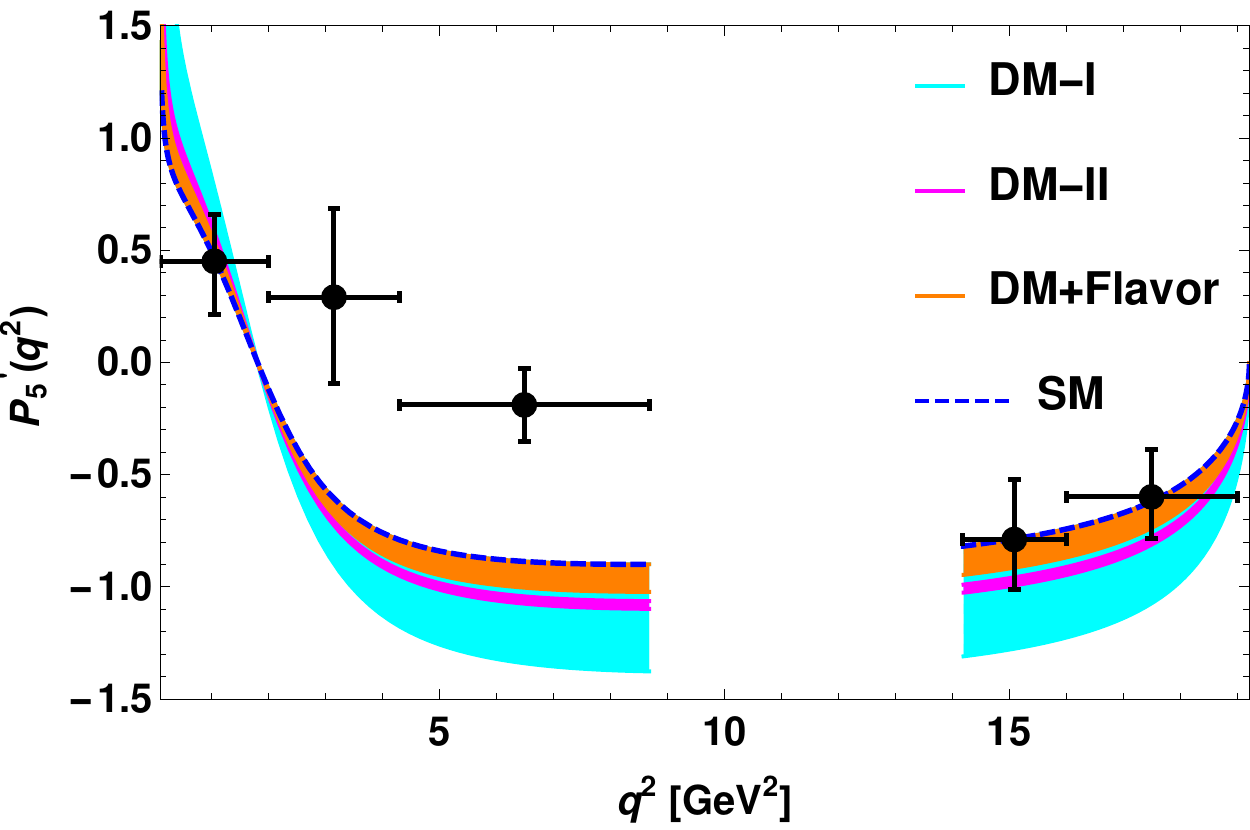}
\quad
\includegraphics[scale=0.6]{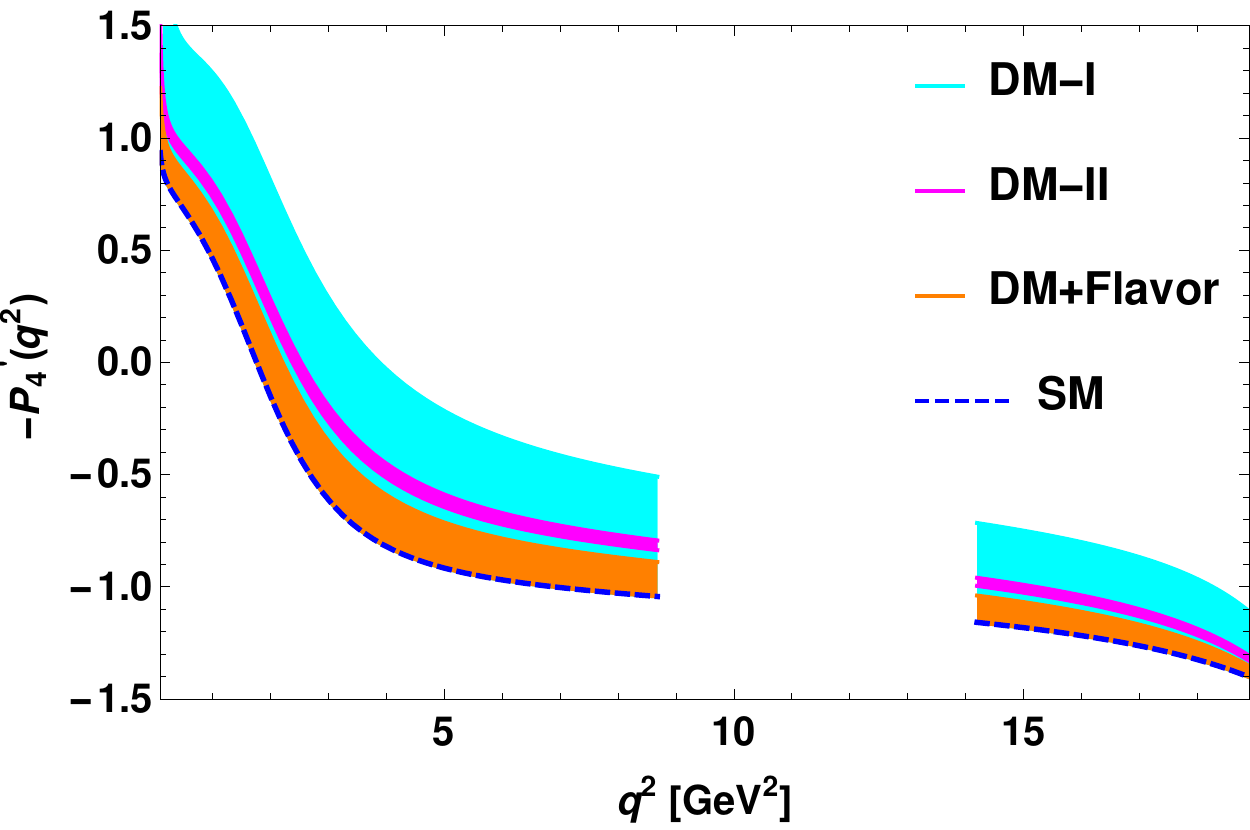}
\quad
\includegraphics[scale=0.6]{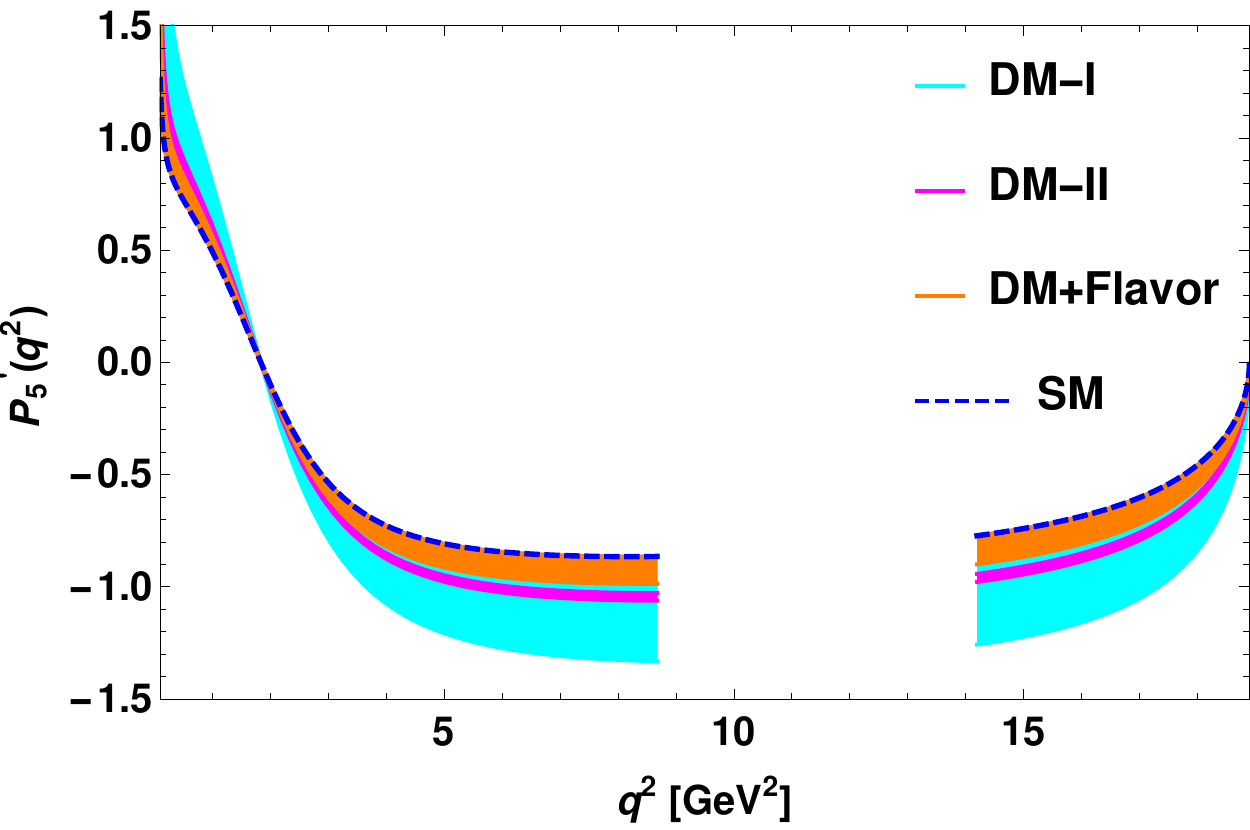}
\caption{Top panel represents the variation of $P_4^\prime$ (left panel) and $P_5^\prime$ (right panel) observables of $ B^0 \to K^{*0} \mu^+ \mu^-$ process with respect to $q^2$. The behaviour of $P_4^\prime$ (left panel) and $P_5^\prime$ (right panel) for $B_s \to \phi \mu^+ \mu^-$ are shown in the bottom panel. The bin-wise experimental data points with error bars  are shown in black  \cite{Aaij:2013qta}. Note that $P_4^\prime \big |^{\rm LHCb}=-P_4^\prime$.}
\label{Fig:P5p}
\end{figure}

After getting familiar with the different observables and the allowed values of the new parameters, we now proceed for numerical analysis in the full dilepton mass region i.e., $4m^2_l \leqslant q^2 \leqslant \left(M_B - M_{V}\right)^2$,
 leaving the regions around $q^2 \sim m_{J/\psi}^2$ and $m_{\psi^\prime}^2$. The cuts are employed to remove the dominant charmonium resonance $\left(c\bar{c}\right) = J/\psi, \psi^\prime $ backgrounds 
from $B \rightarrow V \left(c\bar{c}\right) \rightarrow V l^+ l^-$.  In Fig. \ref{RKstar}\,, we show the behaviour of $R_{K}$ (top-left panel), $R_{K^*}$ (top-right panel) and $R_\phi$ (bottom panel) with respect to $q^2$ in the  full kinematically accessible physical region. In these figures,  the blue dashed lines stand for the SM contribution, the orange bands are due to the  allowed  region of parameters  shown in Table \ref{allowed par}\,,  favored by both DM and flavor (DM+Flavor) and cyan (magenta) bands for only DM case i.e., DM-I (DM-II). The bin-wise experimental values of $R_{K^{(*)}}$ are shown in black. From the top-left panel of Fig. \ref{RKstar}\,, it can be seen that the result  obtained in the $q^2\in [1,6]~{\rm GeV}^2$ bin, by using the constraint from only  DM observable is consistent with the experimental data and can be explained within $1\sigma$ for DM+Flavor case.   The measured value of $R_{K^*}$ in the $q^2\in [0.045,1.1]~{\rm GeV}^2$ bin can be accommodated  within $1\sigma$ (DM-I)  and $2\sigma$  (DM-II and DM+Flavor),  $q^2\in [1.1,6]~{\rm GeV}^2$ bin result can be described within $1\sigma$ (DM-II)  and $2\sigma$  (DM+Flavor) and  is consistent with experimental data for DM-I case.  Though there is no experimental evidence for $R_\phi$ parameter, the additional NP   contribution arising from the allowed  parameter space of all cases (DM-I,II and DM+Flavor) provide significant deviation from the SM prediction,  implying the presence of lepton universality violation in the $B_s \to\phi \mu^+ \mu^-$ process.   In Table  \ref{results}\,, we present our  predicted values of $R_{K^{(*)}}$ and $R_\phi$  for different bins. The $q^2$ variation of famous optimized observables $-P_4^\prime$ (top-left panel) and $P_5^\prime$ (top-right panel)  of $ B^0 \to K^{*0} \mu^+ \mu^-$ process  are depicted in Fig. \ref{Fig:P5p}\,. The bottom panel of this figure describes analogous plots for $B_s \to \phi \mu^+ \mu^-$ process in both the high and low recoil limit. It should be noted that $P_4^\prime \big |^{\rm LHCb}=-P_4^\prime$.  In the low $q^2$ region, our predictions on $-P_4^\prime$ observable of $ B^0 \to K^{*0} \mu^+ \mu^-$ process   is in very good agreement with the LHCb data. For $ B^0 \to K^{*0} \mu^+ \mu^-$ decay mode, the model can accommodate the $P_5^\prime$ observable within  $1\sigma$ range  of the experimental data,  in the high and very low $q^2$ region. We notice considerable deviation between the results of SM  and  the presented $L_\mu-L_\tau$ model on the  $P_{4,5}^\prime$ observables for $B_s \to \phi \mu^+ \mu^-$ decay modes.   The numerical values of all these observables are given in Table \ref{results}\,.   We found that our results on the angular observables of  $ B \to V ll$ process,  obtained from DM-I parameter space  are almost consistent with
the corresponding measured experimental data.

\begin{table}[htb]
\centering
\begin{tabular}{|c|c|c|c|c|c|}
\hline
~&~Observables~&~ Values for SM~&~Values for DM-I~&~Values for DM-II~&~Values for DM+Flavor~\\
\hline
\hline
\multirow{7}{*}{} ~&$R_{K^*}|_{q^2\in [0.045,1.1]~{\rm GeV}^2}$~&~$0.949$ ~&~$0.793-0.949$ ~&~$0.852-0.863$ ~&~$0.881-0.949$~ \\ 
$B^0$~&$R_{K^*}|_{q^2\in [1.1,6]~{\rm GeV}^2}$~&~$0.993$ ~&~$0.671-0.993$ ~&~$0.786-0.811$~&~$0.848-0.993$~ \\
$\downarrow$~&$R_{K^*}|_{q^2\geq 14.18~{\rm GeV}^2}$~&~$0.998$ ~&~$0.754-0.998$ ~&~$0.832-0.851$~&~$0.879-0.998$~ \\ \cline{2-6}
$ K^{*0}$~&$P_5^\prime|_{q^2\in [1,6]~{\rm GeV}^2}$~&~$-0.542 \pm 0.044$ ~&~$-0.82\to -0.542$ ~&~$-0.655\to -0.635$~&~ $-0.612\to -0.542$~\\
$\mu^+$~&$P_5^\prime|_{q^2\geq 14.18~{\rm GeV}^2}$~&~$-0.695 \pm 0.056$ ~&~$-1.165\to -0.695$ ~&~$-0.894\to -0.86$ ~&~$-0.818 \to -0.695$ ~\\ 
$\mu^-$~&$P_4^\prime|_{q^2\in [1,6]~{\rm GeV}^2}$~&~ $0.563\pm 0.045$~&~$-0.239\to 0.563$ ~&~ $0.195-0.257$~&~$0.334\to 0.563$~ \\
&$P_4^\prime|_{q^2\geq 14.18~{\rm GeV}^2}$~&~$1.2 \pm 0.096$ ~&~$0.787-1.2$ ~&~$1.03-1.06$ ~&~$1.098-1.2$~ \\
\hline
\multirow{7}{*}{}~&$R_{\phi}|_{q^2\in [0.045,1.1]~{\rm GeV}^2}$~&~$0.9499$ ~&~$0.753-0.9499$ ~&~$0.828-0.844$ ~&~$0.866-0.9499$~ \\
$B_s$~&$R_{\phi}|_{q^2\in [1.1,6]~{\rm GeV}^2}$~&~$0.994$ ~&~$0.642-0.994$ ~&~$0.77-0.8$ ~&~$0.839-0.994$ ~\\
$\downarrow$~&$R_{\phi}|_{q^2\geq 14.18~{\rm GeV}^2}$~&~$0.998$ ~&~$0.731-0.998$ ~&~$0.82-0.84$ ~&~$0.871-0.998$~ \\  \cline{2-6}
$\phi$~&~$P_5^\prime|_{q^2\in [1,6]~{\rm GeV}^2}$~&~$-0.511 \pm 0.041$ ~&~$-0.766\to -0.551$ ~&~$-0.618\to -0.599$ ~&~ $-0.578\to -0.511$~\\
$\mu^+$~&$P_5^\prime|_{q^2\geq 14.18~{\rm GeV}^2}$~&~$-0.645 \pm 0.052$ &$-1.1 \to -0.645$ &$-0.839\to -0.805$ &$-0.765 \to -0.645$\\ 
$\mu^-$&$P_4^\prime|_{q^2\in [1,6]~{\rm GeV}^2}$& $0.596\pm 0.048$&$-0.202\to 0.596$ &$0.228-0.29$ &$0.367\to 0.596$ \\
&$P_4^\prime|_{q^2\geq 14.18~{\rm GeV}^2}$&$1.233 \pm 0.099$ &$0.822-1.233$ &$1.059-1.092$ &$1.13-1.233$\\

\hline
~&$R_{K}|_{q^2\in [1,6]~{\rm GeV}^2}$~&~$1.0004$~&~$0.643-1.0004$ ~&~$0.71-0.766$~&~$0.834-1.0004$~ \\
~&$R_{K}|_{q^2\geq 14.18~{\rm GeV}^2}$~&~$1.002$ ~&~$0.645-1.002$ ~&~$0.71-0.768$~&~$0.836-1.002$~\\
\hline
\end{tabular}
\caption{Predicted numerical values of LNU parameters $(R_V)$ and $P_{4,5}^\prime$ observables of $ B_{(s)} \to V \mu^+ \mu^-, ~V=K^*,\phi $ processes in the high and low recoil limits.  The  values of $R_K$ parameter are also presented in both low and high $q^2$ regime}. \label{results}
\end{table}

\section{Summary and Conclusion}

Summarizing the article, we have studied Majorana dark matter in a new version of $U(1)_{L_\mu-L_\tau}$ gauge extension of the standard model. The model is free from triangle gauge anomalies with the inclusion of three neutral fermions with $L_{\mu}-L_{\tau}$ charges $0,1$ and $-1$.  A scalar singlet, charged $+2$ under the new $U(1)$ is added to spontaneously break the $L_{\mu}-L_{\tau}$ gauge symmetry, thereby giving masses to the new fermions and the neutral boson $Z^\prime$ associated with gauge extension. In addition, the scalar sector is enriched with an inert doublet and a $(\bar{3},1,1/3)$ scalar leptoquark to obtain the neutrino mass at one-loop level and address the flavor anomalies respectively. All the new fermions, leptoquark and inert doublet are assigned with charge $-1$ under $Z_2$ symmetry. Choosing the lightest mass eigenstate of the new fermion spectrum as dark matter, we made a thorough study of Majorana dark matter in relic density and direct detection perspective. The channels contributing to relic density are mediated by the scalar leptoquark, $Z^\prime$ and inert doublet components. As $Z^\prime$ doesn't couple to quarks, the $Z^\prime$-mediated tree level interaction for direct detection is not permitted. Only leptoquark portal channels contribute to spin-dependent WIMP-nucleon cross section. Imposing Planck limit on relic density and well known PICO-60, LUX bounds on spin-dependent cross section, we have constrained the new parameters of the model. We have also computed the spin-dependent and spin-independent contributions of one-loop diagrams involving scalar leptoquark, but found to give zero impact on the model parameters. We have also showed the mechanism of generating light neutrino mass radiatively using the inert doublet. A note on the viable parameter region consistent with neutrino oscillation data is also addressed.

We have further restricted the new parameters from quark and lepton sectors i.e., by comparing the theoretical predictions of ${\rm Br}(\tau \to \mu \nu_\tau \bar \nu_\mu)$, ${\rm Br}( B \to X_s \gamma)$, ${\rm Br}(B^+ \to K^+ \tau^+ \tau^-)$, ${\rm Br}(\bar B^0 \to \bar K^0 \mu^+ \mu^-)$ and $B_s-\bar B_s$ mixing with their corresponding $3\sigma$  experimental data.  The  neutral and charged lepton flavor violating decay processes  are absent due to zero $Z^\prime \tau \mu$  coupling. And also the vanishing $Z^\prime q \bar q$ coupling restricts the involvement of $Z^\prime$  in 
$B_s -\bar B_s$ mixing, $b \to s \gamma$ processes  at one-loop level. We have then investigated the implication on $P^\prime_{4,5}$, $R_{K^{(*)}}$ and $R_\phi$ observables of $ B_{(s)} \to  K^{(*)}(\phi) l^+ l^-$ decay modes in the full kinematically allowed $q^2$ region for two cases i.e., dark matter and flavor allowed, only dark matter allowed parameter space.   We observed that our model can explain the $R_K$ LNU parameter very well.  The $R_{K^*}$ observable obtained from the allowed parameter space consistent with only dark matter is found to be within its $1\sigma$ and from both dark matter and flavor is within $2\sigma$ experimental limit. In the presence of new physics, the violation of lepton universality is observed in $B_s \to \phi \mu^+ \mu^-$ process,  thus, can be probed in LHCb 
experiment. We  noticed that the proposed $L_\mu-L_\tau$ model is also able to explain the LHCb experimental data of the   famous optimized $P_{4,5}^\prime$ observables of $ B^0 \to  K^{*0} \mu^+ \mu^-$ process. We also perceived that the form factor independent  observables  for $B_s \to \phi \mu^+ \mu^-$  decay modes have sizeable deviation from the standard model. We observed that the parameter region satisfying only dark matter observables for $M_- \le 560$ GeV have a good impact on the flavor anomalies. To conclude, we have made a comprehensive study of Majorana dark matter, neutrino phenomenology and flavor anomalies in a $U(1)_{L_{\mu}-L_{\mu}}$ gauge extended model. This simple framework survives all the current experimental limits on dark matter and flavor observables, can be probed in upcoming high luminosity experiments.

\acknowledgments 
SS would like to thank Department of Science and Technology (DST) - Inspire Fellowship division, Govt. of India for the financial support through ID No. IF130927.  RM would like to thank Science and Engineering Research Board (SERB), Government of India for financial support through grant Nos. SB/S2/HEP-017/2013 and EMR/2017/001448. 

\appendix

\section{Loop functions}
The loop functions required to compute the  $b \to s ll$ decays are given by \cite{Baek:2017sew, Hisano:1995cp}
\bea
 \mathcal{V}_{sb}(\chi_-, \chi_+)&=&\cos^2\alpha\sin^2\alpha\left(1+4\sqrt{\chi_-\chi_+}j\left(\chi_-,\chi_+\right)-2k\left(\chi_-,\chi_+\right)\right)\nn \\&+&2\sin^2\alpha I(\chi_+)+2\cos^2\alpha I(\chi_-)\,,
\eea
where 
\bea \label{A:loop-2}
f(\chi_1, \chi_2, \chi_3,\cdots)\equiv\frac{f(\chi_1, \chi_3,\cdots)-f(\chi_2, \chi_3,\cdots)}{\chi_1-\chi_2},~~~~~~~~f=j,\kappa\,,
\eea
with
\bea \label{A:loop-3}
j(\chi)&=&\frac{\chi \log \chi}{\chi-1}\,,\\\label{A:loop-4}
\kappa(\chi)&=&\frac{\chi^2 \log \chi}{\chi-1}\,,\\\label{A:loop-5}
I(\chi)&=&\frac{-3\chi^2+4\chi-1+2\chi^2\log \chi}{8(\chi-1)^2}\,.
\eea 
Eqns. (\ref{A:loop-2}, \ref{A:loop-3}, \ref{A:loop-4}) are used to investigate $B_s-\bar B_s$ mixing.

\section{$J_i$ coefficients of $ B_{(s)} \to  V l l$ processes}
The  expressions for the $J_i$ coefficients of $ B_{(s)} \to  V l l$ process in terms of transversity amplitudes are given by \cite{Egede:2010zc, Altmannshofer:2008dz} 
 \bea
 J^s_1 &=& \frac{\left(2+\beta ^2_l\right)}{4}\Bigg[|A_\perp ^L|^2 + |A_\parallel ^L|^2 + \left(L\rightarrow R\right)\Bigg] + \frac{4m^2_l}{q^2} 
{\rm Re}\left(A_\perp ^L A_\perp ^{R^*} + A_\parallel ^L A_\parallel ^{R^*}\right),~~ \\
  J^c_1 &=& |A_0^L|^2 + |A_0^R|^2 +\frac{4m^2_l}{q^2}\Bigg[|A_t|^2 + 2{\rm Re}\left(A_0^L A_0^{R^*}\right)\Bigg] + \beta^2_l |A_S|^2, \\
  J^s_2 &=& \frac{\beta^2_l}{4}\left[|A_\perp ^L|^2 + |A_\parallel ^L|^2 + \left(L\rightarrow R\right)\right], \\
  J^c_2 &=& -\beta^2_l\left[|A_0^L|^2 +\left(L\rightarrow R\right)\right], \\
 J_3 &=& \frac{1}{2}\beta^2_l\left[|A_\perp ^L|^2 - |A_\parallel ^L|^2 + \left(L \rightarrow R\right)\right], \\
 J_4 &=& \frac{1}{\sqrt{2}}\beta^2_l\left[{\rm Re}\left(A_0^L A_\parallel ^{L^*}\right) + \left(L \rightarrow R\right)\right], \\
 J_5 &=& \sqrt{2}\beta_l\left[{\rm Re}\left(A_0^L A_\perp ^{L^*}\right) - \left(L \rightarrow R\right) - 
\frac{m_l}{\sqrt{q^2}} {\rm Re}\left(A_\parallel ^L A^* _S + A_\parallel ^R A^* _S\right)\right], \\
  J^s_6 &=& 2\beta_l\left[{\rm Re}\left(A_\parallel ^L A_\perp ^{L^*}\right) - \left(L \rightarrow R\right)\right], 
  \eea
  \bea
  J^c_6 &=& 4\beta_l\frac{m_l}{\sqrt{q^2}} {\rm Re}\left[A_0^L A_S^* + \left(L \rightarrow R\right)\right], \\
 J_7 &=& \sqrt{2}\beta_l\left[{\rm Im}\left(A_0^L A_\parallel ^{L^*}\right) - \left(L \rightarrow R\right) + \frac{m_l}{\sqrt{q^2}} 
{\rm Im}\left(A_\perp ^L A^* _S + A_\perp ^R A^* _S\right)\right], \\
  J_8 &=& \frac{1}{\sqrt{2}}\beta^2_l\left[{\rm Im}\left(A_0^L A_\perp ^{L^*}\right) + \left(L \rightarrow R\right)\right], \\
 J_9 &=& \beta^2_l\left[{\rm Im}\left(A_\parallel^{L^*} A_\perp ^L\right) + \left(L \rightarrow R\right)\right],
 \eea
 where 
 \bea
 A_i A_j^* = A_{i}^{ L}\left(q^2\right) A^{* L}_{j}\left(q^2\right) + A_{i}^{ R}\left(q^2\right) A^{* R}_{j}\left(q^2\right) 
\hspace{1cm} \left(i,j = 0, \parallel, \perp\right),
 \eea
 in shorthand notation.

\bibliography{BL}

\begin{thebibliography}{94}
\expandafter\ifx\csname natexlab\endcsname\relax\def\natexlab#1{#1}\fi
\expandafter\ifx\csname bibnamefont\endcsname\relax
  \def\bibnamefont#1{#1}\fi
\expandafter\ifx\csname bibfnamefont\endcsname\relax
  \def\bibfnamefont#1{#1}\fi
\expandafter\ifx\csname citenamefont\endcsname\relax
  \def\citenamefont#1{#1}\fi
\expandafter\ifx\csname url\endcsname\relax
  \def\url#1{\texttt{#1}}\fi
\expandafter\ifx\csname urlprefix\endcsname\relax\def\urlprefix{URL }\fi
\providecommand{\bibinfo}[2]{#2}
\providecommand{\eprint}[2][]{\url{#2}}

\bibitem[{\citenamefont{Aaij et~al.}(2017)}]{Aaij:2017vbb}
\bibinfo{author}{\bibfnamefont{R.}~\bibnamefont{Aaij}} \bibnamefont{et~al.}
  (\bibinfo{collaboration}{LHCb}), \bibinfo{journal}{JHEP}
  \textbf{\bibinfo{volume}{08}}, \bibinfo{pages}{055} (\bibinfo{year}{2017}),
  \eprint{1705.05802}.

\bibitem[{\citenamefont{Aaij et~al.}(2014{\natexlab{a}})}]{Aaij:2014ora}
\bibinfo{author}{\bibfnamefont{R.}~\bibnamefont{Aaij}} \bibnamefont{et~al.}
  (\bibinfo{collaboration}{LHCb}), \bibinfo{journal}{Phys. Rev. Lett.}
  \textbf{\bibinfo{volume}{113}}, \bibinfo{pages}{151601}
  (\bibinfo{year}{2014}{\natexlab{a}}), \eprint{1406.6482}.

\bibitem[{\citenamefont{Aaij et~al.}(2013{\natexlab{a}})}]{Aaij:2013aln}
\bibinfo{author}{\bibfnamefont{R.}~\bibnamefont{Aaij}} \bibnamefont{et~al.}
  (\bibinfo{collaboration}{LHCb}), \bibinfo{journal}{JHEP}
  \textbf{\bibinfo{volume}{07}}, \bibinfo{pages}{084}
  (\bibinfo{year}{2013}{\natexlab{a}}), \eprint{1305.2168}.

\bibitem[{\citenamefont{Aaij et~al.}(2013{\natexlab{b}})}]{Aaij:2013qta}
\bibinfo{author}{\bibfnamefont{R.}~\bibnamefont{Aaij}} \bibnamefont{et~al.}
  (\bibinfo{collaboration}{LHCb}), \bibinfo{journal}{Phys. Rev. Lett.}
  \textbf{\bibinfo{volume}{111}}, \bibinfo{pages}{191801}
  (\bibinfo{year}{2013}{\natexlab{b}}), \eprint{1308.1707}.

\bibitem[{\citenamefont{Aaij et~al.}(2014{\natexlab{b}})}]{Aaij:2014pli}
\bibinfo{author}{\bibfnamefont{R.}~\bibnamefont{Aaij}} \bibnamefont{et~al.}
  (\bibinfo{collaboration}{LHCb}), \bibinfo{journal}{JHEP}
  \textbf{\bibinfo{volume}{06}}, \bibinfo{pages}{133}
  (\bibinfo{year}{2014}{\natexlab{b}}), \eprint{1403.8044}.

\bibitem[{\citenamefont{Langenbruch}(2015)}]{Langenbruch:2015dqz}
\bibinfo{author}{\bibfnamefont{C.}~\bibnamefont{Langenbruch}}
  (\bibinfo{collaboration}{LHCb}), in \emph{\bibinfo{booktitle}{{Proceedings,
  50th Rencontres de Moriond Electroweak Interactions and Unified Theories: La
  Thuile, Italy, March 14-21, 2015}}} (\bibinfo{year}{2015}), pp.
  \bibinfo{pages}{317--324}, \eprint{1505.04160},
  \urlprefix\url{http://inspirehep.net/record/1370442/files/arXiv:1505.04160.pdf}.

\bibitem[{\citenamefont{Bobeth et~al.}(2007)\citenamefont{Bobeth, Hiller, and
  Piranishvili}}]{Bobeth:2007dw}
\bibinfo{author}{\bibfnamefont{C.}~\bibnamefont{Bobeth}},
  \bibinfo{author}{\bibfnamefont{G.}~\bibnamefont{Hiller}}, \bibnamefont{and}
  \bibinfo{author}{\bibfnamefont{G.}~\bibnamefont{Piranishvili}},
  \bibinfo{journal}{JHEP} \textbf{\bibinfo{volume}{12}}, \bibinfo{pages}{040}
  (\bibinfo{year}{2007}), \eprint{0709.4174}.

\bibitem[{\citenamefont{Capdevila et~al.}(2018)\citenamefont{Capdevila,
  Crivellin, Descotes-Genon, Matias, and Virto}}]{Capdevila:2017bsm}
\bibinfo{author}{\bibfnamefont{B.}~\bibnamefont{Capdevila}},
  \bibinfo{author}{\bibfnamefont{A.}~\bibnamefont{Crivellin}},
  \bibinfo{author}{\bibfnamefont{S.}~\bibnamefont{Descotes-Genon}},
  \bibinfo{author}{\bibfnamefont{J.}~\bibnamefont{Matias}}, \bibnamefont{and}
  \bibinfo{author}{\bibfnamefont{J.}~\bibnamefont{Virto}},
  \bibinfo{journal}{JHEP} \textbf{\bibinfo{volume}{01}}, \bibinfo{pages}{093}
  (\bibinfo{year}{2018}), \eprint{1704.05340}.

\bibitem[{\citenamefont{He et~al.}(1991{\natexlab{a}})\citenamefont{He, Joshi,
  Lew, and Volkas}}]{He:1990pn}
\bibinfo{author}{\bibfnamefont{X.~G.} \bibnamefont{He}},
  \bibinfo{author}{\bibfnamefont{G.~C.} \bibnamefont{Joshi}},
  \bibinfo{author}{\bibfnamefont{H.}~\bibnamefont{Lew}}, \bibnamefont{and}
  \bibinfo{author}{\bibfnamefont{R.~R.} \bibnamefont{Volkas}},
  \bibinfo{journal}{Phys. Rev.} \textbf{\bibinfo{volume}{D43}},
  \bibinfo{pages}{22} (\bibinfo{year}{1991}{\natexlab{a}}).

\bibitem[{\citenamefont{He et~al.}(1991{\natexlab{b}})\citenamefont{He, Joshi,
  Lew, and Volkas}}]{He:1991qd}
\bibinfo{author}{\bibfnamefont{X.-G.} \bibnamefont{He}},
  \bibinfo{author}{\bibfnamefont{G.~C.} \bibnamefont{Joshi}},
  \bibinfo{author}{\bibfnamefont{H.}~\bibnamefont{Lew}}, \bibnamefont{and}
  \bibinfo{author}{\bibfnamefont{R.~R.} \bibnamefont{Volkas}},
  \bibinfo{journal}{Phys. Rev.} \textbf{\bibinfo{volume}{D44}},
  \bibinfo{pages}{2118} (\bibinfo{year}{1991}{\natexlab{b}}).

\bibitem[{\citenamefont{Crivellin
  et~al.}(2015{\natexlab{a}})\citenamefont{Crivellin, D'Ambrosio, and
  Heeck}}]{Crivellin:2015mga}
\bibinfo{author}{\bibfnamefont{A.}~\bibnamefont{Crivellin}},
  \bibinfo{author}{\bibfnamefont{G.}~\bibnamefont{D'Ambrosio}},
  \bibnamefont{and} \bibinfo{author}{\bibfnamefont{J.}~\bibnamefont{Heeck}},
  \bibinfo{journal}{Phys. Rev. Lett.} \textbf{\bibinfo{volume}{114}},
  \bibinfo{pages}{151801} (\bibinfo{year}{2015}{\natexlab{a}}),
  \eprint{1501.00993}.

\bibitem[{\citenamefont{Patra et~al.}(2017)\citenamefont{Patra, Rao, Sahoo, and
  Sahu}}]{Patra:2016shz}
\bibinfo{author}{\bibfnamefont{S.}~\bibnamefont{Patra}},
  \bibinfo{author}{\bibfnamefont{S.}~\bibnamefont{Rao}},
  \bibinfo{author}{\bibfnamefont{N.}~\bibnamefont{Sahoo}}, \bibnamefont{and}
  \bibinfo{author}{\bibfnamefont{N.}~\bibnamefont{Sahu}},
  \bibinfo{journal}{Nucl. Phys.} \textbf{\bibinfo{volume}{B917}},
  \bibinfo{pages}{317} (\bibinfo{year}{2017}), \eprint{1607.04046}.

\bibitem[{\citenamefont{Biswas et~al.}(2016)\citenamefont{Biswas, Choubey, and
  Khan}}]{Biswas:2016yan}
\bibinfo{author}{\bibfnamefont{A.}~\bibnamefont{Biswas}},
  \bibinfo{author}{\bibfnamefont{S.}~\bibnamefont{Choubey}}, \bibnamefont{and}
  \bibinfo{author}{\bibfnamefont{S.}~\bibnamefont{Khan}},
  \bibinfo{journal}{JHEP} \textbf{\bibinfo{volume}{09}}, \bibinfo{pages}{147}
  (\bibinfo{year}{2016}), \eprint{1608.04194}.

\bibitem[{\citenamefont{Kamada et~al.}(2018)\citenamefont{Kamada, Kaneta,
  Yanagi, and Yu}}]{Kamada:2018zxi}
\bibinfo{author}{\bibfnamefont{A.}~\bibnamefont{Kamada}},
  \bibinfo{author}{\bibfnamefont{K.}~\bibnamefont{Kaneta}},
  \bibinfo{author}{\bibfnamefont{K.}~\bibnamefont{Yanagi}}, \bibnamefont{and}
  \bibinfo{author}{\bibfnamefont{H.-B.} \bibnamefont{Yu}},
  \bibinfo{journal}{JHEP} \textbf{\bibinfo{volume}{06}}, \bibinfo{pages}{117}
  (\bibinfo{year}{2018}), \eprint{1805.00651}.

\bibitem[{\citenamefont{Bauer et~al.}(2018)\citenamefont{Bauer, Diefenbacher,
  Plehn, Russell, and Camargo}}]{Bauer:2018egk}
\bibinfo{author}{\bibfnamefont{M.}~\bibnamefont{Bauer}},
  \bibinfo{author}{\bibfnamefont{S.}~\bibnamefont{Diefenbacher}},
  \bibinfo{author}{\bibfnamefont{T.}~\bibnamefont{Plehn}},
  \bibinfo{author}{\bibfnamefont{M.}~\bibnamefont{Russell}}, \bibnamefont{and}
  \bibinfo{author}{\bibfnamefont{D.~A.} \bibnamefont{Camargo}}
  (\bibinfo{year}{2018}), \eprint{1805.01904}.

\bibitem[{\citenamefont{Crivellin
  et~al.}(2015{\natexlab{b}})\citenamefont{Crivellin, D'Ambrosio, and
  Heeck}}]{Crivellin:2015lwa}
\bibinfo{author}{\bibfnamefont{A.}~\bibnamefont{Crivellin}},
  \bibinfo{author}{\bibfnamefont{G.}~\bibnamefont{D'Ambrosio}},
  \bibnamefont{and} \bibinfo{author}{\bibfnamefont{J.}~\bibnamefont{Heeck}},
  \bibinfo{journal}{Phys. Rev.} \textbf{\bibinfo{volume}{D91}},
  \bibinfo{pages}{075006} (\bibinfo{year}{2015}{\natexlab{b}}),
  \eprint{1503.03477}.

\bibitem[{\citenamefont{Das et~al.}(2017)\citenamefont{Das, Nomura, Okada, and
  Roy}}]{Das:2017ski}
\bibinfo{author}{\bibfnamefont{A.}~\bibnamefont{Das}},
  \bibinfo{author}{\bibfnamefont{T.}~\bibnamefont{Nomura}},
  \bibinfo{author}{\bibfnamefont{H.}~\bibnamefont{Okada}}, \bibnamefont{and}
  \bibinfo{author}{\bibfnamefont{S.}~\bibnamefont{Roy}},
  \bibinfo{journal}{Phys. Rev.} \textbf{\bibinfo{volume}{D96}},
  \bibinfo{pages}{075001} (\bibinfo{year}{2017}), \eprint{1704.02078}.

\bibitem[{\citenamefont{Das et~al.}(2018{\natexlab{a}})\citenamefont{Das,
  Okada, and Raut}}]{Das:2017deo}
\bibinfo{author}{\bibfnamefont{A.}~\bibnamefont{Das}},
  \bibinfo{author}{\bibfnamefont{N.}~\bibnamefont{Okada}}, \bibnamefont{and}
  \bibinfo{author}{\bibfnamefont{D.}~\bibnamefont{Raut}},
  \bibinfo{journal}{Eur. Phys. J.} \textbf{\bibinfo{volume}{C78}},
  \bibinfo{pages}{696} (\bibinfo{year}{2018}{\natexlab{a}}),
  \eprint{1711.09896}.

\bibitem[{\citenamefont{Das et~al.}(2018{\natexlab{b}})\citenamefont{Das,
  Okada, and Raut}}]{Das:2017flq}
\bibinfo{author}{\bibfnamefont{A.}~\bibnamefont{Das}},
  \bibinfo{author}{\bibfnamefont{N.}~\bibnamefont{Okada}}, \bibnamefont{and}
  \bibinfo{author}{\bibfnamefont{D.}~\bibnamefont{Raut}},
  \bibinfo{journal}{Phys. Rev.} \textbf{\bibinfo{volume}{D97}},
  \bibinfo{pages}{115023} (\bibinfo{year}{2018}{\natexlab{b}}),
  \eprint{1710.03377}.

\bibitem[{\citenamefont{Baek}(2018)}]{Baek:2017sew}
\bibinfo{author}{\bibfnamefont{S.}~\bibnamefont{Baek}}, \bibinfo{journal}{Phys.
  Lett.} \textbf{\bibinfo{volume}{B781}}, \bibinfo{pages}{376}
  (\bibinfo{year}{2018}), \eprint{1707.04573}.

\bibitem[{\citenamefont{Mandal}(2018)}]{Mandal:2018czf}
\bibinfo{author}{\bibfnamefont{R.}~\bibnamefont{Mandal}}
  (\bibinfo{year}{2018}), \eprint{1808.07844}.

\bibitem[{\citenamefont{Arcadi et~al.}(2017)\citenamefont{Arcadi, Dutra, Ghosh,
  Lindner, Mambrini, Pierre, Profumo, and Queiroz}}]{Arcadi:2017kky}
\bibinfo{author}{\bibfnamefont{G.}~\bibnamefont{Arcadi}},
  \bibinfo{author}{\bibfnamefont{M.}~\bibnamefont{Dutra}},
  \bibinfo{author}{\bibfnamefont{P.}~\bibnamefont{Ghosh}},
  \bibinfo{author}{\bibfnamefont{M.}~\bibnamefont{Lindner}},
  \bibinfo{author}{\bibfnamefont{Y.}~\bibnamefont{Mambrini}},
  \bibinfo{author}{\bibfnamefont{M.}~\bibnamefont{Pierre}},
  \bibinfo{author}{\bibfnamefont{S.}~\bibnamefont{Profumo}}, \bibnamefont{and}
  \bibinfo{author}{\bibfnamefont{F.~S.} \bibnamefont{Queiroz}}
  (\bibinfo{year}{2017}), \eprint{1703.07364}.

\bibitem[{\citenamefont{Allahverdi et~al.}(2018)\citenamefont{Allahverdi, Dev,
  and Dutta}}]{Allahverdi:2017edd}
\bibinfo{author}{\bibfnamefont{R.}~\bibnamefont{Allahverdi}},
  \bibinfo{author}{\bibfnamefont{P.~S.~B.} \bibnamefont{Dev}},
  \bibnamefont{and} \bibinfo{author}{\bibfnamefont{B.}~\bibnamefont{Dutta}},
  \bibinfo{journal}{Phys. Lett.} \textbf{\bibinfo{volume}{B779}},
  \bibinfo{pages}{262} (\bibinfo{year}{2018}), \eprint{1712.02713}.

\bibitem[{\citenamefont{Georgi and Glashow}(1974)}]{Georgi:1974sy}
\bibinfo{author}{\bibfnamefont{H.}~\bibnamefont{Georgi}} \bibnamefont{and}
  \bibinfo{author}{\bibfnamefont{S.~L.} \bibnamefont{Glashow}},
  \bibinfo{journal}{Phys. Rev. Lett.} \textbf{\bibinfo{volume}{32}},
  \bibinfo{pages}{438} (\bibinfo{year}{1974}).

\bibitem[{\citenamefont{Georgi}(1975)}]{Georgi:1974my}
\bibinfo{author}{\bibfnamefont{H.}~\bibnamefont{Georgi}}, \bibinfo{journal}{AIP
  Conf. Proc.} \textbf{\bibinfo{volume}{23}}, \bibinfo{pages}{575}
  (\bibinfo{year}{1975}).

\bibitem[{\citenamefont{Langacker}(1981)}]{Langacker:1980js}
\bibinfo{author}{\bibfnamefont{P.}~\bibnamefont{Langacker}},
  \bibinfo{journal}{Phys. Rept.} \textbf{\bibinfo{volume}{72}},
  \bibinfo{pages}{185} (\bibinfo{year}{1981}).

\bibitem[{\citenamefont{Fritzsch and Minkowski}(1975)}]{Fritzsch:1974nn}
\bibinfo{author}{\bibfnamefont{H.}~\bibnamefont{Fritzsch}} \bibnamefont{and}
  \bibinfo{author}{\bibfnamefont{P.}~\bibnamefont{Minkowski}},
  \bibinfo{journal}{Annals Phys.} \textbf{\bibinfo{volume}{93}},
  \bibinfo{pages}{193} (\bibinfo{year}{1975}).

\bibitem[{\citenamefont{Pati and Salam}(1974)}]{Pati:1974yy}
\bibinfo{author}{\bibfnamefont{J.~C.} \bibnamefont{Pati}} \bibnamefont{and}
  \bibinfo{author}{\bibfnamefont{A.}~\bibnamefont{Salam}},
  \bibinfo{journal}{Phys. Rev.} \textbf{\bibinfo{volume}{D10}},
  \bibinfo{pages}{275} (\bibinfo{year}{1974}), \bibinfo{note}{[Erratum: Phys.
  Rev.D11,703(1975)]}.

\bibitem[{\citenamefont{Pati and Salam}(1973{\natexlab{a}})}]{Pati:1973uk}
\bibinfo{author}{\bibfnamefont{J.~C.} \bibnamefont{Pati}} \bibnamefont{and}
  \bibinfo{author}{\bibfnamefont{A.}~\bibnamefont{Salam}},
  \bibinfo{journal}{Phys. Rev.} \textbf{\bibinfo{volume}{D8}},
  \bibinfo{pages}{1240} (\bibinfo{year}{1973}{\natexlab{a}}).

\bibitem[{\citenamefont{Pati and Salam}(1973{\natexlab{b}})}]{Pati:1973rp}
\bibinfo{author}{\bibfnamefont{J.~C.} \bibnamefont{Pati}} \bibnamefont{and}
  \bibinfo{author}{\bibfnamefont{A.}~\bibnamefont{Salam}},
  \bibinfo{journal}{Phys. Rev. Lett.} \textbf{\bibinfo{volume}{31}},
  \bibinfo{pages}{661} (\bibinfo{year}{1973}{\natexlab{b}}).

\bibitem[{\citenamefont{Shanker}(1982{\natexlab{a}})}]{Shanker:1981mj}
\bibinfo{author}{\bibfnamefont{O.~U.} \bibnamefont{Shanker}},
  \bibinfo{journal}{Nucl. Phys.} \textbf{\bibinfo{volume}{B206}},
  \bibinfo{pages}{253} (\bibinfo{year}{1982}{\natexlab{a}}).

\bibitem[{\citenamefont{Shanker}(1982{\natexlab{b}})}]{Shanker:1982nd}
\bibinfo{author}{\bibfnamefont{O.~U.} \bibnamefont{Shanker}},
  \bibinfo{journal}{Nucl. Phys.} \textbf{\bibinfo{volume}{B204}},
  \bibinfo{pages}{375} (\bibinfo{year}{1982}{\natexlab{b}}).

\bibitem[{\citenamefont{Schrempp and Schrempp}(1985)}]{Schrempp:1984nj}
\bibinfo{author}{\bibfnamefont{B.}~\bibnamefont{Schrempp}} \bibnamefont{and}
  \bibinfo{author}{\bibfnamefont{F.}~\bibnamefont{Schrempp}},
  \bibinfo{journal}{Phys. Lett.} \textbf{\bibinfo{volume}{153B}},
  \bibinfo{pages}{101} (\bibinfo{year}{1985}).

\bibitem[{\citenamefont{Gripaios}(2010)}]{Gripaios:2009dq}
\bibinfo{author}{\bibfnamefont{B.}~\bibnamefont{Gripaios}},
  \bibinfo{journal}{JHEP} \textbf{\bibinfo{volume}{02}}, \bibinfo{pages}{045}
  (\bibinfo{year}{2010}), \eprint{0910.1789}.

\bibitem[{\citenamefont{Kaplan}(1991)}]{Kaplan:1991dc}
\bibinfo{author}{\bibfnamefont{D.~B.} \bibnamefont{Kaplan}},
  \bibinfo{journal}{Nucl. Phys.} \textbf{\bibinfo{volume}{B365}},
  \bibinfo{pages}{259} (\bibinfo{year}{1991}).

\bibitem[{\citenamefont{Alok et~al.}(2017)\citenamefont{Alok, Bhattacharya,
  Datta, Kumar, Kumar, and London}}]{Alok:2017sui}
\bibinfo{author}{\bibfnamefont{A.~K.} \bibnamefont{Alok}},
  \bibinfo{author}{\bibfnamefont{B.}~\bibnamefont{Bhattacharya}},
  \bibinfo{author}{\bibfnamefont{A.}~\bibnamefont{Datta}},
  \bibinfo{author}{\bibfnamefont{D.}~\bibnamefont{Kumar}},
  \bibinfo{author}{\bibfnamefont{J.}~\bibnamefont{Kumar}}, \bibnamefont{and}
  \bibinfo{author}{\bibfnamefont{D.}~\bibnamefont{London}},
  \bibinfo{journal}{Phys. Rev.} \textbf{\bibinfo{volume}{D96}},
  \bibinfo{pages}{095009} (\bibinfo{year}{2017}), \eprint{1704.07397}.

\bibitem[{\citenamefont{Bečirević and Sumensari}(2017)}]{Becirevic:2017jtw}
\bibinfo{author}{\bibfnamefont{D.}~\bibnamefont{Bečirević}} \bibnamefont{and}
  \bibinfo{author}{\bibfnamefont{O.}~\bibnamefont{Sumensari}},
  \bibinfo{journal}{JHEP} \textbf{\bibinfo{volume}{08}}, \bibinfo{pages}{104}
  (\bibinfo{year}{2017}), \eprint{1704.05835}.

\bibitem[{\citenamefont{Hiller and Nisandzic}(2017)}]{Hiller:2017bzc}
\bibinfo{author}{\bibfnamefont{G.}~\bibnamefont{Hiller}} \bibnamefont{and}
  \bibinfo{author}{\bibfnamefont{I.}~\bibnamefont{Nisandzic}},
  \bibinfo{journal}{Phys. Rev.} \textbf{\bibinfo{volume}{D96}},
  \bibinfo{pages}{035003} (\bibinfo{year}{2017}), \eprint{1704.05444}.

\bibitem[{\citenamefont{D'Amico et~al.}(2017)\citenamefont{D'Amico, Nardecchia,
  Panci, Sannino, Strumia, Torre, and Urbano}}]{DAmico:2017mtc}
\bibinfo{author}{\bibfnamefont{G.}~\bibnamefont{D'Amico}},
  \bibinfo{author}{\bibfnamefont{M.}~\bibnamefont{Nardecchia}},
  \bibinfo{author}{\bibfnamefont{P.}~\bibnamefont{Panci}},
  \bibinfo{author}{\bibfnamefont{F.}~\bibnamefont{Sannino}},
  \bibinfo{author}{\bibfnamefont{A.}~\bibnamefont{Strumia}},
  \bibinfo{author}{\bibfnamefont{R.}~\bibnamefont{Torre}}, \bibnamefont{and}
  \bibinfo{author}{\bibfnamefont{A.}~\bibnamefont{Urbano}},
  \bibinfo{journal}{JHEP} \textbf{\bibinfo{volume}{09}}, \bibinfo{pages}{010}
  (\bibinfo{year}{2017}), \eprint{1704.05438}.

\bibitem[{\citenamefont{Bečirević et~al.}(2016)\citenamefont{Bečirević,
  Fajfer, Košnik, and Sumensari}}]{Becirevic:2016yqi}
\bibinfo{author}{\bibfnamefont{D.}~\bibnamefont{Bečirević}},
  \bibinfo{author}{\bibfnamefont{S.}~\bibnamefont{Fajfer}},
  \bibinfo{author}{\bibfnamefont{N.}~\bibnamefont{Košnik}}, \bibnamefont{and}
  \bibinfo{author}{\bibfnamefont{O.}~\bibnamefont{Sumensari}},
  \bibinfo{journal}{Phys. Rev.} \textbf{\bibinfo{volume}{D94}},
  \bibinfo{pages}{115021} (\bibinfo{year}{2016}), \eprint{1608.08501}.

\bibitem[{\citenamefont{Bauer and Neubert}(2016)}]{Bauer:2015knc}
\bibinfo{author}{\bibfnamefont{M.}~\bibnamefont{Bauer}} \bibnamefont{and}
  \bibinfo{author}{\bibfnamefont{M.}~\bibnamefont{Neubert}},
  \bibinfo{journal}{Phys. Rev. Lett.} \textbf{\bibinfo{volume}{116}},
  \bibinfo{pages}{141802} (\bibinfo{year}{2016}), \eprint{1511.01900}.

\bibitem[{\citenamefont{Li et~al.}(2016)\citenamefont{Li, Yang, and
  Zhang}}]{Li:2016vvp}
\bibinfo{author}{\bibfnamefont{X.-Q.} \bibnamefont{Li}},
  \bibinfo{author}{\bibfnamefont{Y.-D.} \bibnamefont{Yang}}, \bibnamefont{and}
  \bibinfo{author}{\bibfnamefont{X.}~\bibnamefont{Zhang}},
  \bibinfo{journal}{JHEP} \textbf{\bibinfo{volume}{08}}, \bibinfo{pages}{054}
  (\bibinfo{year}{2016}), \eprint{1605.09308}.

\bibitem[{\citenamefont{Calibbi et~al.}(2015)\citenamefont{Calibbi, Crivellin,
  and Ota}}]{Calibbi:2015kma}
\bibinfo{author}{\bibfnamefont{L.}~\bibnamefont{Calibbi}},
  \bibinfo{author}{\bibfnamefont{A.}~\bibnamefont{Crivellin}},
  \bibnamefont{and} \bibinfo{author}{\bibfnamefont{T.}~\bibnamefont{Ota}},
  \bibinfo{journal}{Phys. Rev. Lett.} \textbf{\bibinfo{volume}{115}},
  \bibinfo{pages}{181801} (\bibinfo{year}{2015}), \eprint{1506.02661}.

\bibitem[{\citenamefont{Freytsis et~al.}(2015)\citenamefont{Freytsis, Ligeti,
  and Ruderman}}]{Freytsis:2015qca}
\bibinfo{author}{\bibfnamefont{M.}~\bibnamefont{Freytsis}},
  \bibinfo{author}{\bibfnamefont{Z.}~\bibnamefont{Ligeti}}, \bibnamefont{and}
  \bibinfo{author}{\bibfnamefont{J.~T.} \bibnamefont{Ruderman}},
  \bibinfo{journal}{Phys. Rev.} \textbf{\bibinfo{volume}{D92}},
  \bibinfo{pages}{054018} (\bibinfo{year}{2015}), \eprint{1506.08896}.

\bibitem[{\citenamefont{Dumont et~al.}(2016)\citenamefont{Dumont, Nishiwaki,
  and Watanabe}}]{Dumont:2016xpj}
\bibinfo{author}{\bibfnamefont{B.}~\bibnamefont{Dumont}},
  \bibinfo{author}{\bibfnamefont{K.}~\bibnamefont{Nishiwaki}},
  \bibnamefont{and} \bibinfo{author}{\bibfnamefont{R.}~\bibnamefont{Watanabe}},
  \bibinfo{journal}{Phys. Rev.} \textbf{\bibinfo{volume}{D94}},
  \bibinfo{pages}{034001} (\bibinfo{year}{2016}), \eprint{1603.05248}.

\bibitem[{\citenamefont{Doršner et~al.}(2016)\citenamefont{Doršner, Fajfer,
  Greljo, Kamenik, and Košnik}}]{Dorsner:2016wpm}
\bibinfo{author}{\bibfnamefont{I.}~\bibnamefont{Doršner}},
  \bibinfo{author}{\bibfnamefont{S.}~\bibnamefont{Fajfer}},
  \bibinfo{author}{\bibfnamefont{A.}~\bibnamefont{Greljo}},
  \bibinfo{author}{\bibfnamefont{J.~F.} \bibnamefont{Kamenik}},
  \bibnamefont{and} \bibinfo{author}{\bibfnamefont{N.}~\bibnamefont{Košnik}},
  \bibinfo{journal}{Phys. Rept.} \textbf{\bibinfo{volume}{641}},
  \bibinfo{pages}{1} (\bibinfo{year}{2016}), \eprint{1603.04993}.

\bibitem[{\citenamefont{de~Medeiros~Varzielas and
  Hiller}(2015)}]{Varzielas:2015iva}
\bibinfo{author}{\bibfnamefont{I.}~\bibnamefont{de~Medeiros~Varzielas}}
  \bibnamefont{and} \bibinfo{author}{\bibfnamefont{G.}~\bibnamefont{Hiller}},
  \bibinfo{journal}{JHEP} \textbf{\bibinfo{volume}{06}}, \bibinfo{pages}{072}
  (\bibinfo{year}{2015}), \eprint{1503.01084}.

\bibitem[{\citenamefont{Dorsner et~al.}(2011)\citenamefont{Dorsner, Drobnak,
  Fajfer, Kamenik, and Kosnik}}]{Dorsner:2011ai}
\bibinfo{author}{\bibfnamefont{I.}~\bibnamefont{Dorsner}},
  \bibinfo{author}{\bibfnamefont{J.}~\bibnamefont{Drobnak}},
  \bibinfo{author}{\bibfnamefont{S.}~\bibnamefont{Fajfer}},
  \bibinfo{author}{\bibfnamefont{J.~F.} \bibnamefont{Kamenik}},
  \bibnamefont{and} \bibinfo{author}{\bibfnamefont{N.}~\bibnamefont{Kosnik}},
  \bibinfo{journal}{JHEP} \textbf{\bibinfo{volume}{11}}, \bibinfo{pages}{002}
  (\bibinfo{year}{2011}), \eprint{1107.5393}.

\bibitem[{\citenamefont{Davidson et~al.}(1994)\citenamefont{Davidson, Bailey,
  and Campbell}}]{Davidson:1993qk}
\bibinfo{author}{\bibfnamefont{S.}~\bibnamefont{Davidson}},
  \bibinfo{author}{\bibfnamefont{D.~C.} \bibnamefont{Bailey}},
  \bibnamefont{and} \bibinfo{author}{\bibfnamefont{B.~A.}
  \bibnamefont{Campbell}}, \bibinfo{journal}{Z. Phys.}
  \textbf{\bibinfo{volume}{C61}}, \bibinfo{pages}{613} (\bibinfo{year}{1994}),
  \eprint{hep-ph/9309310}.

\bibitem[{\citenamefont{Saha et~al.}(2010)\citenamefont{Saha, Misra, and
  Kundu}}]{Saha:2010vw}
\bibinfo{author}{\bibfnamefont{J.~P.} \bibnamefont{Saha}},
  \bibinfo{author}{\bibfnamefont{B.}~\bibnamefont{Misra}}, \bibnamefont{and}
  \bibinfo{author}{\bibfnamefont{A.}~\bibnamefont{Kundu}},
  \bibinfo{journal}{Phys. Rev.} \textbf{\bibinfo{volume}{D81}},
  \bibinfo{pages}{095011} (\bibinfo{year}{2010}), \eprint{1003.1384}.

\bibitem[{\citenamefont{Mohanta}(2014)}]{Mohanta:2013lsa}
\bibinfo{author}{\bibfnamefont{R.}~\bibnamefont{Mohanta}},
  \bibinfo{journal}{Phys. Rev.} \textbf{\bibinfo{volume}{D89}},
  \bibinfo{pages}{014020} (\bibinfo{year}{2014}), \eprint{1310.0713}.

\bibitem[{\citenamefont{Sahoo and Mohanta}(2016{\natexlab{a}})}]{Sahoo:2015fla}
\bibinfo{author}{\bibfnamefont{S.}~\bibnamefont{Sahoo}} \bibnamefont{and}
  \bibinfo{author}{\bibfnamefont{R.}~\bibnamefont{Mohanta}},
  \bibinfo{journal}{New J. Phys.} \textbf{\bibinfo{volume}{18}},
  \bibinfo{pages}{013032} (\bibinfo{year}{2016}{\natexlab{a}}),
  \eprint{1509.06248}.

\bibitem[{\citenamefont{Sahoo and Mohanta}(2016{\natexlab{b}})}]{Sahoo:2015pzk}
\bibinfo{author}{\bibfnamefont{S.}~\bibnamefont{Sahoo}} \bibnamefont{and}
  \bibinfo{author}{\bibfnamefont{R.}~\bibnamefont{Mohanta}},
  \bibinfo{journal}{Phys. Rev.} \textbf{\bibinfo{volume}{D93}},
  \bibinfo{pages}{114001} (\bibinfo{year}{2016}{\natexlab{b}}),
  \eprint{1512.04657}.

\bibitem[{\citenamefont{Sahoo and Mohanta}(2016{\natexlab{c}})}]{Sahoo:2015qha}
\bibinfo{author}{\bibfnamefont{S.}~\bibnamefont{Sahoo}} \bibnamefont{and}
  \bibinfo{author}{\bibfnamefont{R.}~\bibnamefont{Mohanta}},
  \bibinfo{journal}{Phys. Rev.} \textbf{\bibinfo{volume}{D93}},
  \bibinfo{pages}{034018} (\bibinfo{year}{2016}{\natexlab{c}}),
  \eprint{1507.02070}.

\bibitem[{\citenamefont{Sahoo and Mohanta}(2015)}]{Sahoo:2015wya}
\bibinfo{author}{\bibfnamefont{S.}~\bibnamefont{Sahoo}} \bibnamefont{and}
  \bibinfo{author}{\bibfnamefont{R.}~\bibnamefont{Mohanta}},
  \bibinfo{journal}{Phys. Rev.} \textbf{\bibinfo{volume}{D91}},
  \bibinfo{pages}{094019} (\bibinfo{year}{2015}), \eprint{1501.05193}.

\bibitem[{\citenamefont{Kosnik}(2012)}]{Kosnik:2012dj}
\bibinfo{author}{\bibfnamefont{N.}~\bibnamefont{Kosnik}},
  \bibinfo{journal}{Phys. Rev.} \textbf{\bibinfo{volume}{D86}},
  \bibinfo{pages}{055004} (\bibinfo{year}{2012}), \eprint{1206.2970}.

\bibitem[{\citenamefont{Chauhan et~al.}(2018)\citenamefont{Chauhan, Kindra, and
  Narang}}]{Chauhan:2017ndd}
\bibinfo{author}{\bibfnamefont{B.}~\bibnamefont{Chauhan}},
  \bibinfo{author}{\bibfnamefont{B.}~\bibnamefont{Kindra}}, \bibnamefont{and}
  \bibinfo{author}{\bibfnamefont{A.}~\bibnamefont{Narang}},
  \bibinfo{journal}{Phys. Rev.} \textbf{\bibinfo{volume}{D97}},
  \bibinfo{pages}{095007} (\bibinfo{year}{2018}), \eprint{1706.04598}.

\bibitem[{\citenamefont{Bečirević et~al.}(2018)\citenamefont{Bečirević,
  Doršner, Fajfer, Košnik, Faroughy, and Sumensari}}]{Becirevic:2018afm}
\bibinfo{author}{\bibfnamefont{D.}~\bibnamefont{Bečirević}},
  \bibinfo{author}{\bibfnamefont{I.}~\bibnamefont{Doršner}},
  \bibinfo{author}{\bibfnamefont{S.}~\bibnamefont{Fajfer}},
  \bibinfo{author}{\bibfnamefont{N.}~\bibnamefont{Košnik}},
  \bibinfo{author}{\bibfnamefont{D.~A.} \bibnamefont{Faroughy}},
  \bibnamefont{and}
  \bibinfo{author}{\bibfnamefont{O.}~\bibnamefont{Sumensari}},
  \bibinfo{journal}{Phys. Rev.} \textbf{\bibinfo{volume}{D98}},
  \bibinfo{pages}{055003} (\bibinfo{year}{2018}), \eprint{1806.05689}.

\bibitem[{\citenamefont{Angelescu et~al.}(2018)\citenamefont{Angelescu,
  Bečirević, Faroughy, and Sumensari}}]{Angelescu:2018tyl}
\bibinfo{author}{\bibfnamefont{A.}~\bibnamefont{Angelescu}},
  \bibinfo{author}{\bibfnamefont{D.}~\bibnamefont{Bečirević}},
  \bibinfo{author}{\bibfnamefont{D.~A.} \bibnamefont{Faroughy}},
  \bibnamefont{and}
  \bibinfo{author}{\bibfnamefont{O.}~\bibnamefont{Sumensari}},
  \bibinfo{journal}{JHEP} \textbf{\bibinfo{volume}{10}}, \bibinfo{pages}{183}
  (\bibinfo{year}{2018}), \eprint{1808.08179}.

\bibitem[{\citenamefont{Singirala et~al.}(2017)\citenamefont{Singirala,
  Mohanta, Patra, and Rao}}]{Singirala:2017cch}
\bibinfo{author}{\bibfnamefont{S.}~\bibnamefont{Singirala}},
  \bibinfo{author}{\bibfnamefont{R.}~\bibnamefont{Mohanta}},
  \bibinfo{author}{\bibfnamefont{S.}~\bibnamefont{Patra}}, \bibnamefont{and}
  \bibinfo{author}{\bibfnamefont{S.}~\bibnamefont{Rao}} (\bibinfo{year}{2017}),
  \eprint{1710.05775}.

\bibitem[{\citenamefont{Nanda and Borah}(2017)}]{Nanda:2017bmi}
\bibinfo{author}{\bibfnamefont{D.}~\bibnamefont{Nanda}} \bibnamefont{and}
  \bibinfo{author}{\bibfnamefont{D.}~\bibnamefont{Borah}},
  \bibinfo{journal}{Phys. Rev.} \textbf{\bibinfo{volume}{D96}},
  \bibinfo{pages}{115014} (\bibinfo{year}{2017}), \eprint{1709.08417}.

\bibitem[{\citenamefont{Aghanim et~al.}(2018)}]{Aghanim:2018eyx}
\bibinfo{author}{\bibfnamefont{N.}~\bibnamefont{Aghanim}} \bibnamefont{et~al.}
  (\bibinfo{collaboration}{Planck}) (\bibinfo{year}{2018}),
  \eprint{1807.06209}.

\bibitem[{\citenamefont{Agrawal et~al.}(2010)\citenamefont{Agrawal, Chacko,
  Kilic, and Mishra}}]{Agrawal:2010fh}
\bibinfo{author}{\bibfnamefont{P.}~\bibnamefont{Agrawal}},
  \bibinfo{author}{\bibfnamefont{Z.}~\bibnamefont{Chacko}},
  \bibinfo{author}{\bibfnamefont{C.}~\bibnamefont{Kilic}}, \bibnamefont{and}
  \bibinfo{author}{\bibfnamefont{R.~K.} \bibnamefont{Mishra}}
  (\bibinfo{year}{2010}), \eprint{1003.1912}.

\bibitem[{\citenamefont{Amole et~al.}(2017)}]{Amole:2017dex}
\bibinfo{author}{\bibfnamefont{C.}~\bibnamefont{Amole}} \bibnamefont{et~al.}
  (\bibinfo{collaboration}{PICO}), \bibinfo{journal}{Phys. Rev. Lett.}
  \textbf{\bibinfo{volume}{118}}, \bibinfo{pages}{251301}
  (\bibinfo{year}{2017}), \eprint{1702.07666}.

\bibitem[{\citenamefont{Akerib et~al.}(2017{\natexlab{a}})}]{Akerib:2017kat}
\bibinfo{author}{\bibfnamefont{D.~S.} \bibnamefont{Akerib}}
  \bibnamefont{et~al.} (\bibinfo{collaboration}{LUX}), \bibinfo{journal}{Phys.
  Rev. Lett.} \textbf{\bibinfo{volume}{118}}, \bibinfo{pages}{251302}
  (\bibinfo{year}{2017}{\natexlab{a}}), \eprint{1705.03380}.

\bibitem[{\citenamefont{Ibarra et~al.}(2016)\citenamefont{Ibarra, Yaguna, and
  Zapata}}]{Ibarra:2016dlb}
\bibinfo{author}{\bibfnamefont{A.}~\bibnamefont{Ibarra}},
  \bibinfo{author}{\bibfnamefont{C.~E.} \bibnamefont{Yaguna}},
  \bibnamefont{and} \bibinfo{author}{\bibfnamefont{O.}~\bibnamefont{Zapata}},
  \bibinfo{journal}{Phys. Rev.} \textbf{\bibinfo{volume}{D93}},
  \bibinfo{pages}{035012} (\bibinfo{year}{2016}), \eprint{1601.01163}.

\bibitem[{\citenamefont{Herrero-Garcia
  et~al.}(2018)\citenamefont{Herrero-Garcia, Molinaro, and
  Schmidt}}]{Herrero-Garcia:2018koq}
\bibinfo{author}{\bibfnamefont{J.}~\bibnamefont{Herrero-Garcia}},
  \bibinfo{author}{\bibfnamefont{E.}~\bibnamefont{Molinaro}}, \bibnamefont{and}
  \bibinfo{author}{\bibfnamefont{M.~A.} \bibnamefont{Schmidt}},
  \bibinfo{journal}{Eur. Phys. J.} \textbf{\bibinfo{volume}{C78}},
  \bibinfo{pages}{471} (\bibinfo{year}{2018}), \eprint{1803.05660}.

\bibitem[{\citenamefont{Cui et~al.}(2017)}]{Cui:2017nnn}
\bibinfo{author}{\bibfnamefont{X.}~\bibnamefont{Cui}} \bibnamefont{et~al.}
  (\bibinfo{collaboration}{PandaX-II}), \bibinfo{journal}{Phys. Rev. Lett.}
  \textbf{\bibinfo{volume}{119}}, \bibinfo{pages}{181302}
  (\bibinfo{year}{2017}), \eprint{1708.06917}.

\bibitem[{\citenamefont{Aprile et~al.}(2017)}]{Aprile:2017iyp}
\bibinfo{author}{\bibfnamefont{E.}~\bibnamefont{Aprile}} \bibnamefont{et~al.}
  (\bibinfo{collaboration}{XENON}) (\bibinfo{year}{2017}), \eprint{1705.06655}.

\bibitem[{\citenamefont{Akerib et~al.}(2017{\natexlab{b}})}]{Akerib:2016vxi}
\bibinfo{author}{\bibfnamefont{D.~S.} \bibnamefont{Akerib}}
  \bibnamefont{et~al.} (\bibinfo{collaboration}{LUX}), \bibinfo{journal}{Phys.
  Rev. Lett.} \textbf{\bibinfo{volume}{118}}, \bibinfo{pages}{021303}
  (\bibinfo{year}{2017}{\natexlab{b}}), \eprint{1608.07648}.

\bibitem[{\citenamefont{Ma}(2006)}]{Ma:2006km}
\bibinfo{author}{\bibfnamefont{E.}~\bibnamefont{Ma}}, \bibinfo{journal}{Phys.
  Rev.} \textbf{\bibinfo{volume}{D73}}, \bibinfo{pages}{077301}
  (\bibinfo{year}{2006}), \eprint{hep-ph/0601225}.

\bibitem[{\citenamefont{Capozzi et~al.}(2016)\citenamefont{Capozzi, Lisi,
  Marrone, Montanino, and Palazzo}}]{Capozzi:2016rtj}
\bibinfo{author}{\bibfnamefont{F.}~\bibnamefont{Capozzi}},
  \bibinfo{author}{\bibfnamefont{E.}~\bibnamefont{Lisi}},
  \bibinfo{author}{\bibfnamefont{A.}~\bibnamefont{Marrone}},
  \bibinfo{author}{\bibfnamefont{D.}~\bibnamefont{Montanino}},
  \bibnamefont{and} \bibinfo{author}{\bibfnamefont{A.}~\bibnamefont{Palazzo}},
  \bibinfo{journal}{Nucl. Phys.} \textbf{\bibinfo{volume}{B908}},
  \bibinfo{pages}{218} (\bibinfo{year}{2016}), \eprint{1601.07777}.

\bibitem[{\citenamefont{Ade et~al.}(2016)}]{Ade:2015xua}
\bibinfo{author}{\bibfnamefont{P.~A.~R.} \bibnamefont{Ade}}
  \bibnamefont{et~al.} (\bibinfo{collaboration}{Planck}),
  \bibinfo{journal}{Astron. Astrophys.} \textbf{\bibinfo{volume}{594}},
  \bibinfo{pages}{A13} (\bibinfo{year}{2016}), \eprint{1502.01589}.

\bibitem[{\citenamefont{Bobeth et~al.}(2000)\citenamefont{Bobeth, Misiak, and
  Urban}}]{Bobeth:1999mk}
\bibinfo{author}{\bibfnamefont{C.}~\bibnamefont{Bobeth}},
  \bibinfo{author}{\bibfnamefont{M.}~\bibnamefont{Misiak}}, \bibnamefont{and}
  \bibinfo{author}{\bibfnamefont{J.}~\bibnamefont{Urban}},
  \bibinfo{journal}{Nucl. Phys.} \textbf{\bibinfo{volume}{B574}},
  \bibinfo{pages}{291} (\bibinfo{year}{2000}), \eprint{hep-ph/9910220}.

\bibitem[{\citenamefont{Bobeth et~al.}(2002)\citenamefont{Bobeth, Buras,
  Kruger, and Urban}}]{Bobeth:2001jm}
\bibinfo{author}{\bibfnamefont{C.}~\bibnamefont{Bobeth}},
  \bibinfo{author}{\bibfnamefont{A.~J.} \bibnamefont{Buras}},
  \bibinfo{author}{\bibfnamefont{F.}~\bibnamefont{Kruger}}, \bibnamefont{and}
  \bibinfo{author}{\bibfnamefont{J.}~\bibnamefont{Urban}},
  \bibinfo{journal}{Nucl. Phys.} \textbf{\bibinfo{volume}{B630}},
  \bibinfo{pages}{87} (\bibinfo{year}{2002}), \eprint{hep-ph/0112305}.

\bibitem[{\citenamefont{Hou et~al.}(2014)\citenamefont{Hou, Kohda, and
  Xu}}]{Hou:2014dza}
\bibinfo{author}{\bibfnamefont{W.-S.} \bibnamefont{Hou}},
  \bibinfo{author}{\bibfnamefont{M.}~\bibnamefont{Kohda}}, \bibnamefont{and}
  \bibinfo{author}{\bibfnamefont{F.}~\bibnamefont{Xu}}, \bibinfo{journal}{Phys.
  Rev.} \textbf{\bibinfo{volume}{D90}}, \bibinfo{pages}{013002}
  (\bibinfo{year}{2014}), \eprint{1403.7410}.

\bibitem[{\citenamefont{Inami and Lim}(1981)}]{Inami:1980fz}
\bibinfo{author}{\bibfnamefont{T.}~\bibnamefont{Inami}} \bibnamefont{and}
  \bibinfo{author}{\bibfnamefont{C.~S.} \bibnamefont{Lim}},
  \bibinfo{journal}{Prog. Theor. Phys.} \textbf{\bibinfo{volume}{65}},
  \bibinfo{pages}{297} (\bibinfo{year}{1981}), \bibinfo{note}{[Erratum: Prog.
  Theor. Phys.65,1772(1981)]}.

\bibitem[{\citenamefont{Tanabashi et~al.}(2018)}]{Tanabashi:2018oca}
\bibinfo{author}{\bibfnamefont{M.}~\bibnamefont{Tanabashi}}
  \bibnamefont{et~al.} (\bibinfo{collaboration}{Particle Data Group}),
  \bibinfo{journal}{Phys. Rev.} \textbf{\bibinfo{volume}{D98}},
  \bibinfo{pages}{030001} (\bibinfo{year}{2018}).

\bibitem[{\citenamefont{Charles et~al.}(2015)}]{Charles:2015gya}
\bibinfo{author}{\bibfnamefont{J.}~\bibnamefont{Charles}} \bibnamefont{et~al.},
  \bibinfo{journal}{Phys. Rev.} \textbf{\bibinfo{volume}{D91}},
  \bibinfo{pages}{073007} (\bibinfo{year}{2015}), \eprint{1501.05013}.

\bibitem[{\citenamefont{Ball and Zwicky}(2005)}]{Ball:2004ye}
\bibinfo{author}{\bibfnamefont{P.}~\bibnamefont{Ball}} \bibnamefont{and}
  \bibinfo{author}{\bibfnamefont{R.}~\bibnamefont{Zwicky}},
  \bibinfo{journal}{Phys. Rev.} \textbf{\bibinfo{volume}{D71}},
  \bibinfo{pages}{014015} (\bibinfo{year}{2005}), \eprint{hep-ph/0406232}.

\bibitem[{\citenamefont{Hisano et~al.}(1996)\citenamefont{Hisano, Moroi, Tobe,
  and Yamaguchi}}]{Hisano:1995cp}
\bibinfo{author}{\bibfnamefont{J.}~\bibnamefont{Hisano}},
  \bibinfo{author}{\bibfnamefont{T.}~\bibnamefont{Moroi}},
  \bibinfo{author}{\bibfnamefont{K.}~\bibnamefont{Tobe}}, \bibnamefont{and}
  \bibinfo{author}{\bibfnamefont{M.}~\bibnamefont{Yamaguchi}},
  \bibinfo{journal}{Phys. Rev.} \textbf{\bibinfo{volume}{D53}},
  \bibinfo{pages}{2442} (\bibinfo{year}{1996}), \eprint{hep-ph/9510309}.

\bibitem[{\citenamefont{Colangelo et~al.}(1997)\citenamefont{Colangelo,
  De~Fazio, Santorelli, and Scrimieri}}]{Colangelo:1996ay}
\bibinfo{author}{\bibfnamefont{P.}~\bibnamefont{Colangelo}},
  \bibinfo{author}{\bibfnamefont{F.}~\bibnamefont{De~Fazio}},
  \bibinfo{author}{\bibfnamefont{P.}~\bibnamefont{Santorelli}},
  \bibnamefont{and}
  \bibinfo{author}{\bibfnamefont{E.}~\bibnamefont{Scrimieri}},
  \bibinfo{journal}{Phys. Lett.} \textbf{\bibinfo{volume}{B395}},
  \bibinfo{pages}{339} (\bibinfo{year}{1997}), \eprint{hep-ph/9610297}.

\bibitem[{\citenamefont{Amhis et~al.}(2017)}]{Amhis:2016xyh}
\bibinfo{author}{\bibfnamefont{Y.}~\bibnamefont{Amhis}} \bibnamefont{et~al.}
  (\bibinfo{collaboration}{HFLAV}), \bibinfo{journal}{Eur. Phys. J.}
  \textbf{\bibinfo{volume}{C77}}, \bibinfo{pages}{895} (\bibinfo{year}{2017}),
  \eprint{1612.07233}.

\bibitem[{\citenamefont{Misiak et~al.}(2015)}]{Misiak:2015xwa}
\bibinfo{author}{\bibfnamefont{M.}~\bibnamefont{Misiak}} \bibnamefont{et~al.},
  \bibinfo{journal}{Phys. Rev. Lett.} \textbf{\bibinfo{volume}{114}},
  \bibinfo{pages}{221801} (\bibinfo{year}{2015}), \eprint{1503.01789}.

\bibitem[{\citenamefont{Altmannshofer
  et~al.}(2014{\natexlab{a}})\citenamefont{Altmannshofer, Gori, Pospelov, and
  Yavin}}]{Altmannshofer:2014cfa}
\bibinfo{author}{\bibfnamefont{W.}~\bibnamefont{Altmannshofer}},
  \bibinfo{author}{\bibfnamefont{S.}~\bibnamefont{Gori}},
  \bibinfo{author}{\bibfnamefont{M.}~\bibnamefont{Pospelov}}, \bibnamefont{and}
  \bibinfo{author}{\bibfnamefont{I.}~\bibnamefont{Yavin}},
  \bibinfo{journal}{Phys. Rev.} \textbf{\bibinfo{volume}{D89}},
  \bibinfo{pages}{095033} (\bibinfo{year}{2014}{\natexlab{a}}),
  \eprint{1403.1269}.

\bibitem[{\citenamefont{Mishra et~al.}(1991)}]{Mishra:1991bv}
\bibinfo{author}{\bibfnamefont{S.~R.} \bibnamefont{Mishra}}
  \bibnamefont{et~al.} (\bibinfo{collaboration}{CCFR}), \bibinfo{journal}{Phys.
  Rev. Lett.} \textbf{\bibinfo{volume}{66}}, \bibinfo{pages}{3117}
  (\bibinfo{year}{1991}).

\bibitem[{\citenamefont{Altmannshofer
  et~al.}(2014{\natexlab{b}})\citenamefont{Altmannshofer, Gori, Pospelov, and
  Yavin}}]{Altmannshofer:2014pba}
\bibinfo{author}{\bibfnamefont{W.}~\bibnamefont{Altmannshofer}},
  \bibinfo{author}{\bibfnamefont{S.}~\bibnamefont{Gori}},
  \bibinfo{author}{\bibfnamefont{M.}~\bibnamefont{Pospelov}}, \bibnamefont{and}
  \bibinfo{author}{\bibfnamefont{I.}~\bibnamefont{Yavin}},
  \bibinfo{journal}{Phys. Rev. Lett.} \textbf{\bibinfo{volume}{113}},
  \bibinfo{pages}{091801} (\bibinfo{year}{2014}{\natexlab{b}}),
  \eprint{1406.2332}.

\bibitem[{\citenamefont{Ali et~al.}(2000)\citenamefont{Ali, Ball, Handoko, and
  Hiller}}]{Ali:1999mm}
\bibinfo{author}{\bibfnamefont{A.}~\bibnamefont{Ali}},
  \bibinfo{author}{\bibfnamefont{P.}~\bibnamefont{Ball}},
  \bibinfo{author}{\bibfnamefont{L.~T.} \bibnamefont{Handoko}},
  \bibnamefont{and} \bibinfo{author}{\bibfnamefont{G.}~\bibnamefont{Hiller}},
  \bibinfo{journal}{Phys. Rev.} \textbf{\bibinfo{volume}{D61}},
  \bibinfo{pages}{074024} (\bibinfo{year}{2000}), \eprint{hep-ph/9910221}.

\bibitem[{\citenamefont{Wirbel et~al.}(1985)\citenamefont{Wirbel, Stech, and
  Bauer}}]{Wirbel:1985ji}
\bibinfo{author}{\bibfnamefont{M.}~\bibnamefont{Wirbel}},
  \bibinfo{author}{\bibfnamefont{B.}~\bibnamefont{Stech}}, \bibnamefont{and}
  \bibinfo{author}{\bibfnamefont{M.}~\bibnamefont{Bauer}}, \bibinfo{journal}{Z.
  Phys.} \textbf{\bibinfo{volume}{C29}}, \bibinfo{pages}{637}
  (\bibinfo{year}{1985}).

\bibitem[{\citenamefont{Bobeth et~al.}(2008)\citenamefont{Bobeth, Hiller, and
  Piranishvili}}]{Bobeth:2008ij}
\bibinfo{author}{\bibfnamefont{C.}~\bibnamefont{Bobeth}},
  \bibinfo{author}{\bibfnamefont{G.}~\bibnamefont{Hiller}}, \bibnamefont{and}
  \bibinfo{author}{\bibfnamefont{G.}~\bibnamefont{Piranishvili}},
  \bibinfo{journal}{JHEP} \textbf{\bibinfo{volume}{07}}, \bibinfo{pages}{106}
  (\bibinfo{year}{2008}), \eprint{0805.2525}.

\bibitem[{\citenamefont{Egede et~al.}(2010)\citenamefont{Egede, Hurth, Matias,
  Ramon, and Reece}}]{Egede:2010zc}
\bibinfo{author}{\bibfnamefont{U.}~\bibnamefont{Egede}},
  \bibinfo{author}{\bibfnamefont{T.}~\bibnamefont{Hurth}},
  \bibinfo{author}{\bibfnamefont{J.}~\bibnamefont{Matias}},
  \bibinfo{author}{\bibfnamefont{M.}~\bibnamefont{Ramon}}, \bibnamefont{and}
  \bibinfo{author}{\bibfnamefont{W.}~\bibnamefont{Reece}},
  \bibinfo{journal}{JHEP} \textbf{\bibinfo{volume}{10}}, \bibinfo{pages}{056}
  (\bibinfo{year}{2010}), \eprint{1005.0571}.

\bibitem[{\citenamefont{Egede et~al.}(2008)\citenamefont{Egede, Hurth, Matias,
  Ramon, and Reece}}]{Egede:2008uy}
\bibinfo{author}{\bibfnamefont{U.}~\bibnamefont{Egede}},
  \bibinfo{author}{\bibfnamefont{T.}~\bibnamefont{Hurth}},
  \bibinfo{author}{\bibfnamefont{J.}~\bibnamefont{Matias}},
  \bibinfo{author}{\bibfnamefont{M.}~\bibnamefont{Ramon}}, \bibnamefont{and}
  \bibinfo{author}{\bibfnamefont{W.}~\bibnamefont{Reece}},
  \bibinfo{journal}{JHEP} \textbf{\bibinfo{volume}{11}}, \bibinfo{pages}{032}
  (\bibinfo{year}{2008}), \eprint{0807.2589}.

\bibitem[{\citenamefont{Altmannshofer et~al.}(2009)\citenamefont{Altmannshofer,
  Ball, Bharucha, Buras, Straub, and Wick}}]{Altmannshofer:2008dz}
\bibinfo{author}{\bibfnamefont{W.}~\bibnamefont{Altmannshofer}},
  \bibinfo{author}{\bibfnamefont{P.}~\bibnamefont{Ball}},
  \bibinfo{author}{\bibfnamefont{A.}~\bibnamefont{Bharucha}},
  \bibinfo{author}{\bibfnamefont{A.~J.} \bibnamefont{Buras}},
  \bibinfo{author}{\bibfnamefont{D.~M.} \bibnamefont{Straub}},
  \bibnamefont{and} \bibinfo{author}{\bibfnamefont{M.}~\bibnamefont{Wick}},
  \bibinfo{journal}{JHEP} \textbf{\bibinfo{volume}{01}}, \bibinfo{pages}{019}
  (\bibinfo{year}{2009}), \eprint{0811.1214}.

\bibitem[{\citenamefont{Descotes-Genon
  et~al.}(2013)\citenamefont{Descotes-Genon, Matias, Ramon, and
  Virto}}]{DescotesGenon:2012zf}
\bibinfo{author}{\bibfnamefont{S.}~\bibnamefont{Descotes-Genon}},
  \bibinfo{author}{\bibfnamefont{J.}~\bibnamefont{Matias}},
  \bibinfo{author}{\bibfnamefont{M.}~\bibnamefont{Ramon}}, \bibnamefont{and}
  \bibinfo{author}{\bibfnamefont{J.}~\bibnamefont{Virto}},
  \bibinfo{journal}{JHEP} \textbf{\bibinfo{volume}{01}}, \bibinfo{pages}{048}
  (\bibinfo{year}{2013}), \eprint{1207.2753}.

\end{thebibliography}

\end{document}